\documentclass[rmp,twocolumn,groupaddress,showpacs]{revtex4}
\usepackage{epsfig,amsmath,amssymb,graphics,color,calc}

\newcommand{\be}{\begin{equation}}
\newcommand{\ee}{\end{equation}}
\newcommand{\ba}{\begin{eqnarray}}
\newcommand{\ea}{\end{eqnarray}}

\newcommand{\w}{_{\mathrm{w}}}
\newcommand{\tw}{t\w}
\newcommand{\Chi}{\chi}

\newcommand{\bq}{{\bf q}} 
\newcommand{\bx}{{\bf x}} 
\newcommand{\by}{{\bf y}}

\begin{document}

\title{Theoretical perspective on the glass transition 
and amorphous materials}

\author{Ludovic Berthier}
\affiliation{Laboratoire Charles Coulomb, UMR 5521
CNRS and Universit\'e Montpellier 2, Montpellier, France}

\author{Giulio Biroli}
\affiliation{Institut de Physique 
Th\'eorique, CEA, IPhT, 91191 Gif sur Yvette, France \&
CNRS URA 2306}

\date{\today}

\begin{abstract}
We provide a theoretical perspective on 
the glass transition in molecular liquids at thermal equilibrium, 
on the spatially heterogeneous and aging dynamics 
of disordered materials, and on the rheology of soft glassy materials.
We start with a broad introduction to the field and 
emphasize its connections with other subjects and its relevance.  
The important role played
by computer simulations to study and understand the dynamics 
of systems close to the glass transition at the molecular level 
is spelled out. We review the recent progress 
on the subject of the spatially heterogeneous dynamics 
that characterizes structural relaxation in materials with slow dynamics.
We then present the main theoretical approaches 
describing the glass transition in supercooled liquids, focusing  
on theories that have a microscopic, statistical mechanics basis. 
We describe both successes and 
failures, and critically assess the current status of each of these 
approaches. The physics of aging dynamics in disordered materials
and the rheology of soft glassy materials are then discussed, 
and recent theoretical progress is described.   
For each section, we give an extensive overview of the most recent 
advances, but we also describe in some detail the important open problems
that, we believe, will occupy a central place 
in this field in the coming years. 
\end{abstract}



\maketitle

\tableofcontents

\newpage

\section{Incipit}

Glasses belong to a well-known state of matter~\cite{tabor}: we easily design 
glasses with desired mechanical or optical properties on an 
industrial scale, they are widely present in our daily life. 
Yet, a deep microscopic understanding 
of the glassy state of matter remains a challenge for condensed matter
physicists~\cite{angellscience,reviewnature}. 

Glasses share similarities with crystalline
solids since they are both mechanically rigid, but also with liquids because 
they both have similar disordered structures at the molecular level.
It is mainly this mixed character that makes them fascinating
even to non-scientists~\cite{zanotto}. 
Given that glasses are neither normal liquids nor standard solids, they 
are quite often not described in any detail by 
standard textbooks. For instance, 
glasses are not described in textbooks on condensed 
matter~\cite{chaikin-book}, or solid state physics~\cite{ashcroft}, 
they only made it in the latest edition 
of the reference textbook on liquids~\cite{hansen}, while 
statistical mechanics textbooks usually culminate with a presentation
of our current understanding of phase transitions in pure 
materials using renormalization group concepts 
\cite{chandler-green,sethna}, leaving out disordered systems.

As we shall describe in detail in this review, modern
statistical mechanics approaches to the glass transition
involve good knowledge of 
advanced liquid state theory, field theory,  
renormalization group, solution of lattice models, percolation,
replica calculations, and concepts developed for far-from-equilibrium, 
driven systems~\cite{youngbook,leszouches,binderkob}. 
These development are all posterior to the mid-70's important breakthroughs 
on phase transitions: the canonical spin glass Hamiltonian 
was introduced in 1975 ~\cite{ea}, to be solved in infinite
dimension only several years later \cite{parisi80,beyond}, mode-coupling theory
was developed in the mid-80's \cite{gotze}, just before 
kinetic lattice glass models were introduced~\cite{FA}. 
The aging and rheology of disordered systems
such as spin glasses or soft materials emerged as broad research
fields during the 90's.  In this paper, 
we shall review the fruits that have grown
out of these important seeds. Given none of these advances 
has allowed the derivation of a complete, well-accepted
theory of amorphous media, we present a large number of different
approaches. We try to discuss both successes and failures, we
explain similarities and differences between them,  
and we present the current status of each approach. Thus, the article 
takes at times a somewhat subjective tone.

A glass can be obtained by cooling the temperature of a liquid 
below its glass temperature, $T_g$. The quench must be fast enough 
that the more standard first order phase transition towards the crystalline
phase is avoided. The glass `transition' is not a thermodynamic 
transition at all, since $T_g$ is only empirically defined as the temperature
below which the material has become too viscous to flow on a `reasonable' 
timescale (and it is hard to define the word `reasonable' in
any reasonable manner). Therefore, $T_g$ does not play a fundamental role, as 
a phase transition temperature would. 
It is simply the temperature below which the material looks solid. 
When quenched in the glass phase below $T_g$, liquids slowly 
evolve towards an equilibrium state they cannot reach on experimental 
timescales. Physical properties  are then found to evolve slowly
with time in far from equilibrium states, a process known as 
`aging'~\cite{struik}.  

\begin{figure}
\psfig{file=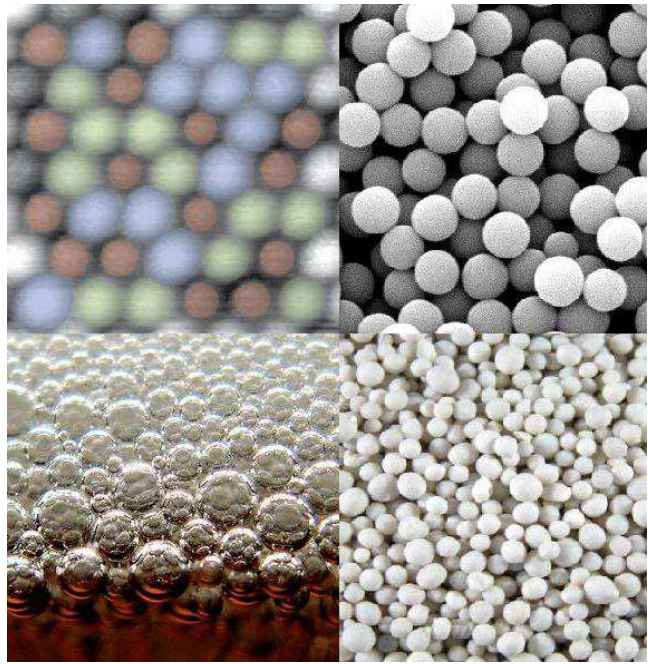,width=8.5cm}
\caption{\label{joli} 
Glassy phases occur at low temperature or large density
in many different systems spanning a broad range of lengthscales, such as 
atomic (top left, atomic force spectroscopy image of an alloy linear size 
4.3 nm~\cite{sugimoto}), colloidal (top right) systems, 
foams (bottom left, a beer foam with bubbles of 
submillimeter size) and 
granular materials (bottom right, a fertilizer made of 
millimeter size grains).}
\end{figure}

The subject of the glass transition has quite broad implications. 
A material is said 
to be `glassy' when its typical relaxation timescale becomes of the order of,
and often much larger than, the typical duration of an experiment
or a numerical simulation. With this generic definition, a large number
of systems can be considered as glassy materials~\cite{youngbook}.
One can be interested in the physics of liquids (window glasses are then
the archetype), in `hard' condensed matter (for instance 
type II superconductors in the presence of disorder such as
high-$T_c$ superconducting materials), in charge density waves or spin glasses, 
in `soft' condensed matter with numerous complex fluids such as
colloidal assemblies, emulsions, foams, but also granular materials, proteins, 
etc. Glass physics thus covers a remarkably broad range of 
timescales and lengthscales, as illustrated in Fig.~\ref{joli}.
All these materials exhibit, in some part of their phase diagrams, 
some sort of glassy dynamics characterized by a very rich 
phenomenology with effects such as aging, hysteresis, creep, memory, 
effective temperatures, 
rejuvenation, dynamic heterogeneity, non-linear response, etc.  
 
These long enumerations explain why this research field has received
increasing attention from physicists in the last two decades.   
`Glassy' topics now go much beyond the physics of simple liquids (glass
transition physics) and models and concepts developed for one
system often find applications elsewhere in physics, from 
algorithmics to biophysics~\cite{complexbook}. 
Motivations to study glassy materials 
are numerous. Glassy materials are everywhere around us 
and therefore obviously attract interest beyond academic research. 
At the same time, the glass conundrum 
provides theoretical physicists with 
deep fundamental questions since standard statistical mechanics tools are 
sometimes  not sufficient to properly account for the glass state. 
Additionally, simulating in the computer the dynamics of microscopically 
realistic material on timescales that are experimentally relevant 
is not an easy task, even with modern computers. 
Finally, the field is constantly stimulated by
new, and sometimes quite beautiful, experimental developments 
to produce new types of disordered materials, or to 
obtain more microscopic information on the structure and
dynamics of glassy systems.

The outline of the article is as follows. Sec.~\ref{broadintro}
provides a broad introduction to glassy materials.
The issue of dynamic heterogeneity is tackled in Sec.~\ref{dh}, while
the main theoretical perspectives, characterized by a 
microscopic, statistical mechanics basis 
are summarized in Sec.~\ref{theory}.
Aging and nonequilibrium phenomena occupy Sec.~\ref{aging}.
Finally, we present a set of general and concluding remarks
in Sec.~\ref{nofuture}.  

\section{A broad introduction about glasses}

\label{broadintro}

\subsection{Some phenomenology}

\label{phenomenology}

\subsubsection{The basic phenomenon}

A vast majority of liquids (molecular liquids, polymeric liquids, etc.)
form a glass if cooled fast enough in order to avoid
the crystallization transition~\cite{angellscience}. 
Typical values of cooling rate 
in laboratory experiments are $0.1-100$~K/min. 
The metastable phase reached in this way is called `supercooled phase'.
In this regime the typical timescales increase in a dramatic way
and they end up being many orders of magnitudes larger than microscopic
timescales at $T_g$, the glass transition temperature. 

 For example, around the melting temperature $T_m$,
the typical timescale $\tau_\alpha$ on which density fluctuations relax,
is of the order of  $\sqrt{m a^2/k_BT}$, which corresponds to few picoseconds ($m$ is the molecular
mass, $T$ the temperature, $k_B$ the Boltzmann constant [which will often 
be set to unity in the later theoretical sections], and $a$ a 
typical distance between molecules).     
At $T_g$ the typical timescale has become of the 
order of $100$~s, i.e. $14$ orders of magnitude larger!
This increase is even more remarkable because the corresponding 
temperature decrease is, as a rule of thumb, about $\frac{1}{3} T_m$--- 
not a large value if one considers $k_BT_m$ as the typical energy scale. 
The  increasing of $\tau_\alpha$ is accompanied by a concomitant increase
of the shear viscosity $\eta$. This can be understood by a 
simple Maxwell model in which 
$\eta$ and $\tau_\alpha$ are related by $\eta=G_{\infty}\tau_\alpha$, where
$G_{\infty}$ is the instantaneous (elastic) shear modulus which does not vary
considerably in the supercooled regime. In fact,
viscosities at the glass transition temperature are of the order of 
$10^{12}$~Pa.s.
In order to grasp  how viscous this is, recall 
that the typical viscosity 
of water at ambient temperature is of the order of $10^{-2}$ Pa.s. 
How long would one have to wait to drink a glass of 
water with a viscosity $10^{14}$ times larger, 
or how long would it take for cathedral glasses to flow~\cite{zanotto}?

As a matter of fact,
the temperature at which the liquid does not flow anymore and becomes an
amorphous solid, called a `glass', is protocol-dependent. It depends 
on the cooling 
rate and on the patience of the people carrying out the experiment: 
solidity is 
a timescale-dependent notion~\cite{SaussetBiroliKurchan}. 
Pragmatically, $T_g$ is defined as the 
temperature
at which the shear viscosity is equal to $10^{13}$ Poise (also 
$10^{12}$~Pa.s).

\subsubsection{Kinetic fragility}

\begin{figure}
\psfig{file=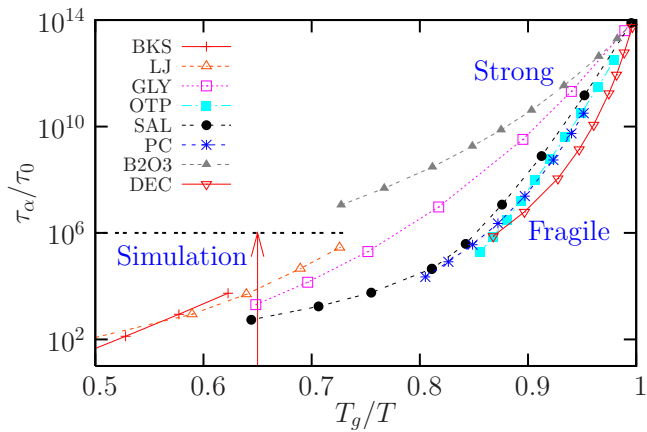,width=8.5cm}
\caption{\label{angellfig} Arrhenius plot of the relaxation 
time of several glass-forming liquids approaching the glass 
temperature $T_g$.
For `strong' glasses, $\tau_\alpha$ increases in an Arrhenius 
manner as temperature is decreased, $\log \tau_\alpha 
\sim E/(k_B T)$, where
$E$ is an activation energy and the plot is a straight line. 
For `fragile' liquids, the plot is bent and the effective 
activation energy increases when $T$ is decreased towards $T_g$.
Timescales accessible to numerical simulations are indicated.
BKS: numerical model of silica;
LJ: numerical model of a binary Lennard-Jones mixture;
GLY: glycerol; OTP: ortho-terphenyl; SAL: salol; PC: propylene carbonate;
DEC: decaline.}
\end{figure}

The increase of the relaxation timescale of supercooled liquids is remarkable
not only because of the large number of decades involved but 
also because of its 
temperature dependence. This is vividly demonstrated by
plotting the logarithm of the 
viscosity (or the relaxation time) as a function 
of $T_g/T$, as in Fig.~\ref{angellfig}.
This is called the `Angell' plot~\cite{angellscience} 
and is very helpful in classifying
supercooled liquids. 
A liquid is called strong or fragile depending on its position in the 
Angell plot. Straight lines correspond
to `strong' glass-formers and to an Arrhenius behaviour,
\be
\tau_\alpha = \tau_\infty \exp \left( \frac{E}{k_B T} \right).
\ee
In this case, one can 
extract from the plot an effective activation energy, $E$,
suggesting quite a simple mechanism for relaxation by `breaking' 
locally for instance a chemical bond.
The typical relaxation time is then dominated by the energy barrier to activate
this process and, hence, has an Arrhenius behaviour. 
Window glasses generically fall in this category. The terminology `strong' 
and `fragile' is not related to the mechanical properties of the glass but 
was introduced in relation  to the evolution of the short-range 
order close to $T_g$. Strong liquids, such as SiO$_2$, typically 
have a locally tetrahedric  structure which persists both below 
and above the glass transition, contrary to fragile liquids. 
Nowadays, the notion of fragility appears in connection with
the temperature evolution of transport properties. 

If one tries to define an effective activation energy for 
fragile glass-formers 
using the slope of the curve in Fig.~\ref{angellfig}, then one 
finds that this energy scale increases
when the temperature decreases, a `super-Arrhenius' behaviour. 
This increase of energy 
barriers immediately suggests that the glass formation is a 
collective phenomenon for fragile supercooled liquids.  
Support for this interpretation is provided by the 
fact that a good fit of the 
relaxation time or the viscosity is given by the 
Vogel-Fulcher-Tamman law (VFT): 
\be 
\tau_\alpha = \tau_0 \exp \left[ \frac{DT_0}{(T-T_0)}\right], 
\label{vft}
\ee
which suggests a 
divergence of the relaxation time, and therefore perhaps
a phase transition of some kind, at a finite temperature $T_0$. 
A smaller $D$ in the VFT law corresponds to a more fragile 
glass. 
Note that there are other comparably good fits of these curves, such as the 
B\"assler law~\cite{bassler}, 
\be
\tau_\alpha = \tau_0 
\exp \left[ K \left( \frac{T_*}{T} \right)^2 \right],
\label{basslerlaw}
\ee 
that only lead to a divergence at zero temperature.
Another fit proposed and tested recently consists in replacing 
$1/T$ by $1/T-1/T_{\rm on}$ in the B\"assler law. This form, where 
$T_{\rm on}$ 
represents some onset temperature for slow dynamics, is motivated 
in part by dynamical facilitation models~\cite{elmatad},
see Sec.~\ref{KCMsection}. 
Actually, although the relaxation time increases by $14$ 
orders of magnitude, the increase of its logarithm, and therefore 
of the effective activation energy is very modest, and experimental
data do not allow one to unambiguously determine 
the true underlying functional law without any 
reasonable doubt \cite{dyrenature}. Several comparisons between fits 
aimed at determining which one works better can be found in the literature, 
see e.g. the recent work \cite{elmatad2010}. Although these are 
certainly useful, one should not forget 
that since the increase of the  effective activation energy is very 
modest, pre-asymptotic 
effects to the `true' limiting behavior can play an important role. 
Hence, comparisons of simple 
fits with a small number of parameters could be misleading.  
For this and other reasons, physical interpretations in terms of 
a finite temperature phase transition must always 
be taken with a grain of salt. It is recommended to use the 
same grain of salt to deal with fits 
supposedly demonstrating the absence of
finite temperature singularities \cite{dyrenature,elmatad}. 

\subsubsection{Thermodynamic aspects}

\begin{table}
\begin{tabular}{|c|c|c|c|c|c|}
        \hline
        \hline
{\bf Substance}    & OTP & 2-MTH & 
N-PROP & 3-BP & 12PD\\
        \hline 
$T_g$   & 246 & 91 & 97 & 108 & 172\\
$T_0$  & 202.4 & 69.6 & 70.2 & 82.9 & 114  \\
$T_K$   & 204.2 & 69.3 & 72.2 & 82.5  &  127\\
$T_K/T_0$   & 1.009 & 0.996 & 1.028 & 0.995 &1.11 \\
        \hline
        \hline
\end{tabular}
\caption{\label{table} Values of glass transition temperature, VFT 
singularity and Kauzmann temperatures for five supercooled 
liquids~\cite{angellrichert}.
OTP: o-terphenyl; 2-MTH: 2-methyltetra-hydrofuran; 
N-PROP: n-propanol; 3-BP: 3-bromopentane. 12PD: 1-2 prop-diol.}
\end{table}

There are additional experimental facts that 
shed some light and might reinforce the interpretation of data
in terms of a finite temperature singularity.  Among them, 
is an empirical connection found between kinetic 
and thermodynamic behaviours.
Consider the part of the entropy of the liquids, $S_{\rm exc}$, which is 
in excess compared to the entropy of the corresponding
crystal. Once this quantity, normalized by its value at the melting 
temperature, 
is plotted as a function of $T$, a remarkable connection with the 
dynamics, in particular the VFT law,  
emerges (see \cite{martinez} for a compilation of experimental data and 
\cite{reviewnature} for a discussion). As for the
relaxation time, one cannot follow this curve below $T_g$ in thermal 
equilibrium. However, extrapolating the curve below $T_g$ 
apparently indicates 
that the excess entropy vanishes linearly at some finite 
temperature, called $T_K$, which is very close to zero for strong glasses
and, generically, very close to $T_0$, 
the temperature at which a VFT fit diverges.
This coincidence is quite remarkable: for materials with glass transition 
temperatures that vary from $50$~K to $1000$~K
the ratio $T_K/T_0$ remains close to 1. Some
examples are provided in Table~\ref{table}, see  \cite{angellrichert} 
for a more extensive list. For the majority of liquids the ratio is 
close to one up 
to few percent. Note, however, there are some liquids where $T_K$ and $T_0$
differ by as much as $20\%$, and so a perfect correlation 
between the two temperatures is not established experimentally 
\cite{tanaka03}.  

The chosen subscript for $T_K$ stands for
Kauzmann~\cite{kauzmann} who recognized $T_K$ as a very important 
temperature for the physics of glasses.
Kauzmann further suggested that some change of behaviour 
(phase transition, crystal nucleation, etc.) must take place
above $T_K$, because below $T_K$ the entropy of the liquid, a disordered 
state of matter,
becomes less than the entropy of the crystal, an ordered state of matter. 
This situation, that seemed perhaps paradoxical 
at that time, is in fact not a serious problem. 
There is no general principle that would constrain
the entropy of the liquid to be larger than that of the crystal. 
As a matter of fact, the crystallization
transition for hard spheres takes place precisely because the crystal 
becomes the state with the largest entropy at sufficiently high 
density \cite{alder}. 

On the other hand, the importance of $T_K$ stands,
partially because it is experimentally very close 
to $T_0$. Additionally,  the quantity $S_{\rm exc}$ which 
vanishes at $T_K$, is thought to be a proxy
for the so-called configurational entropy, $S_c$, 
which quantifies the number of metastable states (actually, 
its logarithm, see below). A popular  
physical picture due to Goldstein~\cite{Goldstein} is that close to $T_g$ 
the system explores a part of the energy landscape 
(or configuration space) which is full of minima separated by barriers that 
increase when temperature decreases.
The dynamic evolution in the energy landscape
 would then consist in a rather short equilibration 
inside the minima followed by `jumps' between different 
minima that are well-separated in time. 
At $T_g$ the barriers have become  so large that the system 
remains trapped in one minimum, identified as one of the
possible microscopic amorphous configurations of a glass. 
Following this interpretation, one can split 
the entropy into two parts. A first contribution is due to the fast relaxation 
inside one minimum, a second one counting
the number of metastable states, $S_c=\log N_{\rm 
metastable}$, which 
is called the `configurational' entropy. 
Assuming that the contribution to the entropy due to the `vibrations' 
around an amorphous glass configuration 
is not very different from the entropy of the crystal, one finds that 
$S_{\rm exc} \approx S_c$.  Within this approximation
$T_K$ corresponds to the temperature at which the configurational
entropy vanishes. 
Since the configurational contribution 
to the specific heat is given by $T d S_{\rm exc} / dT$,
a linear vanishing of $S_{\rm exc}$ near $T_K$ 
(suggested both by experimental observations 
and theoretical arguments, see Sec.~\ref{theory}) would lead to
a discontinuity (a downward jump) 
of the specific heat and thus to a thermodynamic phase transition. 
Note, however, that the above assumptions should not be taken for granted, 
see for instance the recent discussions in \cite{richert,dyrermp,wyart}. 
Furthermore, locating the transition temperature requires an 
extrapolation that is not well controlled, as is also required 
for the relaxation time.

\subsubsection{Static and dynamic correlation functions}

At this point the reader might have 
reached---despite our numerous warnings---the conclusion that 
the glass transition may not be such a difficult problem: 
there are experimental indications of a diverging timescale and 
a concomitant singularity in the thermodynamics. 
It simply remains to find static correlation functions displaying a 
diverging correlation length related to the emergence
of `amorphous order', which would indeed classify the glass transition as 
a standard second order phase transition. 
Remarkably, this remains an open and debated question despite several 
decades of research. 
Simple static correlation function are quite featureless in the supercooled
regime, notwithstanding the dramatic changes in the dynamics.
A simple static quantity is the structure factor defined by  
\be
S(q)= \left\langle 
\frac{1}{N} \rho_{\bf q} \rho_{\bf -q} \right\rangle, 
\label{sofq}
\ee
where the Fourier component of the density reads
\be
\rho_{\bf q} = \sum_{j=1}^N e^{i {\bf q} \cdot {\bf r}_j},
\ee
with $N$ is the number of particles, $V$ the volume, 
and ${\bf r}_j$ is the position of particle $j$. The structure
factor measures the spatial correlations of 
particle positions, but it 
does not show any diverging peak in contrast to what 
happens, for example, at the liquid-gas critical point where 
there is a divergence at small ${\bf q}$.  
A snapshot of a supercooled liquid configuration in fact 
just looks like a glass configuration, despite their widely different
dynamic properties.
More complicated static correlation functions have been 
studied~\cite{debenedetti}, especially in numerical work, 
but until now there are no strong indications of a diverging static 
lengthscale, although this issue is constantly 
debated \cite{static1,Nelson,static2,cavagna-review,tanaka}. 
Recent results do suggest that it is possible to identify some 
growing static lengthscales. 
We will come back to this point in Sec.~\ref{nofuture} and in other sections.  

\begin{figure}
\psfig{file=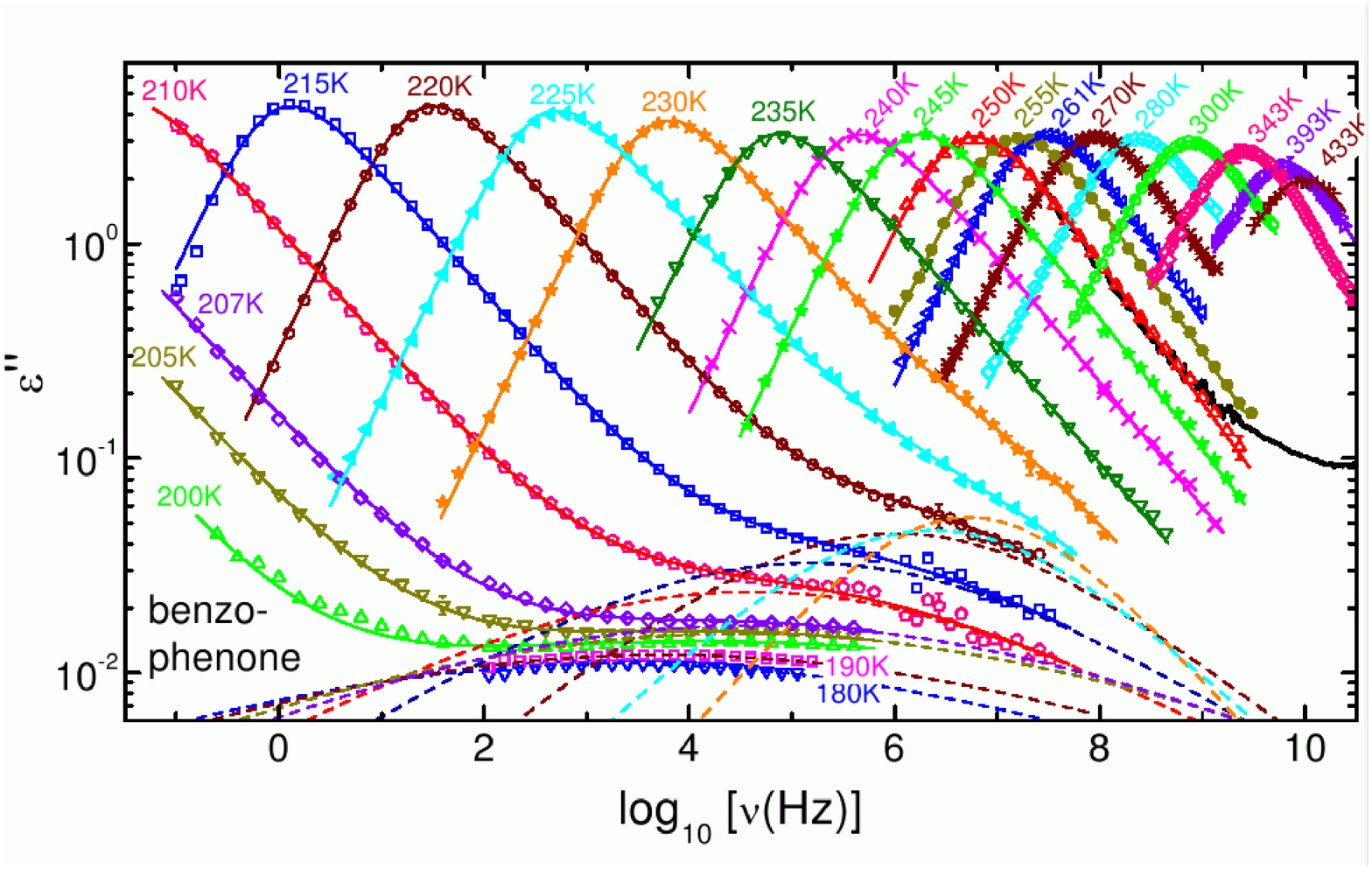,width=8.5cm}
\caption{\label{lunk} Temperature evolution of the dielectric 
susceptibility of the glass-former benzophenone measured over more than 
10 decades of relaxation times~\cite{lunkenheimer}. 
Dynamics slows down dramatically as 
temperature is decreased and relaxation spectra become very broad 
at low temperature and reveal the existence of additional `secondary'
relaxation processes.}
\end{figure}

The difficulty in finding a signature of the glass transition 
disappears if one 
focuses on dynamic correlations or response functions. 
\begin{figure}
\psfig{file=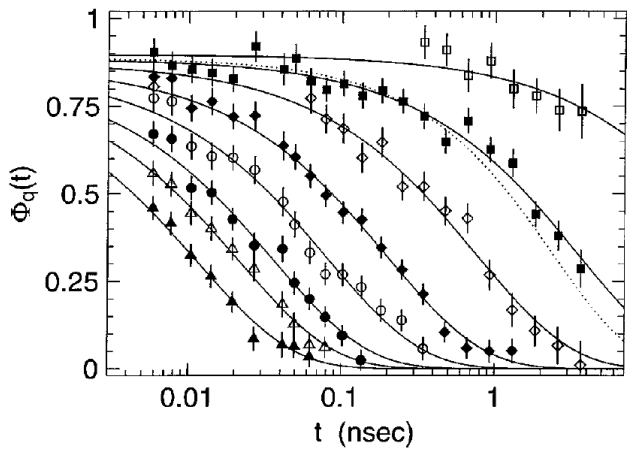,width=8.5cm}
\caption{\label{fqt} Temperature evolution of the intermediate scattering
  function normalized by its value at time equal to zero for supercooled
  glycerol~\cite{glycerol}. 
  Temperatures decrease from 413~K to 270~K from left to right.
  The solid lines are fit with a stretched exponential with
  exponent $\beta=0.7$. The dotted line represents another 
  fit with $\beta=0.82$.}
\end{figure}
For instance, a dynamic observable studied in light and neutron scattering 
experiments is the
intermediate scattering function, 
\begin{equation}
F({\bf q},t) = \left\langle \frac 1 N \rho_{\bf q}(t) \rho_{\bf -q}(0) 
\right\rangle.
\label{isf}
\end{equation}
Different $F({\bf q},t)$ measured 
by neutron scattering in supercooled
glycerol~\cite{glycerol} 
are shown for different temperatures in Fig.~\ref{fqt}.
These curves suggest a first, rather fast
(and hence not accessible in this experiment), relaxation to a plateau followed
by a second, much slower, relaxation. 
The plateau is due to the fraction of density fluctuations that 
are frozen on intermediate timescales, but
eventually relax during the second relaxation. The latter is called 
`alpha-relaxation', and corresponds to the structural relaxation 
of the liquid. This plateau is akin to the Edwards-Anderson order parameter, 
$q_{EA}$, defined for spin glasses which measures the fraction 
of frozen spin fluctuations~\cite{beyond,binderkob}. 
Note that $q_{EA}$ continuously increases from zero 
below the spin glass transition. Instead, for structural glasses,  
a finite plateau already appears above any putative transition.
Figures \ref{lunk}, \ref{si:fig} and \ref{schematic:fig} 
below contain more illustration of the different
time regimes observed in time correlators.

The intermediate scattering function can be probed only on a 
relatively small regime of temperatures. 
In order to track the dynamic slowing down from microscopic 
to macroscopic timescales, other correlators have been studied. 
A popular one is obtained from 
the dielectric susceptibility, which is related 
by the fluctuation-dissipation theorem to 
the time correlation of polarization fluctuations.
It is generally admitted that different dynamic probes 
reveal similar temperature dependencies for the relaxation 
time \cite{niss}. 
The temperature 
evolution of the imaginary part of the dielectric susceptibility, 
$\epsilon''(\omega)$, is shown in Fig.~\ref{lunk} 
for glass-former benzophenone, where a 
a very wide temperature window is covered~\cite{lunkenheimer}. 
At high temperature,
a good representation of the data is given by a Debye law,  
$\epsilon(\omega) = \epsilon(\infty) + \Delta \epsilon / 
(1 + i \omega \tau_\alpha)$, which corresponds to an 
exponential relaxation in the time domain. When temperature is decreased, 
however, 
the relaxation spectra become very broad and strongly non-Debye, 
which is the frequency analog of the streching of
the relaxation observed in the time-domain. 
Indeed one particularly well-known feature of the spectra 
is that they are well fitted, in the time domain,  
for times corresponding to the alpha-relaxation 
with a stretched exponential, 
$\exp ( -( t /\tau_\alpha )^{\beta} ).$
In the Fourier domain, forms such as the Havriliak-Negami
law are used, 
$\epsilon(\omega) = \epsilon(\infty) + \Delta \epsilon / 
(1 + (i \omega \tau_\alpha)^\alpha)^\gamma$,
which generalizes the Debye law.  
The exponents $\beta$, $\alpha$ and $\gamma$ 
depend in general on temperature and on the particular 
dynamic probe chosen, but they capture the fact that relaxation
is increasingly non-exponential when $T$ decreases towards $T_g$. 
A connection was  empirically established 
between fragility and degree of non-exponentiality, 
more fragile liquids being characterized by broader relaxation 
spectra~\cite{reviewnature}, although the correlation is---again---not 
very solid \cite{heuer}.
The data of Fig.~\ref{lunk} also make it clear 
that the relaxation spectra are actually quite complex and characterized
by one or several secondary processes, that have been quite 
extensively studied experimentally \cite{JG70,dixon,bloch}.

To sum up, there are many remarkable phenomena that take place when 
a supercooled liquid approaches the 
glass transition. Striking ones have been 
presented, but many others have been left out for lack of 
space~\cite{angellscience,reviewnature,debenedetti,binderkob}.
We have discussed physical behaviours, relationships or
empirical correlations observed in a broad class of 
materials. This is quite remarkable and suggests that 
there is some interesting physics to be studied in the 
problem of the glass transition, which 
we see as a collective (critical?) 
phenomenon which is relatively independent of 
microscopic details. 

\subsection{Other `glasses' in science}
\label{beyondglass}

We now introduce some other systems whose 
phenomenological behaviour is close or, at least, related, 
to the one of glass-forming liquids, showing that 
glassiness is truly ubiquitous. 
It does not only appear 
in many different physical situations but
also in more abstract contexts, 
such as computer science.

\subsubsection{The jamming transition of colloids and grains}

\label{jammingsubsection}

Colloidal suspensions consist of big particles suspended in a 
solvent~\cite{larson}. 
The typical radii of the particles are in the range $R=1-500$~nm.
The solvent, which is at equilibrium at temperature $T$, 
acts as a source of noise on the particles whose 
short-time dynamics is better
described as being Brownian rather than Newtonian.
The microscopic 
timescale for this diffusion 
is given by $\tau = R^2/D$ where $D$ is the short-time 
self-diffusion coefficient. Typical values are of the order of
$\tau \sim 1$~ms, 
and thus are much larger than the ones for molecular liquids
(in the picosecond regime). 
The interaction potential between particles depends on the systems, 
and this large tunability makes colloids very attractive 
objects for technical applications.

A particularly relevant case, on which we will focus in the 
following, is a purely hard sphere potential, which is zero when 
particles do not overlap and 
infinite otherwise. In this case the temperature becomes 
irrelevant, apart from a trivial rescaling of the 
microscopic timescale.
Colloidal hard spheres systems 
have been intensively studied~\cite{larson} in experiments, simulations 
and theory varying their density $\rho$, 
or their volume fraction $\phi=\frac{4}{3} \pi R^3 \rho $. 
Hard spheres display a fluid phase from $\phi=0$ 
to intermediate volume fractions,
a  freezing-crystallization transition at $\phi \simeq 0.494$, 
and a melting transition at $\phi\simeq 0.545$. 
Above this latter value the system can be compressed 
until the close packing point $\phi\simeq 0.74$, 
which corresponds to the FCC crystal. Interestingly for our purposes, 
a small amount of polydispersity (particles with slightly different sizes)
efficiently prevents crystallization.
In this case, the system can be more easily `supercompressed' above the 
freezing transition without nucleating
the crystal, at least on experimental timescales. 
In this regime the relaxation timescale increases 
very rapidly with $\phi$~\cite{naturepusey}. At a packing fraction $\phi_g 
\simeq  0.57-0.59$ it becomes so large compared to typical 
experimental timescales that the system does not relax anymore: 
it is `jammed'. 
This `jamming' transition is obviously reminiscent of the 
glass transition of molecular systems. 
In particular, the location $\phi_g$ of the 
colloidal glass transition is as ill-defined as 
the glass temperature $T_g$. 

Actually, the phenomena that take place increasing the volume 
fraction are analogous to the ones seen in molecular 
supercooled liquid \cite{naturepusey}: the viscosity
increases very rapidly and can be 
fitted~\cite{chaikin} by a VFT 
law in density as in Eq.~(\ref{vft});
dynamical correlation functions display a broad spectrum of timescales 
and develop a plateau \cite{vanmegen},
no static growing correlation length has been found, etc.
Also the phenomenon of dynamic heterogeneity that will be addressed 
in Sec.~\ref{dh} seems similar in colloids and atomic systems
\cite{kegel,weeks}. However, it is important to underline a 
major difference: because the microscopic
timescale for colloids is so large, experiments can only track at best
the first 5-6 decades of slowing 
down \cite{LucaaboveMCT}. 
A major consequence is that 
the comparison between the glass and colloidal transitions 
must be performed by focusing 
in both cases on the first 5 decades of the slowing down, 
which corresponds to relatively high temperatures 
in molecular liquids. 
Understanding how much and to what extent the glassiness of colloidal
suspensions is related to the one of molecular liquids 
remains an active domain
of research.

The glassiness of driven granular media has recently been
thoroughly analyzed.
Grains are macroscopic objects and, as a
consequence, do not have any thermal motion. 
A granular material is 
therefore frozen in a given configuration 
if no energy is injected into the system~\cite{grainsbook}. However, 
it can be forced in a steady state by an external drive, such as shearing
or tapping. The dynamics in this steady state shows remarkable
similarities (and differences) with simple fluids. The physics
of granular materials is a very wide subject~\cite{grainsbook}. 
In the following we only
address briefly what happens to a polydisperse granular fluid at 
very high packing fractions.
As for colloids, the timescales for relaxation or diffusion increase very
fast when density is increased, 
without any noticeable change in structural properties. 
Again, it is now established~\cite{gremaud,dauchot,durian,pinaki,candelier1} 
that many 
phenomenological properties of the granular glass transition also occur
in granular assemblies. As for colloids, going beyond the mere analogy
and understanding how much these different physical systems are related 
is a very active domain of research~\cite{jamming}. 

This very question has been asked in a visual manner by
Liu and Nagel~\cite{liunagel} who rephrased it in a single picture, 
now known as a `jamming phase diagram'. By building a common phase 
diagram for glasses, colloids and grains, they asked whether
the glass and jamming transitions of molecular liquids, 
colloids and granular media 
are different facets of the same `jammed' phase. 
In this unifying phase diagram, a jammed 
`phase' (or jammed phases) can be reached either 
by lowering the temperature as in molecular liquids, or increasing the packing
fraction or decreasing the external drive in colloids
and granular media.

\subsubsection{More `glasses' in physics and beyond}

There are many other physical contexts in which glassiness plays an 
important role~\cite{youngbook}. 
One of the most famous examples is the field of spin glasses. 
Experimentally, spin glasses are composed of magnetic impurities 
interacting by quenched random couplings. At low
temperatures, their dynamics become extremely slow and they freeze in 
an amorphous spin configuration dubbed a `spin glass' by P. W. Anderson.
Experiments on spin glasses, in particular aging studies,
have played an important role in the context of amorphous
materials.
There are many other physical systems, often characterized by quenched
disorder, that show glassy behaviour, like Coulomb glasses, 
Bose glasses, frustrated magnets, etc.
In many cases, however, 
one finds quite a different physics from structural 
glasses: the similarity between these systems is 
therefore only superficial from the phenomenological point of view,
but the theoretical techniques and ideas 
developed in particular in the 
field of spin glasses are highly relevant in thoeretical studies 
of the glass transition. 

Finally, and quite remarkably, glassiness emerges even 
in other branches of science~\cite{complexbook}. 
In particular, it has been discovered
recently that concepts and techniques developed for glassy systems 
turn out to apply and be very useful tools in the field 
of neural networks and computer science.  
Problems like combinatorial optimization 
display phenomena completely analogous to phase transitions, actually, to 
glassy phase transitions. {\it A posteriori}, this is quite natural, 
because a typical optimization problem 
consists in finding a solution in a presence 
of a large number of constraints.
This can be defined, for instance, as a set 
of $N$ Boolean variables that satisfies $M$ constraints. 
For $N$ and $M$ very large at fixed 
$\alpha=M/N$, this problem very much resembles 
finding a ground state in a statistical mechanics problem with quenched
disorder. Indeed one can 
define an energy function (a Hamiltonian) 
as the number of unsatisfied constraints, that has to be minimized, 
as in a $T=0$ statistical mechanics problem.
The connection with glassy systems lies in the fact that in both cases
the energy landscape is extremely complicated, full of minima and saddles.
The fraction of constraints per degree of freedom, $\alpha$, plays a role
similar to the density in a hard sphere system.  
A detailed presentation of the relationship between optimization problems 
and glassy systems is clearly 
out of the scope of the present review. We simply illustrate it 
pointing out that a central problem in optimization, 
random $k$-satisfiability, has been shown to undergo a glass transition when 
$\alpha$ increases that is reminiscent of the one of structural 
glasses and can be treated analytically using similar
tools~\cite{PNAS}.   

\subsection{Computer simulations of molecular glass-formers}
\label{simu}

Studying the glass transition of molecular liquids 
at a microscopic level is in principle straightforward since one has to 
 answer a very simple question: how do particles move in 
a liquid close to $T_g$? It is of course 
a daunting task to attempt answering this question 
experimentally because one should then 
resolve the dynamics of single molecules 
to be able to follow the trajectories of objects that are a few Angstroms
large on timescales of tens or hundreds of seconds, which sounds like
eternity when compared to typical molecular dynamics usually
lying in the picosecond regime. 
In recent years, such direct experimental investigations have been 
attempted using time and space resolved techniques such as atomic
force microscopy~\cite{israeloff} 
or single molecule spectroscopy~\cite{single,laura}, 
but this remains a very difficult task.

\begin{figure}
\psfig{file=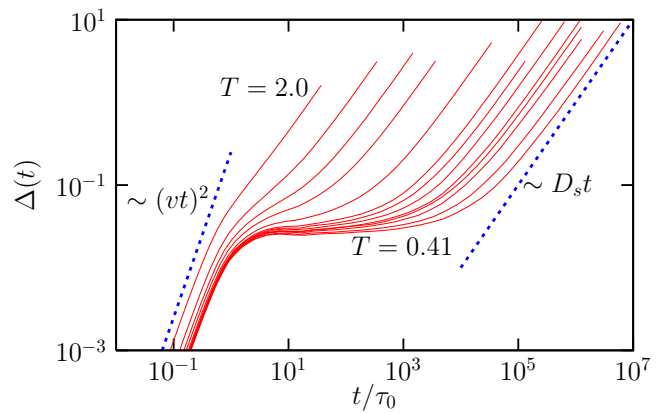,width=8.5cm}
\caption{\label{msd} Mean-squared displacements of individual 
particles in a simple model of a glass-forming liquid composed 
of Lennard-Jones particles observed 
on a wide time window. When temperature decreases (from left to right), 
the particle displacements become increasingly slow with several distinct
time regimes corresponding, in this order, to ballistic,
localized, and diffusive regimes.}
\end{figure}

In numerical simulations, by contrast,
the trajectory of each particle in the system can,
by construction, be followed at all times.
This allows one to quantify easily single particle dynamics, as
proved  in Fig.~\ref{msd} where the averaged 
mean-squared displacement $\Delta(t)$ 
measured in a simple Lennard-Jones glass-former
is shown. The mean-squared displacement is defined by
\begin{equation}
\Delta(t) = \left\langle \frac{1}{N} \sum_{i=1}^N  
| {\bf r}_i(t) -{\bf r}_i(0) |^2 \right\rangle,
\end{equation} 
where ${\bf r}_i(t)$ represents the position of particle $i$ at time
$t$ in a system composed of $N$ particles; the brackets 
indicate an ensemble average over initial conditions 
weighted with the Boltzmann distribution. 
The main observation from the data shown in Fig.~\ref{msd}
is that single particle displacements 
considerably slow down when $T$ is decreased. 
This can be quantified by measuring the 
self-diffusion constant, $D_s$, formally defined as 
$D_s = \lim_{t \to \infty} \Delta(t) / (6 t)$.
The data in Fig.~\ref{msd} show that 
$D_s$ decreases by orders of magnitude when 
temperature decreases, and thus mirrors the behaviour of the 
(inverse of the)
viscosity shown in Fig.~\ref{angellfig} for real systems.
Therefore, to explain the 
phenomenon of the glass transition, one must equivalently explain
why molecular motions become so slow at low temperatures. 

Additionally, a very rich dynamics is 
observed in Fig~\ref{msd}, with a plateau regime at intermediate timescales, 
corresponding to an extended time window during which 
particles vibrate around their initial positions, 
as in a crystalline solid. The difference with a crystal
is of course that this transient localization does not correspond to a 
well-defined position in an ordered structure, and it is 
only transient so that all particles
eventually escape and, concomitantly, the structure relaxes at long times.
Describing the molecular motions responsible 
for this very broad spectrum of relaxation timescales is 
a challenge in this field.

In recent years, computer experiments have played an increasingly 
important role in glass transition studies \cite{andersen}. 
It could almost be said 
that particle trajectories in numerical work have been studied 
under so many different angles that probably very little remains to be learnt
from such studies in the regime that is currently
accessible using present day computers. Unfortunately, this 
does not imply complete knowledge of the physics 
of supercooled liquids. As shown in Fig.~\ref{msd}, 
it is presently possible to follow the 
dynamics of a simple glass-forming liquid 
over more than eight decades of time, and over a temperature 
window in which average relaxation timescales 
increase by more than five decades. 
This might sound impressive, but a quick 
look at Fig.~\ref{angellfig}
shows, however, that at the lowest
temperatures studied in the computer, 
the relaxation timescales are still orders of magnitude 
faster than in experiments performed close to the glass 
transition temperature. 
Simulations can be directly compared to experiments performed 
in this high temperature regime, but 
this also implies that simulations
focus on a relaxation regime that is about eight to ten decades 
of times faster than in experiments performed close to $T_g$. 
Whether numerical works are useful to understand 
the glass transition itself at all is therefore an open, 
widely debated, question. We believe that it is now possible to
numerically access temperatures which are low enough that 
many features associated to the
glass transition physics can be observed: 
strong decoupling phenomena (see Sec.~\ref{dh}), clear deviations
from fits to the mode-coupling theory (which 
are experimentally known to hold only at high
temperatures, see Sec.~\ref{secmct}), 
and crossovers towards truly activated dynamics.
  
Classical computer simulations of supercooled liquids 
usually proceed by solving a cleverly discretized version 
of Hamilton's equations for the particles' positions and momenta and 
a given potential interaction 
between particles~\cite{allen}:
\be
\frac{\partial {\bf r}_i}{\partial t} = \frac{\partial H}{\partial 
{\bf p}_i}, \quad    \frac{\partial {\bf p}_i}{\partial t}
= -  \frac{\partial H}{\partial 
{\bf r}_i},
\label{ham}
\ee 
where
\be
H(\{ {\bf p}_i, {\bf r}_i \} ) 
= \sum_{i=1}^N \frac{{{\bf p}_i}^2}{2m_i} + V(\{ {\bf r}_i \}) 
\label{ham2}
\ee
is the system's Hamiltonian composed of a kinetic part and an
interaction term $V(\{ {\bf r}_i \} )$.
We have written Eqs.~(\ref{ham}) and (\ref{ham2})
in terms of the center of mass trajectories, as is appropriate
for atoms although, of course, numerical simulations can deal with molecular 
degrees of freedom as well~\cite{allen}. Since the 
equations of motion are energy conserving, they describe the dynamics
of atomistic systems in the microcanonical ensemble. Constant
temperature or constant pressure schemes have been developed allowing
simulations to be performed in any desired statistical 
ensemble~\cite{allen}. Similarly, nonequilibrium 
simulation techniques exist that allow, for instance, computer studies
of the aging dynamics or the nonlinear rheology 
of supercooled fluids~\cite{evans}, see also Sec.~\ref{aging}.

If quantitative agreement with experimental data
on an existing specific material is sought, the interaction must be 
carefully chosen in order to reproduce reality, for instance
by combining classical to {\it ab-initio} simulations. 
From the more fundamental perspective adopted here,
one rather seeks the simplest
model that is still able to reproduce qualitatively the phenomenology 
of real glass-formers, while being considerably simpler to study. 
The implicit, but quite strong, hypothesis is that molecular details are
not needed to explain the behaviour of supercooled liquids, so that
the glass transition is indeed a topic for statistical mechanics, 
with little influence from chemical details.
A considerable amount of work has therefore been dedicated to 
studying models where point particles interact via
a simple pair potential such as Lennard-Jones interactions:
\be
V(\{ {\bf r}_i \})  = \sum_{i=1}^N \sum_{j=i}^N \epsilon
\left[ 
\left(\frac{\sigma}{r_{ij}} \right)^{12} -  
\left( \frac{\sigma}{r_{ij}}\right)^6
\right],
\label{LJ}
\ee 
where $r_{ij} = |{\bf r}_i - {\bf r}_j|$, $\epsilon$ and $\sigma$
represent an energy scale and the particle diameter, respectively.
Other popular models are soft spheres, where only the steep short 
range repulsion in Eq.~(\ref{LJ}) is considered, or even hard spheres
where the repulsion is made infinitely steep. If the system
is too simple, such as the one defined in (\ref{LJ}),
the glass transition cannot be studied because crystallization
takes place when temperature is lowered. Some frustration 
must be introduced to devise numerical models with good 
glass-forming abilities. A common solution, inspired by
experimental studies of metallic glasses, is to use
mixtures of different atoms, as in the popular model devised 
by Kob and Andersen~\cite{KALJ} which uses a 
non-additive binary mixture
of Lennard-Jones particles. 

\begin{figure}
\begin{center}
\psfig{file=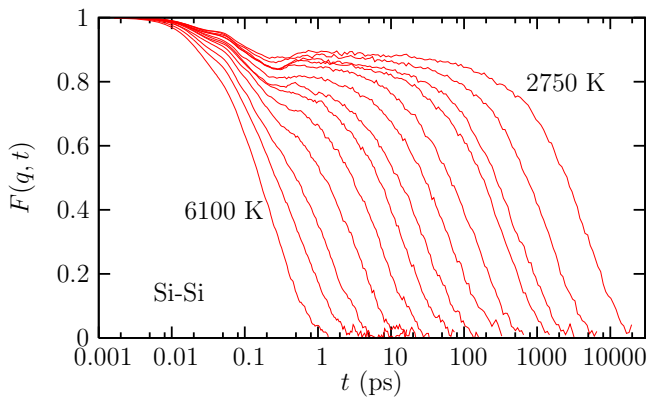,width=8.5cm}
\caption{\label{si:fig} 
Intermediate scattering function at wavevector 
$q=1.7$~\AA$^{-1}$ for the Si atoms from $T=6100$ to $T=2750$~K,  
obtained from molecular
dynamics simulations of silica. Data adapted from 
\cite{horbachkob}.}
\end{center}
\end{figure}

More realistic materials are also studied focusing for instance on 
the physics of network forming materials (such as
silica, SiO$_2$, the main
component of window glasses), multi-component ones, 
anisotropic particles, or molecules with internal degrees of freedom.
Connections to experimental work can be made by computing 
quantities that are experimentally accessible such as the 
intermediate scattering function, Eq.~(\ref{isf}), 
static structure factors, $S({\bf q})$, Eq.~(\ref{sofq}), or 
thermodynamic quantities such as specific heat or configurational 
entropy, which are
directly obtained from particle trajectories
and can be measured in experiments as well. 
As an example we show in Fig.~\ref{si:fig} the 
intermediate scattering function $F({\bf q},t)$ 
obtained from a molecular dynamics simulation 
of a classical model for 
SiO$_2$ as a function of time for different 
temperatures~\cite{horbachkob}. The numerical data compare very favourably 
with the experimental results obtained 
from neutron scattering shown in Fig.~\ref{fqt}.
Note that they actually access the dynamics over 
a broader range of timescales and temperatures.

An important role is played by simulations also because 
a large variety of dynamic and static quantities can be simultaneously 
measured in a single model system. As we shall discuss later, 
there exist scores of different theoretical approaches to 
describe the physics of glass-formers, and they sometimes 
have their own set of predictions that can be readily tested 
by numerical work. For instance, quite a large number of numerical 
papers have been dedicated to testing in detail the predictions
formulated by the mode-coupling theory of the glass transition, 
as reviewed recently in \cite{gotze,gotze2}. Here, computer simulations are
particularly well-suited as the theory specifically 
addresses the relatively high temperature window that is studied in computer
simulations.

While Newtonian dynamics is mainly used in 
numerical work on supercooled liquids, 
a most appropriate choice for atomistic materials, 
it can be interesting to consider 
alternative dynamics that are not deterministic, or which do
not conserve the energy. In colloidal glasses and physical
gels, for instance, particles undergo Brownian motion
arising from collisions with molecules in the solvent, and a
stochastic dynamics is more appropriate. Theoretical 
considerations might also suggest the study of different sorts of
dynamics for a given interaction between particles, for instance, 
to assess the role of conservation laws and
structural information. Of course, if a given dynamics
satisfies detailed balance with respect to the Boltzmann 
distribution, all structural quantities remain unchanged, but the
resulting dynamical behaviour might be very different. 
More generally one can ask the question: how universal 
is the glass transition phenomenon? Does it depend
upon the specific microscopic dynamics?

Several papers
have studied in detail the influence of the
dynamics on the resulting dynamical behaviour in
glass-formers using different types of 
microscopic dynamics. 
Gleim {\it et al.} studied `stochastic dynamics' which generalizes
Newton's equations 
to include non-deterministic forces \cite{gleim}: 
\be
m_i \frac{ d^2 {\bf r}_i}{dt^2} = - \frac{\partial 
V (\{ {\bf r}_i \})}{\partial {\bf r}_i}
- \zeta \frac{\partial {\bf r}_i}{\partial t} + {\bf \eta}_i.
\ee
Specifically, a friction term proportional to the velocity 
with a damping constant $\zeta$ is added to the right hand side, 
as well as a Gaussian distributed white noise ${\bf \eta}_i$, 
whose correlations are related to the damping via the 
fluctuation-dissipation theorem, 
$\langle {\bf \eta}_i(t) {\bf \eta}_j(t') \rangle
= 6 k_B T \zeta \delta_{ij} \delta(t-t')
$, so that the equilibrium distribution at temperature $T$
is indeed recovered. When $\zeta$ gets large the dynamics becomes 
similar to a purely Brownian dynamics, as 
recently studied for 
instance in \cite{szamel}:
\be
\zeta \frac{\partial {\bf r}_i}{\partial t}
= - \frac{\partial 
V (\{ {\bf r}_i \})}{\partial {\bf r}_i}
+ {\bf \eta}_i.
\ee
In that case, particles are described by their positions only, 
and momenta play therefore no role. 
A similar type of description, although numerically more 
efficient~\cite{berthierkob,berthiersio2}, 
is obtained using a standard Monte Carlo approach
where the change in potential energy between two configurations 
is used to accept or reject a trial move. In both cases 
of Brownian and Monte Carlo dynamics, particles have 
diffusive (rather than ballistic) behaviour at very short times
where differences between the different types of dynamics
can therefore be expected.

Quite surprisingly, however, 
the equivalence between these three types of 
stochastic dynamics and the originally
studied Newtonian dynamics 
was quantitatively 
established at the level of the averaged dynamical correlators
for all three types of dynamics mentioned above,
except at very short times where obvious differences 
are indeed expected. 
This equivalence can probably be traced back to the existence of 
fast and slow degrees of freedom. It is reasonable to think that the former act
 as an effective thermal bath for the latter thus making the three 
types of dynamics equivalent on long timescales.  
However, this interpretation has to be taken with a grain of salt since 
important differences were found when fluctuations of 
dynamical correlators were 
considered~\cite{jcpI}, even in the long-time regime corresponding to 
the structural relaxation.

\begin{figure}
\psfig{file=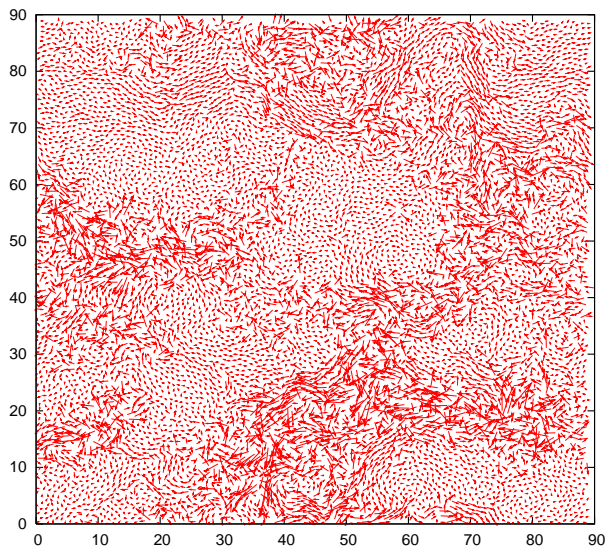,height=8.5cm,angle=-90,clip}
\caption{\label{peter} Spatial map of single particle displacements in 
the simulation of a binary mixture of Lennard-Jones mixture
in two dimensions. Arrows show the displacement
of each particle in a trajectory of length comparable to the structural 
relaxation time. The map reveals the existence of 
particles with different mobilities during relaxation, but 
also the existence of spatial correlations between these 
dynamic fluctuations.}
\end{figure}

Another crucial advantage of molecular simulations 
is illustrated in Fig.~\ref{peter}.
This figure shows a spatial map of single particle
displacements recorded during the simulation of a binary 
Lennard-Jones mixture in two dimensions.
This type of measurement, out of reach of most
experimental techniques that study the liquid state, reveals that 
dynamics might be very different from 
one particle to another. More importantly, Fig.~\ref{peter} also 
unambiguously reveals 
the existence of spatial correlations between these dynamic
fluctuations. The presence of non-trivial spatio-temporal fluctuations 
in supercooled liquids is called `dynamic 
heterogeneity'~\cite{ediger}. The phenomenon 
has become a substantial component of the field of the glass transition, 
and computer simulations have naturally played an important role 
since they reveal the heterogeneous nature of the
relaxation much more directly than experiments.  
We discuss the phenomenon of dynamic heterogeneity
in more detail in the next section.

\section{Dynamic heterogeneity}
\label{dh}

\subsection{Existence of 
spatio-temporal dynamic fluctuations}

A new facet of the relaxational behaviour of supercooled liquids
has emerged in the last decade thanks to a considerable experimental 
and theoretical effort. It is called dynamic heterogeneity,
and plays now a central role in modern descriptions of 
glassy liquids~\cite{ediger}. 
As anticipated in the discussion of Fig.~\ref{peter} in the 
previous section, the 
phenomenon of dynamic heterogeneity is related to the  
spatio-temporal fluctuations of the dynamics. 
Initial motivations stemmed from the search for 
an explanation of the 
non-exponentiality of relaxation processes in supercooled liquids,
related to the existence of a broad relaxation spectrum.
Two natural, but fundamentally different, explanations can be put 
forward. (1) The relaxation is locally exponential,  
but the typical relaxation timescale varies 
spatially. Hence, global correlation
or response functions become non-exponential 
upon spatial averaging over this spatial 
distribution of relaxation times. 
(2) The relaxation is complicated and 
inherently non-exponential, even locally. 
Experimental and theoretical works~\cite{ediger} suggest
that both mechanisms are likely at play, but 
definitely conclude that relaxation is spatially 
heterogeneous, with regions that are 
faster and slower than the average. 
Since supercooled liquids are ergodic systems, 
a slow region will eventually become fast, and vice-versa. 
A physical characterization of dynamic heterogeneity entails the determination 
of the typical lifetime of the heterogeneities, as well as 
their typical lengthscale. 

\begin{figure}
\psfig{file=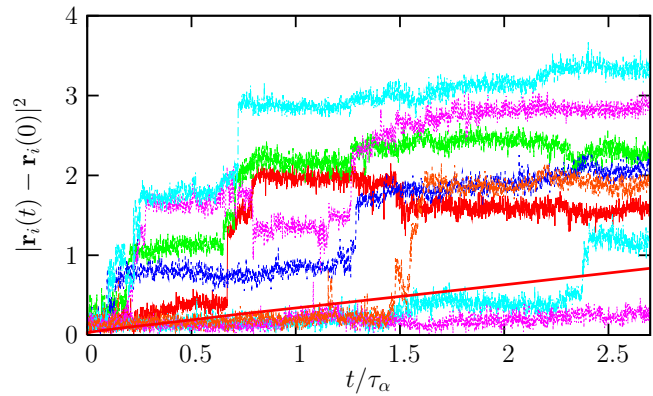,width=8.5cm}
\caption{\label{msd2} Time resolved squared displacements of individual 
particles in a simple model of a glass-forming liquid composed 
of Lennard-Jones particles near the fitted mode-coupling 
critical temperature.
The average is shown as a smooth full line. 
Trajectories are composed of long periods of time during which particles
vibrate around well-defined positions, separated by rapid jumps that
are widely distributed in time, underlying the importance of 
dynamic fluctuations.}
\end{figure}

A clearer and more direct 
confirmation of the heterogeneous character of the dynamics also stems
from simulation studies. For example, whereas the simulated average 
mean-squared displacements are smooth
functions of time (see Fig. \ref{msd}), time signals for individual
particles clearly exhibit specific features that 
are not observed unless dynamics is resolved both in space and time. 
These features are displayed in Fig.~\ref{msd2}. 
What do we see? 
We mainly observe that particle trajectories are not smooth but 
rather composed of a succession of long periods of time 
where particles simply vibrate around 
well-defined locations, separated by rapid `jumps'.  
Vibrations were previously inferred from the plateau observed 
at intermediate times in the mean-squared displacements 
of Fig.~\ref{msd}, but the existence of jumps that are 
clearly statistically widely distributed in time cannot be guessed
from averaged quantities only. The fluctuations in Fig.~\ref{msd2} suggest,
and direct measurements confirm, the importance played by 
fluctuations around the averaged dynamical behaviour. 

A simple type of such 
fluctuations has been studied in much detail. When looking 
at Fig.~\ref{msd2}, 
it is indeed natural to ask, for any given time, what is the distribution 
of particle displacements. This is quantified by the self-part
of the van-Hove function defined as
\begin{equation}
G_s({\bf r},t) = \left\langle 
\frac{1}{N} \sum_{i=1}^N \delta ({\bf r} - [{\bf r}_i(t) - 
{\bf r}_i(0)] ) \right\rangle . 
\label{vanhove}
\end{equation}
For an isotropic Gaussian diffusive process, one gets 
$G_s({\bf r},t) = \exp(-|{\bf r}|^2/(4 D_s t))/(4\pi D_s t)^{3/2}$. 
Simulations reveal instead strong 
deviations from Gaussian behaviour on the timescales 
relevant for structural relaxation~\cite{glotzerkob}. 
In particular they reveal 
tails in the distributions that are `fat', in the sense that they
are much wider than expected 
from the Gaussian approximation. These tails are in fact 
well described by an exponential, rather than 
Gaussian, decay in a wide time window comprising the 
structural relaxation, such that 
$G_s({\bf r},t) \sim \exp(-|{\bf r}|/\lambda(t))$~\cite{pinaki}.
Thus, they reflect the existence of a 
population of particles that moves distinctively further 
than the rest and appears therefore to be much more
mobile. The exponential form of the tail 
originates from the intermittent nature of the particle
trajectories exposed in Fig.~\ref{msd2}, made of a succession
of `jumps' separated by vibrations \cite{pinaki}. Actually, such a tail
would be present in simple jump models for diffusion \cite{hansen}. 
This observation implies that relaxation 
in a viscous liquid differs qualitatively from that of a normal liquid 
where diffusion is close to Gaussian, 
and that a non-trivial statistics of single
particle displacements exists. 

A long series of questions immediately 
follows this seemingly simple observation. Answering them
has been the main occupation of many workers in this field 
over the last decade. What are the particles corresponding to the tails
effectively doing? Why are they faster than the rest? Are 
they located randomly in space or do they cluster? What is
the geometry, time and temperature evolution of the clusters? 
Are these spatial fluctuations correlated to geometric
or thermodynamic properties of the liquids? Do similar correlations 
occur in all glassy materials? Can one predict
these fluctuations theoretically? Can one understand glassy 
phenomenology using fluctuation-based arguments? How can
these fluctuations be detected experimentally?

\begin{figure}
\psfig{file=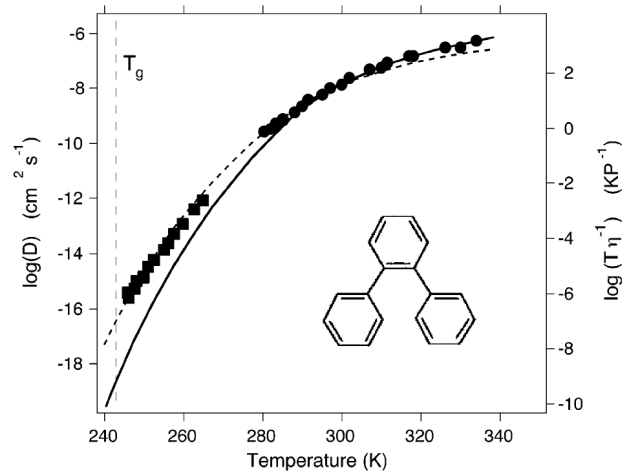,width=8.5cm}
\caption{\label{otp} Decoupling between viscosity (full line) and 
self-diffusion coefficient (symbols) in supercooled 
ortho-terphenyl~\cite{edigerotp}.
The dashed line shows a fit with a `fractional' Stokes-Einstein 
relation, $D_s \sim (T / \eta)^\zeta$ with $\zeta \sim 0.82$
instead of the `normal' value $\zeta=1$ which only holds at high 
temperatures.} 
\end{figure}

Another influential phenomenon that was related early on 
to the existence of dynamic heterogeneity
is the decoupling of self-diffusion ($D_s$) and 
viscosity ($\eta$). In the high temperature
liquid self-diffusion and viscosity are related by the 
Stokes-Einstein relation~\cite{hansen}, 
$D_s \eta / T = const$. 
For a large particle moving in a fluid the constant is equal to $1/(6\pi R)$
where $R$ is the particle radius. Physically, the Stokes-Einstein relation 
means that two different measures of the relaxation time $R^2/D_s$ and $\eta
R^3/T$ lead to the same timescale up to a constant factor. In supercooled
liquids this phenomenological law 
breaks down, as shown in Fig.~\ref{otp} for 
ortho-terphenyl~\cite{edigerotp}. It is commonly found that $D_s^{-1}$ 
does not increase as fast as $\eta/T$ so that,  
at $T_g$, the product $D_s \eta / T$ has increased by 2-3 orders 
of magnitude as compared to its Stokes-Einstein value. 
This phenomenon, although less spectacular than the overall change of 
viscosity, is a
significant indication that different ways to measure relaxation times
lead to different answers and, thus, is a strong hint of the existence
of broad `distributions' of relaxation timescales. 

Indeed, a natural explanation of this effect is that different observables 
probe differently the underlying distribution of relaxation 
times~\cite{ediger}. 
For example, the self-diffusion coefficient of tracer particles is dominated 
by the more mobile particles whereas the viscosity or other measures of
structural relaxation probe the timescale needed for every particle
to move. An unrealistic but instructive example is a model where
there is a small, non-percolative subset of particles that are 
blocked forever, coexisting with a majority of mobile 
particles. In this case, the structure never relaxes but the
self-diffusion coefficient is non-zero because of the mobile particles. 
Of course, in reality all particles move, eventually, but this 
shows how different observables are likely to probe 
different moments of the distribution of
timescales, as explicitly shown within several theoretical 
frameworks~\cite{hodgdon,Gilles,jung}.

The phenomena described above, clearly show  
that the dynamics is spatially heterogeneous. 
However, they are in principle not able to probe whether this is
related to {\it purely local} fluctuations or if there are instead
increasingly {\it spatially correlated} fluctuations. 
This is, however, a fundamental issue both from the experimental and
theoretical points of view.  
How large are the regions that are faster or
slower than the average? How does their size depend on temperature? Are these
regions compact or fractal?    
These important questions were first addressed in pioneering works using 
four-dimensional NMR~\cite{nmr1,nmr2}, 
or by directly 
probing fluctuations at the nanoscopic scale using microscopy techniques. 
In particular, Vidal Russel and
Israeloff using Atomic Force Microscopy techniques \cite{israeloff} measured 
the polarization fluctuations in a volume of size of few tens of nanometers 
in a supercooled polymeric liquid (PVAc) close to $T_g$. 
In this spatially resolved measurement, the hope is to probe a small enough
number of dynamically correlated regions, and detect their dynamics. 
Indeed, the signals shown in \cite{israeloff} reveal that the dynamics
is very intermittent in time: it switches between
moments with intense activity, and moments with no dynamics 
at all, suggesting that extended regions of space 
indeed transiently behave as fast and slow regions.   
A much smoother signal would have been measured 
if such dynamically correlated `domains' 
were not present.
Spatially resolved studies such as NMR experiments or atomic
force microscopy are quite difficult to perform.
They give undisputed information about the typical 
lifetime of the dynamic heterogeneity, but their determination of 
a dynamic correlation lengthscale is 
rather indirect and performed on a small number of 
liquids in a small temperature window.
Nevertheless, the outcome is that a 
non-trivial dynamic correlation length emerges at 
the glass transition, where it reaches a value of the 
order of about $\sim 10$ molecule diameters~\cite{ediger}.     

\subsection{Multi-point correlation functions}
\label{Multi-point}

Recently, substantial progress in characterizing spatio-temporal 
dynamical fluctuations was obtained. 
In particular, it is now 
understood that dynamical fluctuations can be measured 
and characterized through the use of high-order correlation 
and response functions.
These multi-point functions can be 
seen as generalizations of the spin glass 
susceptibility measuring the extent of amorphous 
long-range order in spin glasses. 
We now introduce these dynamical functions and 
summarize the main results obtained from their study.

\subsubsection{Why four-point correlators? The spin glass case} 

\label{spinglass}
 
No simple static correlation has yet been found to reveal any notable feature 
upon approaching the glass 
transition (see Sec.~\ref{nofuture} for recent theoretical 
progress).
As a consequence, it is quite natural to investigate whether a 
growing lengthscale associated  
with the slowing down of the system is hidden in some dynamic  
correlation function. 

Spin glass theory faced a similar conundrum, whose solution we briefly recall 
because it has been instrumental in the developments for glass-forming 
liquids. 
We know that some hidden long-range order 
develops at the spin glass transition~\cite{young}. However, also 
for spin glasses, 
conventional two-point functions are useless. Even if spins $s_{{\bf x}}$ 
and $s_{{\bf x}+{\bf y}}$ have 
non-zero static correlations $\langle s_{{\bf x}} s_{{\bf x}+{\bf y}} \rangle$ 
in the spin glass phase, the average over space, $[ \cdots ]_{\bf x} =
V^{-1} 
\int d {\bf x} \cdots$, for a given distance  
$|{\bf y}|$ vanishes because the pairwise correlations randomly change sign 
whenever ${\bf x}$ changes.  
The insight of Edwards and Anderson is that one should first square 
$\langle s_{{\bf x}} s_{{\bf x}+{\bf y}} \rangle$  
before averaging over space~\cite{ea}.  
In this case, the resulting {\it four-spin} correlation 
function indeed develops long-range tails in the spin glass phase. 
This correlation in fact decays  
so slowly that its volume integral,  
related to the non-linear magnetic susceptibility of the material,  
diverges in the whole spin glass phase~\cite{young}.  
 
The Edwards-Anderson idea can in fact be understood from a dynamical
point of view, which is important for understanding both the physics of
the spin glass just above the transition, and its generalization to
structural glasses.  
Consider, in the language of spins, the following four-point 
correlation function: 
\be 
G_4({\bf y},t) = [\langle s_{{\bf x}}(t=0) s_{{\bf x}+{\bf y}}(t=0) 
s_{{\bf x}}(t) s_{{\bf x}+{\bf y}}(t) \rangle]_{\bf x}. 
\ee 
Suppose 
that spins $s_{{\bf x}}$ and $s_{{\bf x}+{\bf y}}$  
develop static correlations $\langle s_{{\bf x}} s_{{\bf x}+{\bf y}} 
\rangle$ within the glass phase. In this case, $G_4({\bf y},t \to \infty)$ 
will clearly  
converge to the spin glass correlation $[\langle s_{{\bf x}} 
s_{{\bf x}+{\bf y}} \rangle^2]_{\bf x}$. More generally,  
$G_4({\bf y},t)$ for finite $t$ is able to detect {\it transient} 
tendencies to spin glass order, for example slightly above the spin glass  
transition temperature $T_c$. 
Close to the spin glass transition, both the persistence 
time and the dynamic length  
diverge in a critical way: 
\be  
G_4({\bf y},t) \approx y^{2-d-\eta} \hat 
G\left(\frac{y}{\xi},\frac{t}{\tau}\right),  
\ee 
where $\xi \sim (T-T_c)^{-\nu}$ 
and $\tau \sim (T-T_c)^{- z \nu}$. As mentioned above, the static 
non-linear susceptibility diverges as  
$\int d{\bf y} G_4({\bf y},t \to \infty) \sim \xi^{2-\eta}$. 
More generally, one can define a time-dependent dynamic  
susceptibility as: 
\be 
\chi_4(t) \equiv \int d{\bf y} \,\, G_4({\bf y},t), 
\ee 
which defines, provided $G_4({\bf 0},t)$ is a number of order 1, 
a correlation volume, i.e. the typical number 
of spins correlated in dynamic events taking place  
over the time scale $t$. As we shall discuss below, $\chi_4(t)$ 
can also be interpreted as a quantitative measure of the  
dynamic fluctuations. Note however that the precise relation 
between $\chi_4$ and $\xi$ depends on the value of the 
exponent $\eta$, which is physically controlled by 
the detailed spatial structure encoded in $G_4$: 
\be 
\chi_4(t=\tau) \propto \xi^{2-\eta}. 
\ee 
 
Therefore, spin glasses offer a precise example of a system which 
gets slower and slower upon approaching $T_c$ but  
without any detectable long-range order appearing in two-point 
correlation functions. Only more complicated  
four-point functions are sensitive to the genuine amorphous 
long-range order that sets in at $T_c$ and give 
non-trivial information even above $T_c$. In the case of 
spin glasses
it is well established that the transition is related to 
the emergence of a low temperature spin glass phase.

In the case of the glass transition of 
viscous liquids the situation is much less clear. First,
unlike spin glasses where the disorder is quenched,
glass-formers tend to freeze in an amorphous 
state where disorder is instead self-induced. 
Second, there might be no true phase
transition toward a low temperature amorphous phase. 
It is nevertheless still 
reasonable to expect that the dramatic increase of the 
relaxation time is due to 
a transient amorphous order that sets in and whose range 
grows approaching the glass transition.
Growing timescales should be somehow related to growing 
lengthscales~\cite{MS0}. 
A good candidate to unveil the existence of this phenomenon is the
function $G_4({\bf y},t)$ introduced previously, 
since nothing in the above arguments was specific to systems with 
quenched disorder. 
The only difference 
is that although transient order is detected in $G_4({\bf y},t)$ 
or its volume integral $\chi_4(t)$ for times
of the order of the relaxation time, in the long-time limit these 
two functions may not, and indeed do not in the
case of supercooled liquids, show long-range amorphous order. 
This roots back to the different nature of the 
glass and spin glass transitions.

\subsubsection{Four-point functions in supercooled liquids}
 
In the case of liquids, we may consider a certain space 
dependent observable $o({\bf x},t)$, such as, for example, the 
local excess density $\delta \rho({\bf x},t)= \rho({\bf x},t) - \rho_0$, 
where $\rho_0$ is the average density of the liquid, 
or the local dipole moment, the excess energy, etc. 
We will assume in the following that the average of $o({\bf x},t)$ 
is equal to zero, and the variance of $o({\bf x},t)$ normalized to unity. 
The dynamic two-point correlation is defined as: 
\be \label{codef}
C_o({\bf r},t) =  [o({\bf x},t=0) o({\bf x}+ {\bf r},t)]_{\bf x}, 
\ee 
where the normalization ensures that $C_o({\bf r}={\bf 0},t=0) = 1$. 
The Fourier transform of $C_o({\bf r},t)$ defines a generalized  
dynamic structure factor $S_o({\bf k},t)$~\cite{hansen}.  
All experimental and numerical results known to date suggest that as 
the glass transition 
is approached, no spatial anomaly of any kind appears 
in $C_o({\bf r},t)$ (or in $S_o({\bf k},t)$) although of course there  
could still be some signal which is perhaps too small to be measurable. 
The only remarkable feature is that the slowing down of the  
two-point correlation functions often obeys, 
to a good approximation, 
``time-temperature superposition'' in the $\alpha$-relaxation 
regime $t \sim \tau_\alpha$, i.e.: 
\be 
C_o({\bf r},t) \approx q_o({\bf r}) f\left(\frac{t}{\tau_\alpha(T)}\right), 
\ee 
where $q_o$ is often called the non-ergodicity (or Edwards-Anderson) 
parameter, and the scaling function $f(x)$ depends  
only weakly on temperature. This property will be used to simplify 
the following discussions, but it is not a needed ingredient.
 
The spatial correlations of the relaxation process
can be probed studying the distribution 
(over dynamical histories) of the correlation $C_o({\bf r},t)$, 
in particular its covariance. 
Quite generally, one expects that since $C_o({\bf r},t)$ 
is defined as an average over 
some large volume $V$, its variance $\Sigma^2_C$  
is of order $\xi^{2-\eta}/V$, 
where $\xi$ is the lengthscale over which $C_o({\bf r},t)$ is 
significantly correlated. 
More precisely
we define: 
\ba
\label{g4def} 
G_4({\bf y},t) & = & [ o({\bf x},0)  
o({\bf x}+ {\bf r},t)  
o({\bf x} + {\bf y},0) o({\bf x} + {\bf y} + {\bf r},t) ]_{\bf x}  
\nonumber \\ 
& & -  [ o({\bf x},t=0) o({\bf x}+ {\bf r},t) ]_{\bf x}^2 ,  
\ea 
and its space integral, 
\be 
\Sigma^2_C = \frac{1}{V} \int d{\bf y} G_4({\bf y},t), 
\ee 
which is nothing but the variance of the spontaneous 
fluctuations of $C_o({\bf r},t)$ averaged over a volume $V$.
This variance can thus be expressed as 
an integral over space of a four-point correlation 
function, which measures the spatial correlation of the 
temporal correlation. This integral over space is also  
the Fourier transform of $G_4({\bf y},t)$ with respect to ${\bf y}$  
at the wavevector ${\bf q}$ equal to zero. We want to insist  
at this stage that ${\bf r}$ and ${\bf y}$ in the above equations 
play very different roles: the former enters the very 
definition of the 
correlator we are interested in Eq.~(\ref{codef}), whereas the 
latter is associated with the scale over which the dynamics is  
potentially correlated. Correspondingly, great care should be 
devoted to distinguish the wavevector ${\bf k}$, 
conjugate to ${\bf r}$, and ${\bf q}$ conjugate to ${\bf y}$. 
 
Specializing to the case ${\bf r}=0$ (local dynamics), one 
finally obtains: 
\be 
\chi_4(t) \equiv N \Sigma^2_C.
\label{chi4def}
\ee 
The analogy with spin glasses developed above 
suggests that this quantity reveals 
the emergence of transient amorphous long-range order. 
Although, as we shall discuss, the situation is more 
complicated than what was originally surmised, 
this analogy was indeed the main motivation for the first 
numerical investigation of $\chi_4(t)$ in a supercooled 
liquid~\cite{dasgupta}.  It was later realized that $\chi_4(t)$ 
is in fact the natural diverging 
susceptibility in the context of $p$-spin  
descriptions of supercooled 
liquids, where a true dynamical phase transition occurs at a 
certain critical temperature~\cite{KiTh,franzparisi}.   
However, since in real systems no true 
phase transition can be  observed, one expects $\chi_4(t)$ to 
grow until $t \approx \tau_\alpha$ and decay back to zero  
thereafter. Until $\tau_\alpha$, there cannot be strong 
differences between a system with quenched disorder and a  
system where disorder is dynamically self-induced. 

Measuring the `local' relaxation suggests to follow 
the displacement of single particles over distances typically
corresponding to the interparticle distance.
Therefore, $\chi_4(t)$ can be accessed either by measuring 
the fluctuations
of the Fourier transform of $C_o({\bf r},t)$ evaluated at a 
wave-vector, $k_0$, of the order of the 
first peak in the structure factor \cite{berthier}, or 
by performing a 
spatial average $\int d{\bf r} C_o({\bf r},t) w({\bf r})$ where 
$w ({\bf r})$ is an overlap function equal to one for lengths 
of the order of $2\pi/k_0$ and zero otherwise \cite{glotzer2}
or a Gaussian function with a suitable width \cite{dauchotbiroli}. 
The dependence of dynamical correlations on the coarse-graining 
length has been thoroughly studied, both in simulations~\cite{Ck} 
and experiments \cite{dauchotbiroli,luca,abate}, 
showing that the dynamics becomes 
homogeneous when the coarse-graining is made too small 
(where dynamics is dominated by trivial thermal vibrations), or too 
large (because dynamics results in this limit from 
a succession of several uncorrelated rearrangements). 

\begin{figure}
\psfig{file=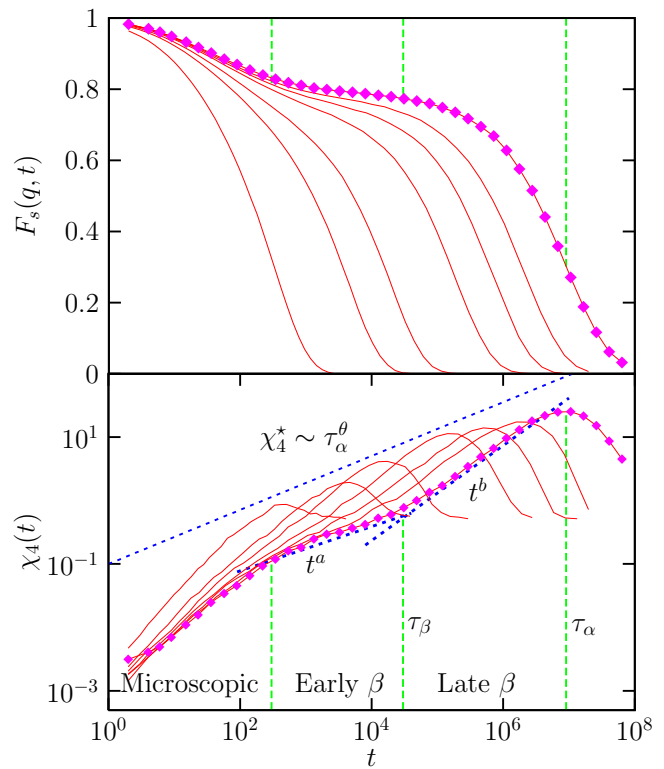,width=8.5cm}
\caption{\label{chi4ludo} 
Time dependence of the self-intermediate scattering function
(top), and its spontaneous fluctutations (bottom), 
for different temperatures decreasing from 
left to right in a Lennard-Jones supercooled liquid in Monte Carlo
simulations.  The lowest temperature is highlighted 
with symbols. 
For each temperature, $\chi_4(t)$ has a maximum 
near the relaxation time $\tau_\alpha$, which shifts to
larger times and has a larger value when $T$ is decreased,
revealing the increasing lengthscale of dynamic heterogeneity in 
supercooled liquids approaching the glass transition.
Moreover, the time dependence of $\chi_4(t)$ is characterized by several
distinct time regimes, corresponding to microscopic, 
early and late $\beta$ regimes, and structural relaxation, as indicated
by the vertical lines.}
\end{figure}

For supercooled liquids, 
the function $\chi_4(t)$ has been measured by molecular dynamics,
Brownian and Monte Carlo simulations in different 
liquids~\cite{glotzerfranzparisi,giorgioaging,glotzer1,glotzer2,glotzerreview,berthier,glotzersilica,berthiersio2,horacio,djamelaging}. 
Moreover, its behaviour has been theoretically investigated using 
various perspectives, as described below
in Sec.~\ref{theory}. 

As suggested by its definition in Eq.~(\ref{chi4def}), 
the four-point dynamic susceptibility is measured 
in practice through the repeated measurement
of a chosen local time-correlation function, $\chi_4(t)$ being 
nothing but the variance of the statistical fluctuations
between different measurements.

A typical example of such a measurement
is shown in Fig.~\ref{chi4ludo}, taken from 
Monte Carlo simulations of a simple Lennard-Jones supercooled liquid.
A similar qualitative behaviour is found in nearly all cases, 
as detailed in \cite{TWBBB}.
At a given temperature, $\chi_4(t)$ is an increasing function
of time at short times reflecting the fact that
dynamic heterogeneities slowly build up with time. 
It then has a  peak on a timescale that tracks the structural relaxation
timescale $\tau_\alpha$, and finally it decreases 
back to zero when $t \to \infty$.  When temperature 
is decreased the observed time evolution becomes slower, 
mimicking the overall slowing down of the dynamics
also seen in averaged two-time correlators, see Fig.~\ref{chi4ludo}.
The decrease at long times constitutes the above mentioned 
major difference with spin glasses. In a spin glass, 
$\chi_4(t)$ would be a monotonically increasing 
function of time whose long-time limit coincides with the static spin
glass susceptibility. 

The most important information extracted from the
temperature evolution of $\chi_4(t)$ is that, at least
in the range available to numerical simulations, 
the value of the peak at $\tau_\alpha$ 
increases typically from a high temperature value of 
order one, and increases by at most 2 orders of magnitude 
down to the lowest temperature  at which the system
can be equilibrated, suggesting that dynamics becomes 
spatially increasingly correlated when $T$ decreases.

As shown in Fig.~\ref{chi4ludo}, the time and temperature
behaviour of $\chi_4(t)$ is very rich. The growth
of $\chi_4(t)$ towards its peak value is composed of several
time regimes, closely reflecting the broad spectrum of relaxation
processes characterizing time correlation functions, 
see Fig.~\ref{chi4ludo}. The short-time dynamics
corresponding to the approach and departure
from the plateau in time correlation functions 
is also reflected in the time dependence of $\chi_4(t)$, 
which grows with distinct temporal power laws 
in the early and late $\beta$ regimes. 
Additionally, the temperature evolution of $\chi_4(t)$, 
and in particular the peak height, can be quantitatively studied.
This peak value, $\chi_4^\star$, measures the volume 
on which the dynamical processes relevant to structural 
relaxation at $t \approx \tau_\alpha$, are correlated.
It is found to increase when the temperature decreases and the 
dynamics slows down. In the temperature regime
above the mode-coupling temperature, the growth
is well-described by an algebraic relation 
between the peak amplitude, $\chi_4^\star$, and 
the relaxation time, $\tau_\alpha$, as shown in Fig.~\ref{chi4ludo}.
The different exponents introduced in Fig.~\ref{chi4ludo}
are discussed in \cite{TWBBB}.

More direct evidences of a growing dynamical correlation length 
can be obtained 
by measuring directly $G_4(\mathbf{y},t)$.  It has been checked 
that the increase of the peak of $\chi_4(t)$
corresponds, as expected, to a growing dynamic 
lengthscale $\xi$~\cite{onukia,glotzer1,glotzer33,dh,glotzer22,glotzer2,steve2,jcpI}. 
However, these 
measurements are much harder than the ones of $\chi_4$, because
very large systems need to be simulated to determine 
$\xi$ unambiguously \cite{AndersenStein,dasguptafss,szamelmct}. Note that if 
the dynamically correlated regions were compact, the peak 
of $\chi_4$ would be proportional to $\xi^3$ in three dimensions, 
directly relating $\chi_4$ measurements to that of the relevant
lengthscale of dynamic heterogeneity.

The study of the growth laws of $\chi_4(t)$, $\xi$ and the evolution 
of $G_4(\mathbf{y},t)$ contains 
very useful information to unveil the complexity
of the relaxation processes and to contrast theoretical 
approaches~\cite{TWBBB}.
In fact, many theories of the glass 
transition assume or predict, in one way or another, that the 
dynamics slows down because there are
increasingly larger regions over which particles have to relax in a  
correlated or cooperative way, see Sec.~\ref{theory}.

Furthermore, the growth of $\xi$ suggests that the glass transition 
is indeed a collective
phenomenon characterized by growing timescales and lengthscales,
reminiscent of critical phenomena. 
A clear and conclusive understanding of the relationship between the 
lengthscale obtained from $G_4(\mathbf{y},t)$ and the relaxation timescale is 
therefore still the focus of an intense research activity. 

\subsubsection{Three-point correlation and response functions} 
 
Although readily accessible in numerical simulations, $\chi_4(t)$ is 
in general very small and difficult to measure directly in 
experiments, except when the range of the dynamic correlation is 
macroscopic, as in granular materials~\cite{dauchot,dauchotbiroli,durian} or in
soft glassy materials \cite{weeks2}, where 
it can reach in some cases the micrometer and 
even millimeter range \cite{mayer,luca,lucasoft}.
To access $\chi_4(t)$ in molecular liquids, one should
perform time-resolved dynamic measurements probing very small volumes,
with a linear size of the order of a few nanometers. Although doable, 
such experiments remain to be performed with the needed 
accuracy.

It was recently realized that 
simpler alternative procedures exist \cite{science}. 
The central idea underpinning
these results is that induced dynamic fluctuations are 
in general more easily accessible than spontaneous 
ones, and both types of fluctuations 
can be related to one another by fluctuation-dissipation 
theorems. The physical motivation is that while four-point 
correlations offer a direct probe of the dynamic heterogeneities, 
other multi-point correlation functions give very useful and direct 
information 
about the microscopic mechanisms leading to these heterogeneities. For 
example, one might expect that the slow part of a local enthalpy 
(or energy, density,...) fluctuation per 
unit volume, $\delta h$, at position ${\bf x}$ 
and time $t=0$
triggers or eases the dynamics in its surroundings, leading to a 
systematic correlation between $\delta h({\bf x},t=0)$ and $o({{\bf 
x} + {\bf y}},t=0) o({{\bf x}}+ {\bf y}+ {\bf r},\tau_\alpha)$. This 
physical intuition suggests the definition of a family of 
three-point correlation functions that relate thermodynamic or 
structural fluctuations to dynamical ones. Interestingly, 
and crucially, some of these 
three-point correlations are both experimentally accessible and give 
bounds or approximations to the four-point dynamic correlations, 
as we now detail.

In the same way that the space integral of the 
four-point correlation function is the variance of the two-point 
correlation, the space integral of the above three-point correlation 
is the covariance of the dynamic correlation with energy 
fluctuations:  
\ba 
\label{cova2}
& &\Sigma_{CH}= \frac{1}{VN} \int d{\bf x}\,d{{\bf 
x'}} o({\bf x'}+ {\bf r},t) o({\bf x'},0) \delta h({{\bf x}},0) 
\nonumber \\ & \equiv & \frac 1 N \int d{\bf y}\, [ o({\bf x}+ {\bf y} + {\bf 
r},t) o({\bf x} + {\bf y},0) \delta h({\bf x},0)]_{\bf x}.   
\ea  
Note that for the enthalpy we use the 
notation $H(t=0)= \frac 1 N 
\int d{\bf x} h({{\bf x}},t=0)$, so that $h$ is an enthalpy per unit volume.
Hence, using the fact that the enthalpy fluctuations per particle are of order 
$\sqrt{c_P} k_B T$ (where $c_P$ is the specific heat in $k_B$ 
units), the quantity $N\Sigma_{CH}/\sqrt{c_P} k_BT$ defines the 
 number of particles over which enthalpy and dynamics are correlated.
Of course, analogous 
identities can be derived for the covariance with, 
say, energy or density fluctuations. 
 
Although interesting in itself, the covariance $\Sigma_{CH}$
is just as hard (or even harder) to measure experimentally
as $\chi_4$. However, $\Sigma_{CH}$ can be related, 
using linear response theory, to a response function which 
is much easier to access in experiments. Let us prove this by considering  
a system in the grand-canonical $NPT$  
ensemble. The probability of a given configuration $\cal C$ is 
given by the Boltzmann weight $\exp( -\beta H[{\cal C}])/Z$, 
where $\beta=1/k_B T$ and $Z$ is the grand-partition function.
Suppose one studies a static observable $O$ with the 
following properties: (i) $O$ only 
depends on the current microscopic 
configuration $\cal C$ of the system and (ii) $O$ can be written as 
a sum of local contributions:  
\be 
O = \frac{1}{V} \int d{\bf y}\; o({\bf y}). 
\ee 
In this case, a well-known static  
fluctuation-dissipation theorem holds~\cite{hansen}: 
\be 
\frac{\partial \langle O \rangle}{\partial \beta} = - 
\int d{\bf y} \, \langle o({\bf y}) \delta h({\bf 0}) \rangle  
\equiv - N\Sigma_{OH}, 
\label{classicfdt}
\ee 
where we decomposed the enthalpy in a sum of local  
contributions as well \cite{hansen}. 
 
Interestingly, in the case of {\it deterministic} 
Hamiltonian dynamics, the value of any local observable $o({\bf x},t)$ can
be seen as a highly complicated function of the initial 
configuration at time $t=0$. Therefore, a two-time correlation   
function, now averaged over both space and 
initial conditions, can be rewritten as 
a thermodynamical average: 
\ba 
C_o({\bf r},t;T) = \frac{1}{Z(\beta)V}  
\int d{\bf x}\; o({\bf x}+{\bf r},t)o({\bf x},t=0) 
\nonumber \\ 
\times  
\exp\left[-\beta \int d{{\bf x}}\; h({{\bf x}},t=0)\right]. 
\ea 
Hence, the derivative of the correlation with respect to 
temperature (at fixed pressure) directly leads, in the case of purely 
conservative Hamiltonian dynamics, to the covariance 
between initial energy fluctuations and the dynamical correlation, 
in direct analogy with Eq.~(\ref{classicfdt}). 
Defining 
\be 
G_T({\bf y},t)=\langle o({\bf y}+{\bf r},t)o({\bf y},0)\delta  
h({\bf 0},0) \rangle, 
\ee
one finds:
\be \label{fdt} 
\chi_T({\bf r},t) 
\equiv  \frac{\partial C_o({\bf r},t;T)}{\partial T}\bigg|_P  
=  \frac{1}{k_B T^2} 
\int d{\bf y}\; G_T({\bf y},t). 
\ee
Hence, the sensitivity of a two-time 
dynamical function to temperature, $\chi_T$, is directly  
related to a three-point spatial correlation function. 
The above result in Eq.~(\ref{fdt}) is extremely general and 
applies to many different situations. However, it 
does {\it not} apply when the dynamics is not 
Newtonian, as for instance for Brownian particles
or in Monte-Carlo numerical simulations~\cite{dh,berthierkob}.  
The reason is that in these cases, not only 
the initial probability but also the transition probability from the 
initial to the final configuration itself explicitly depends on temperature.  
In Brownian dynamics, for example, the noise in the Langevin equation 
depends on temperature. 
Hence, ${\partial C_o({\bf r},t;T)}/{\partial T}$ 
receives extra contributions 
from the whole trajectory, that 
depend on the explicit choice of dynamics.

The equality (\ref{fdt}), although in a sense 
a trivial result obtained from linear response theory, has a deep 
physical consequence,  
which is the {\it growth of a dynamical length upon cooling 
in glassy systems}, as we show now. 
Define $\tau_\alpha(T)$ such that $C_o({\bf
0},t=\tau_\alpha;T)=e^{-1}$ (say). Differentiating this definition
with respect to $T$ gives \be 0 = \frac{ \partial 
\tau_\alpha}{\partial T}
\frac{\partial C_o({\bf 0},t=\tau_\alpha;T)}{\partial t} +
\frac{\partial C_o({\bf 0},t=\tau_\alpha;T)}{\partial T}.  \ee Since
$C_o({\bf 0},t;T)$ decays from $1$ to zero over a time scale
$\tau_\alpha$, one finds that generically, using Eq. (\ref{fdt}): 
\be
\int d{\bf y}\; \frac{\langle o({\bf y},t=\tau_\alpha)o({\bf y},0)
\delta h({\bf 0},0) \rangle}{\rho_0 \sqrt{c_P} k_B T} \sim
\frac{T}{\rho_0 \sqrt{c_P}} \frac{\partial \ln \tau_\alpha}{\partial 
T}. 
\label{18}
\ee 
Now, since $\delta h$ is of order $\rho_0 \sqrt{c_P} k_B T$ and $\langle o^2
\rangle$ is normalized to unity, the quantity $\chi_0 \equiv
G_T({\bf 0},\tau_\alpha)/\rho_0 \sqrt{c_P} k_B T$ is not expected to appreciably
exceed unity. The above integral can be written as $\chi_0 v_T$, which
defines a volume $v_T$ over which enthalpy fluctuations and dynamics
are appreciably correlated. Note that the interpretation of $v_T$ as a
true correlation volume requires that $\chi_0$ be of order one, and its 
increase is only significant if $\chi_0$ is essentially 
temperature independent.  If this is not the case,
then the integral defined in Eq.~(\ref{18}) 
could grow due to a growing $\chi_0$
and not a growing length, which would obviate the notion that $v_T$
is a correlation volume.  

Assuming $\chi_0 \leq 1$, a divergence of the right hand side of 
the equality (\ref{fdt}) necessarily 
requires the growth of $v_T$. More precisely, as soon as $\tau_\alpha$ 
increases faster than an inverse power of temperature,  
the slowing down of a Hamiltonian system must necessarily be 
accompanied by the growth of a dynamic correlation length, which 
is therefore a general, powerful consequence of the use of linear
response theory. This result is thus directly relevant to 
supercooled liquids, where $\tau_\alpha$ typically increases 
in an activated manner, with, possibly, a finite temperature
dynamic singularity. From a theoretical perspective, it also 
implies that any theory predicting a dynamic singularity 
necessarily contains a prediction for diverging dynamic
lengthscales accompanying the glass 
transition~\footnote{An intriguing case, 
which is not fully understood, is the example of 
systems with an Arrhenius behavior at low 
temperature. The general considerations laid out in the text 
suggest that these systems are characterized by a
dynamical correlation length diverging at zero temperature, 
which contrasts with the idea that relaxation in Arrhenius 
systems is a simple, locally activated process. 
However, the present results only hold for energy conserving systems  
for which thermal activation may be more 
collective than usually surmised~\cite{TarziaBiroliTarjus}.}.
 
The study of these three-point correlation and response functions 
is therefore a useful path to characterize dynamical heterogeneity
and dynamical correlations. 
Quantitative results can be obtained studying experimentally $\chi_T$.
This has been done in connection with an inequality on $\chi_4$ that we 
shall describe 
in the following section. 
Another interesting development that will be 
discussed later on
consists in focusing on response functions, like $\chi_T$, but where the 
perturbing 
field is spatially dependent, e.g. with an oscillatory shape \cite{BBMR}. 
This allows one 
to probe directly the size and the shape 
of the dynamically correlated regions. 
 
Before concluding, let us stress that we have considered the response 
of time correlations to a temperature change, but
other perturbing fields may also be 
relevant, such as density, pressure, concentration of  
species in the case of mixtures, etc. For example, for 
hard-sphere colloids, temperature plays very little 
role whereas small changes of density can lead to enormous 
changes in relaxation times~\cite{naturepusey}.  
The derivation of this section can be straightforwardly extended 
to these cases \cite{jcpI}. 

\subsubsection{Inequalities on $\chi_4$ and experimental measurements}

In previous subsections, we argued that $\chi_4(t)$ is a 
fundamental quantity in order to understand
dynamic heterogeneities in supercooled liquids, but we
then proceeded to describe a series of alternative
multi-point susceptibilities, in particular $\chi_T(t)$, 
which contain alternative information on heterogeneities. 
We now close the loop and show that both types of 
susceptibilities are in fact not independent from one another, but
closely related.

This can be done using the general formalism developed long ago 
in \cite{lebo}, which gives expressions 
for the strength of fluctuations of physical 
observables measured in distinct statistical ensembles.
Applied to the spontaneous fluctuations of global 
two-time correlation functions, and considering 
transformation from $NPH$ (where enthalpy is fixed
and temperature fluctuates) 
to $NPT$ (where temperature is fixed but enthalpy fluctuates), 
one obtains: 
\be 
\chi_4^{NPT}(t) = \chi_4^{NPH}(t) + \frac{T^2}{c_P} 
\left(\frac{\partial C_o({\bf 0},t;T)}{\partial T} \bigg|_P
\right)^2, 
\label{leboH}
\ee 
where $\chi_4^{NPH}(t)$ is the variance of the correlation  
function in the $NPH$ ensemble where enthalpy does not fluctuate, a 
manifestly non-negative quantity. This allows us to obtain 
the following inequality: 
\be \label{chi4bound} 
\chi_4(t)  \geq \frac{T^2 \chi_T^2(t)}{c_P} = \frac{T^2}{c_P} 
\left(\frac{\partial C_o({\bf 0},t;T)}{\partial T} \bigg|_P \right)^2. 
\ee
 
Note that there is a simpler way to obtain the above inequality.
In the previous section, $\chi_T(t)$ was shown to be 
related to the covariance of enthalpy and dynamic fluctuations, $\Sigma_{CH}$. 
Since $\chi_4$ is related to the covariance of dynamic fluctuations 
$\Sigma_{CC}$, 
one can easily check that Eq.~(\ref{chi4bound}) is just a rewriting of the 
Cauchy-Schwarz bound: $\Sigma_{CH}^2 \leq \Sigma_C^2 \Sigma_H^2$, where  
$\Sigma_H^2$ is the variance of the enthalpy fluctuations, 
equal to $c_P(k_B T)^2/N$ in the $NPT$ ensemble.

The inequality (\ref{chi4bound}) 
is very interesting because the right hand side is 
an experimentally measurable quantity which therefore provides
a {\it direct lower bound} on $\chi_4$. Thus, if $T^2 \chi_T^2(t) / c_P$ 
increases substantially above one, $\chi_4$ has to increase as least 
as much if not more. 
In particular, as soon as $\chi_T$ increases faster 
than $T^{-1}$ at low
temperatures, $\chi_4$ will eventually exceed unity; since $\chi_4$ 
is the space 
integral of a quantity bounded from above, this again means that the 
lengthscale
over which the four-point correlation $G_4({\bf y},t)$ extends 
{\it has to grow} as the system gets slower and slower. 
Again, more quantitative statements 
require information on the amplitude and shape of 
$G_4({\bf y},t)$, which has to be provided from 
theoretical or numerical calculations.

Equation (\ref{leboH}) makes precise the intuition that dynamic 
fluctuations are partly induced by the fluctuations  
of quantities that physically affect the  
dynamic behavior~\cite{edigertheo,Donth}, 
in that case the enthalpy. 
The inequality (\ref{chi4bound}) provides a correct estimate 
of $\chi_4$ if there are no ``hidden'' variables  which
also contribute to the dynamic fluctuations. 
It is however quite difficult to determine 
whether such additional contributions exist.
Theoretical investigations in the context of
approximate models for the glass transition described 
in Sec.~\ref{theory}, and detailed numerical calculations
where all terms in Eq.~(\ref{leboH}) can be separately 
evaluated, greatly clarified this issue. 
The central conclusion is that the experimentally accessible response
functions which quantify the sensitivity of average
correlation functions to an infinitesimal change in control parameters
can be used in Eq.~(\ref{chi4bound}) 
not only to yield lower bounds to $\chi_4(t)$, 
but in fact to provide useful 
quantitative approximations to four-point functions. 
Although the relative precision on $\chi_4(t)$
provided by the bound at a given temperature 
is modest, the rate of growth is accurately reproduced.
\begin{figure}
\psfig{file=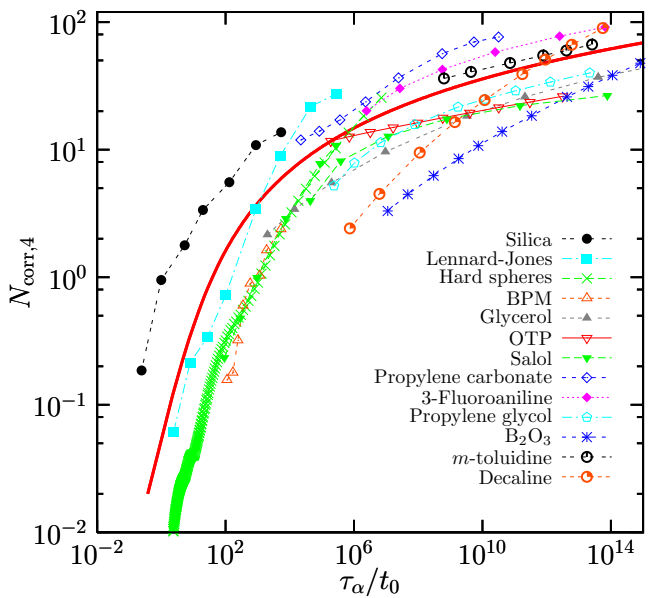,width=8.5cm,clip}
\caption{\label{cecile} Dynamic scaling relation 
between number of dynamically correlated particles, $N_{{\rm corr},4}$, 
and relaxation timescale, $\tau_\alpha$, for a number of 
glass-formers, determined using Eq.~(\ref{chi4bound}).
For all materials, a similar trend is found, with a rapid initial
increase of $N_{{\rm corr},4}$ near the onset of slow dynamics, 
followed by a slower, presumably logarithmic beahvior, closer to the laboratory
glass temperature. Adapted from \cite{cecile}.} 
\end{figure}

The results discussed above have opened the way to experimental
characterization of the growth of $\chi_4$ in molecular glass-formers close
to $T_g$. In order to make use of the inequality (\ref{chi4bound}) 
one must be able to 
detect time correlation functions at constant temperature,
$C_0(t; T)$ with sufficient precision
that dynamics at nearby temperatures, $T$ and $T + \Delta T$
can be resolved, when $\Delta T$ is small enough that 
linear response holds:
\be
\frac{\partial C_0(t; T)}{\partial T} \approx
\frac{C_0(t;T) - C_0(t;T+\Delta T)}{\Delta T}.
\ee  
A simpler alternative, critically discussed in \cite{cecile}, 
is to fit $C_0(t;T)$ with an empirical form containing a few 
number of parameters, and then take 
the temperature derivative of these parameters to 
indirectly estimate $\chi_T(t)$.  
Combining these different methods, the lower bound (\ref{chi4bound})
to $\chi_4$ has been computed for many molecular glass-forming liquids. 
It is easy to convince oneself that the lower bound has also the correct 
time dependence, with a peak developing for times of the order of the 
relaxation 
timescale. The value of this peak is therefore a lower bound to the peak 
of $\chi_4$ and, hence, to the number of dynamically correlated molecules, 
denoted $N_{{\rm corr},4}$. 
We show in Fig.~\ref{cecile} the evolution of $N_{{\rm corr},4}$
for many different glass-formers in the entire supercooled
regime \cite{cecile} as a function of the relaxation
timescale $\tau_\alpha$. 
Note that, actually, $N_{{\rm corr},4}$ is expected to be equal to  the 
number of dynamically correlated molecules 
up to a proportionality constant which is not known
from experiments, probably explaining why
the high temperature values of $N_{{\rm corr},4}$ are smaller than one. 
Figure~\ref{cecile} also indicates that $N_{{\rm corr},4}$ grows faster 
when $\tau_{\alpha}$ is not very large, 
close to the onset of slow dynamics, and 
a power law relationship $N_{{\rm corr},4} \sim \tau_\alpha^\theta$
(see Fig.~\ref{chi4ludo})
is a good description in this regime ($\tau_\alpha /\tau_0 < 10^4$). 
The growth of $N_{{\rm corr},4}$
becomes much slower closer to $T_{g}$. 
A change of 6 decades in time corresponds to a mere increase of a
factor about 4 of $N_{{\rm corr},4}$, suggesting logarithmic
rather than power law growth of dynamic correlations.
This is in agreement with several theories
of the glass transition which are based on 
activated dynamic scaling, see Sec.~\ref{theory}.

Note that all the results above 
can be generalized to cases where the order parameter 
inducing the glass transition is the density. This happens for colloids
and granular media \cite{science,lechenault,LucaaboveMCT}. 
One can obtain a new 
inequality where density fluctuations
play the same role of enthalpy fluctuations and that provide a lower 
bound to $\chi_4$
in terms of the derivative of the correlation function with respect to 
the density. The different inequalities yielding experimentally
accessible ways to quantify the strength of dynamic heterogeneity
certainly have the appeal of simplicity, and are by now 
routinely used in many different systems 
\cite{stevenson,zamponi,cecile2,chen,chit1,chit2,maggi,rolandcoslo}.

\subsection{Current status of dynamic heterogeneity studies}

\label{Multi-point-current}

The present section on dynamic heterogeneity was a very brief
summary of a collective research effort of very large amplitude 
that lasted about 20 years, and which already forms the 
core of several recent reviews \cite{sillescu,ediger,richertreview},
and a book \cite{leidenbook}. 
Progress to characterize, visualize, and quantify 
dynamic heterogeneity as well as an exploration of its detailed 
physical consequences have been truly dramatic in recent years.
The impact of this research is such that tools 
developed to study dynamic heterogeneity in liquids 
are now routinely used in scores of different systems, 
and `dynamic heterogeneity' is a concept that is commonly
employed in a broad range of situations, much beyond the physics 
of the glass transition \cite{leidenbook}. 

Despite this progress, several key questions are still unanswered.
Our discussion above has been focused on the issue
of the characterization of the spatial fluctuations 
involved in the phenomenon of dynamic heterogeneity. This is 
justified because direct measurements of growing dynamic
lengthscales have provided the long-sought evidence 
in favor of the collective nature of the glass transition itself. 
This fact being now established, it remains to 
understand more precisely and quantify the connection between these
growing lengthscales and the increasing viscosity of liquids
approaching the glass transition, which appears as 
a topic for research in the coming years.  

We have described in some detail the physical content 
of multi-point dynamic susceptibility such as $\chi_4(t)$. 
These functions have played a major role in the above story,
but we now understand that they contain also a number 
of embarrassing features. For instance, we mentioned how 
$\chi_4(t)$ retains a dependence on the statistical ensemble
where it is measured, as in Eq.~(\ref{leboH}), and that it
also depends on the microscopic equations of motion for the
system (Newtonian versus Brownian). These subtle issues make the analysis
of four-point susceptibilities somewhat ambiguous, especially
when estimates for lengthscales are sought. We shall describe below,
for instance in 
Sec.~\ref{imct:sec}, that alternative dynamic functions 
now exist that should be easier to analyze, but these have not all been
measured in simulations or experiments yet. Thus,
more detailed studies of a larger family of 
dynamic susceptibilities are certainly most wanted in the future.
A very promising avenue of research consists in studying 
non-linear responses \cite{nonlin}.
A first pioneering experimental measurement of non-linear 
dielectric susceptibility for glass-formers
appeared recently \cite{thibierge}.

Additionally, direct experimental measurements of 
dynamic lengthscales are still not available for molecular glass-formers, 
and are scarce even for colloidal materials. Thus, it would 
be useful to develop new experimental tools to resolve 
the dynamics of molecular glass-formers on small lengthscales
and longer timescales to obtain this much needed information.
It is not yet clear whether molecular dynamics simulations of model 
systems have covered a broad enough range of timescales and  
are thus relevant to understand the physics of real glass-formers
near the experimental glass transition temperature. 
We also believe that further work should be devoted to 
a better characterization of the geometry (and not only
typical lengthscale) of the dynamically heterogeneous 
regions, since contradicting results are available in 
the literature \cite{glotzerstring,kobdemos}. 

\section{Some models and theoretical approaches}
\label{theory}

\subsection{A few key questions}

We now present some theoretical approaches 
to the glass transition. It is impossible to cover all of them 
in this review, simply 
because there are way too many of them. This is perhaps the clearest 
indication that the glass transition 
remains an open theoretical problem. 

We have chosen to present in some detail those approaches that, 
we believe, contain keystone ideas and at the same time 
have a solid statistical mechanics basis. Loosely speaking, they 
have a Hamiltonian, can be simulated numerically, 
or studied analytically with tools from statistical mechanics.
Of course, the choice of Hamiltonian
is crucial and contains very important assumptions about the nature 
of the glass transition. All the 
approaches we present have given rise to 
unexpected results. One finds more in them than what was supposed 
at the beginning, which leads to new, testable predictions.  
Furthermore, with models that are precise enough, one  
can test (and hopefully falsify!) these approaches 
by working out all their predictions in great detail, 
and comparing the outcome to experimental data. Such 
detailed analysis is often not possible with `physical pictures', 
or simpler phenomenological modeling of the problem. 
Our drastic choice of theories leaves behind many 
other approaches that, although interesting, could not 
be covered without increasing the length of this review
beyond reasonable limits. Recent reviews are available
on these and we refer the interested readers to 
\cite{debenedetti,Donth,energylandscape,dyrermp,schweizer}.

Before going into the models and theories, we would like to 
formulate a few important questions that theoreticians
seek to address and that will guide our presentation of theories 
below:

(1) Why do the relaxation time and the viscosity 
increase when $T_g$ is approached? 
Why is this dramatic growth different from an Arrhenius
law?

(2) Can one understand and describe quantitatively the
broad relaxation spectra characterizing the
dynamical behaviour of supercooled liquids, in particular
non-exponential relaxations, and their 
evolution with fragility?

(3) Is there a deep relation between kinetics and 
thermodynamics (such as $T_0 \simeq T_K$), and why? 

(4)  Can one understand and describe quantitatively 
the spatio-temporal fluctuations of 
the dynamics? How and why are these fluctuations related to the 
dynamic slowing down?

(5)
Is the glass transition a collective phenomenon? If yes, 
of which kind? What is the correct `order parameter', and 
the nature of the associated transition?

(6) 
Is the experimental glass transition at $T_g$ 
the manifestation of a finite or zero temperature 
phase transition, sometimes called the `ideal glass transition'?
Or is there instead an avoided, hidden, or inaccessible 
transition?

(7)  Is there a geometric, real space explanation for the dynamic
slowing down that takes into account molecular degrees of freedom? 

(8)
Are there simplified (e.g. lattice) glass models which essentially capture 
the physics of the glass transition of molecular liquids?  
 
 Before embarking in detailed theoretical explanations it is important to 
 stress that the glass transition 
appears as a kind of `intermediate coupling' problem, since 
for instance typical correlation lengthscales are 
found to be at most a few tens of particle long close to $T_g$. 
As a consequence, recognizing and validating `the' correct 
theory is extremely difficult since key signatures
could be buried (and probably are) under preasymptotic, microscopic
details. These are probably useful also to make incorrect theories
appear reasonable. 
To obtain quantitative, testable predictions, one must therefore 
be able to work out also these preasymptotic effects.
This is a particularly difficult task, 
especially in cases where the asymptotic theory itself
has not satisfactorily been worked out yet. 
Therefore, at this time, theories are mainly judged
by their overall predictive power and theoretical consistency.  

\subsection{Mean-field free energy landscapes and Random First Order
Transition (RFOT) theory}

\label{giuliotheory}

\subsubsection{Mean-field glass theory and complex free energy landscapes}

\label{meanfield}

In the last two decades, three independent lines of research,  
Adam-Gibbs theory~\cite{ag}, mode-coupling theory~\cite{gotze2} 
and spin glass theory~\cite{beyond}, have been merged 
in a common framework to produce 
a theoretical ensemble that now goes under the name of 
Random First Order Transition theory (RFOT), a terminology introduced by 
Kirkpatrick, Thirumalai and Wolynes~\cite{KTW} 
who played, among many other
researchers, a key role in its development.
Here, we do not follow the rambling developments as they 
took place, but summarize RFOT theory in a more modern and unified way.
Note that our use of the name `RFOT' is much broader than
the more common, but much narrower meaning often implied 
in the literature. Reviews dedicated to different aspects of RFOT 
theory have appeared recently 
\cite{reviewRFOT,gotze2,cavagna-review,bbreview,reviewRFOTMP}.  

As discussed previously, two hallmarks of the dynamics of glass-formers are
 that (i) 
close to $T_g$ a liquid remain stuck for a very long time in amorphous 
configurations and (ii) the number of these 
configurations is exponentially large in the system size. 
RFOT theory starts as a 
mean-field approach to  these phenomena.
As such it has to be able to capture the right 
kind of symmetry breaking and deal with 
an exponential number of states. 

Broadly speaking, mean-field theory is based on 
the study of the free energy landscape
as a function of the order parameter. For example, for ferromagnets, in the 
Curie-Weiss approach, one computes the free-energy as a function of the 
global magnetization  by a mean-field approximation. 
This yields direct access to the nature and 
properties of the ferromagnetic phase transition, and a 
simple description of the low temperature phase since 
the two minima of the free energy energy correspond to 
the two ferromagnetic states. However, computing 
the free energy as a function of the global energy or density is not 
enough  for the glass transition, because one must 
deal with the existence of 
many different amorphous configurations. As a 
consequence one is forced to compute the free-energy $F$ as a function 
of the entire density field (instead of a single variable
as in the ferromagnetic transition),      
$F$ being defined through the Legendre transform. 

Consider for simplicity 
an interacting particle
lattice model, the generalization to continuum systems is 
straightforward. In the lattice case a given configuration 
is determined by the number of particles, $n_i$, on each site $i$. 
In order to define $F$, one first introduces the thermodynamic `potential'
\begin{equation}  
\label{Wfunction}
W(\{\mu_i\})=-\frac 1 \beta \log \sum_{\{n_i\}}\exp\left(-\beta 
H(\{n_i\})+\sum_i \beta \mu_i n_i\right),
\nonumber
\end{equation}
where $H(\{n_i\})$ is the Hamiltonian.
The free energy function $F(\{\rho_i\})$ is defined as
 \begin{equation}  \label{landscape}
F(\{\rho_i\})=W(\{\mu_i^*\})+\sum_i \mu_i^* \rho_i,
\end{equation}
where the $\mu_i^*$s satisfy the equations $\frac{\partial 
W}{\partial \mu_i}+\rho_i=0$ and, hence,
are functions of $\{ \rho_i \}$, which specifies an 
averaged density profile. Note that this construction 
can be generalized to spin systems replacing 
the positive integer $n_i$ by $\pm 1$ variables $s_i$. In this 
context $F$ is called the `TAP' free energy
since its introduction by 
Thouless, Anderson and Palmer in the context of 
mean-field spin glasses~\cite{TAP}. 
The generalization to continuum system can be also performed by 
replacing the discrete variable $n_i$ by a 
continuum density field $\rho({\bf x})$. In this case $F$ is called 
`density functional'~\cite{DFT}.  
 
The free energy landscape is the hyper-surface generated by 
scanning $F$ over all possible values of $\{\rho_i\}$. Its critical 
points, in particular the minima, play 
a crucial role. In fact, by deriving Eq.~(\ref{landscape}) with 
respect to $\rho_i$ one finds 
\be
\frac{\partial F}{\partial \rho_i}=\mu_i^*.
\ee
Thus, when there are no external fields 
(or local chemical potentials) the solutions of these equations are all 
the stationary points of the free energy landscape\footnote{For particle 
systems there is always a global chemical
potential $\mu$ fixing the number of particles. In this case, one 
includes the global term $\mu \sum_i n_i$ 
in the definition of $F$ so that all $\mu_i^*$ are zero.}.
    
What are the main features of $F$ for a system approaching the 
glass transition? 
Unfortunately, this question cannot be answered exactly for a 
realistic three-dimensional system. 
One has either to make use of approximations (as in the 
Curie-Weiss description of ferromagnets)
or focus on simplified geometries, such as 
mean-field Bethe lattices, which, hopefully, provide a good approximation 
to finite dimensional ones. 
    
A quite large number of such 
studies have led to similar results, and thus to 
a consistent mean-field picture. The free energy 
landscape becomes `rugged' at low temperature and 
characterized by many minima and saddle points. 
Actually, the number of minima is exponentially large in the 
system size, which suggests the definition of 
an entropy, called `configurational entropy' or `complexity': 
\be
s_c= \frac{1}{N} \log {\cal N}(f),
\ee 
where ${\cal N}(f)$ is the 
number of free-energy minima with a given free energy density
$f$ (per unit of free energy density).  
The density profile 
corresponding to one given minimum is amorphous and lacks any type of 
periodic long-range order, and different mimima are very different. 
This is a very welcome theoretical result, as real glasses 
can be found in a large number of different amorphous configurations,
which can be interpreted as mean-field free energy minima.

Assuming that all minima are mutually accessible, 
one can compute the thermodynamic properties, i.e. the partition 
function by summing over all states
with their Boltzmann weights: 
 \begin{equation}  
\label{Zsum}
Z=e^{-\beta W}=\sum_{\alpha}e^{-\beta f_{\alpha}N},
\end{equation}
where the sum runs over the minima.
Formally, one can introduce a free-energy dependent complexity, $s_c(f,T)$, 
that counts the number of free-energy minima with free-energy density $f$ 
at temperature $T$. The partition function of the system then reads:
\be\label{Zeq}
Z(T) = \int d f \, \exp\left[-\frac{Nf}{T} + Ns_c(f,T)\right].
\ee
For large $N$, one can as usual perform a saddle-point
estimate of this integral, that fixes the dominant value of $f$, 
noted $f^*(T)$:
\be
\left.\frac{\partial s_c(f,T)}{\partial f}\right|_{f=f^*(T)} = \frac{1}{T}.
\label{betaf}
\ee

\begin{figure}
\psfig{file=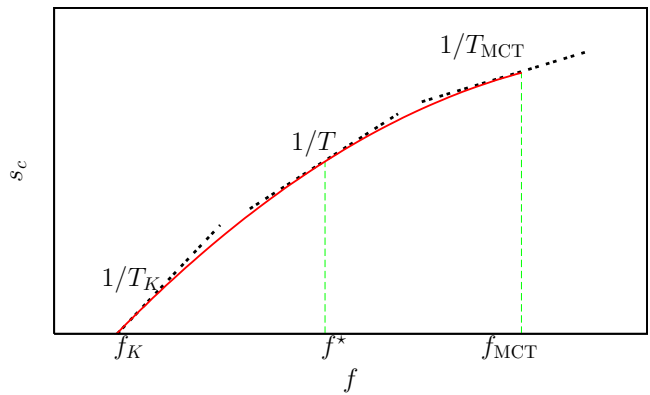,width=8.5cm}
\caption{\label{sc:fig} Typical shape of the configurational entropy, 
$s_c$, as a function of free energy density, $f$ in the range 
$T_K < T <T_{MCT}$ for random first order landscapes. 
A graphic solution of Eq.~(\ref{betaf}) is obtained by
finding the value of $f$ at which the 
slope of the curve is $\beta$. Note that $s_c$ 
is also a function of temperature, 
so this curve in fact changes with $T$.} 
\end{figure}

The temperature dependent complexity is therefore $s_c(T) 
\equiv s_c(f^*(T),T)$. The total free energy of the 
system is $f_p(T)=f^*-Ts_c(T)$. A typical shape 
of the configurational entropy as a function of $f$ and a 
graphic solution of Eq.~(\ref{Zeq}) are plotted in Fig~\ref{sc:fig}.
The analysis of the configurational entropy, or complexity $s_c(T)$, 
reveals that $s_c(T)$ decreases when temperature
decreases, as long as $T$ is above a critical 
temperature, $T_K$, below which $s_c(T)$ vanishes.
There exists also a second, higher temperature, $T_{MCT} > T_K$,
above which $s_c(T)$ drops discontinuously to zero again.
We use the symbol $T_{MCT}$ (as in mode-coupling theory) 
on purpose, and justify 
our choice below. 
Although the complexity vanishes in both regimes, the situations 
below $T_K$ and above $T_{MCT}$ are very different. At these two temperatures 
the part of the
 free-energy landscape relevant for the thermodynamics changes drastically 
in two very different ways.
Above $T_{MCT}$, there is just one minimum dominating the 
equilibrium measure corresponding
to the homogeneous density profile of the high temperature liquid. 
At $T_{MCT}$ the homogeneous liquid state becomes fragmented 
in an exponential number 
of states, or minima. 
At $T_K$ the number of minima becomes sub-exponential in the system size,  
such that $s_c(T < T_K)=0$.

Surprisingly the total free energy 
$f_p(T)$ is not singular at $T_{MCT}$. This is one of the most 
unexpected result consistently 
emerging from analytical solutions. This suggests that 
at $T_{MCT}$ 
the liquid state fractures into an exponential number of amorphous 
states, but that this transition has no thermodynamical
counterpart, and is therefore a purely dynamical phenomenon.
At $T_K$ instead, a thermodynamic phase transition takes place since the 
contribution to the entropy coming from the 
configurational entropy disappears, typically linearly, 
$s_c(T) \sim (T- T_K)$. Therefore, the specific heat is found to 
make a sharp downward jump at $T_K$, thus providing an exact realization
of the `entropy vanishing' mechanism conjectured 
by Kauzmann \cite{kauzmann}.
This is a second welcome result: the thermodynamic
signature of this mean-field transition mirrors the basic experimental finding 
that  the specific heat is nearly discontinuous at the experimental
glass temperature $T_g$. 
   
This rich physical behavior can be derived from a number of perspectives.
A first concrete example is given by `lattice glass 
models'~\cite{BiroliMezard} solved by the Bethe approximation or 
on Bethe lattices \cite{coniglio,bethe,weigt}.
Lattice glass models contain hard particles sitting on the sites of a 
lattice. The Hamiltonian is infinite 
if there is more than one particle on a site and,
more crucially, if the number of occupied neighbors of an occupied 
site is larger than a fixed parameter, $m$. The Hamiltonian is zero 
otherwise. Tuning the parameter $m$, changing 
the type of lattice, in particular its connectivity, can yield different 
models. Lattice glasses are simple 
statistical mechanical models mimicking the physics 
of hard sphere systems. Numerical simulations on cubic lattices have shown
that they seem to 
behave as {\it bona fide} glass-formers~\cite{coniglio,darst}. 

Alternatively, a density functional theory 
analysis of the free energy landscape yields
very similar results~\cite{stoesselwolynes,dasguptadft}. 
This is a more realistic microscopic starting 
point, but it inevitably contains some approximations, 
in particular related to the specific form 
of the free energy functional~\cite{stoesselwolynes,dasgupta}. 
The adopted form is the Ramakrishnan-Youssouf density functional
and most studies focused on hard sphere systems.  
In the first of these investigations \cite{stoesselwolynes} a particular 
amorphous profile,
whose only free parameter was the cage radius over which particles
are free to vibrate was plugged in the density functional. 
Minimization with respect to the cage radius revealed
that amorphous structures become stable, in a variational sense, 
at high enough density. 
More recent investigations performed a full minimization and 
reached qualitatively similar, but much more
detailed conclusions \cite{dasgupta,pinakidft}. 

Finally, other very popular models, are the ones introduced 
in the spin glass literature. Probably the most studied example 
of such spin glasses
is the $p$-spin model, defined by the Hamiltonian~\cite{MezardGross}
\begin{equation}
\label{pspin}
H= - \sum_{i_1,...,i_p}J_{i_1,...,i_p}S_{i_1}...S_{i_{p}},
\end{equation}
where the $S_i$s are Ising or spherical spins, 
$p>2$ and $J_{i_1,...,i_p}$ 
quenched random couplings
with zero mean and variance $p!/(2N^{p-1})$. 

These models are certainly not realistic
in terms of modeling microscopic degrees of freedom in a fluid, 
but they are representative of the class of systems with a random first 
order transition and 
have the advantage that a variety of computations 
can be performed without any approximation, and both their 
dynamic and static properties can be investigated analytically
in full detail, again yielding results as described above.
Their dynamics can be studied
in full detail, including various nonequilibrium 
conditions as described in Sec.~\ref{aging}.
Another fruitful result concerns the interpretation 
of the nature of the low temperature phase in terms 
of replica symmetry breaking, so that connection 
with the field of disordered systems can be made~\cite{beyond,giorgio}.
Technically, the thermodynamics 
of the $p$-spin can be solved, for $p>2$, using
a one-step replica symmetry breaking ansatz, see \cite{hessian}
for a review.
This means that the probability distribution function
of the overlap between states, the Parisi
function $P(q)$, has two peaks below $T_K$, one at 
$q(T)>0$ which quantifies the self-overlap within the states, 
and another one at $q=0$ implying that  different states
are totally uncorrelated~\cite{beyond}.  

Let us now discuss the dynamical behaviour which results from the 
above analysis of the free energy landscape. 
Below $T_{MCT}$, the system is in a metastable state
from which it cannot escape, because free energy barriers
diverge with system size \cite{barratburioni}. 
This divergence is a direct
consequence of the mean-field nature of the 
present set of approximations.
Therefore, the relaxation time
diverges, within mean-field, at $T_{MCT}$. The stability 
of these states can be analyzed  by computing the free energy Hessian
in the minima. One finds that states become 
more `fragile' when $T \to T_{MCT}^-$, are marginally stable 
at $T=T_{MCT}$, unstable for $T>T_{MCT}$. 
As a consequence, one expects that the dynamics slows down approaching 
$T_{MCT}$ from above as the landscape becomes 
more and more `flat' \cite{laloux}. 

\begin{figure}
\psfig{file=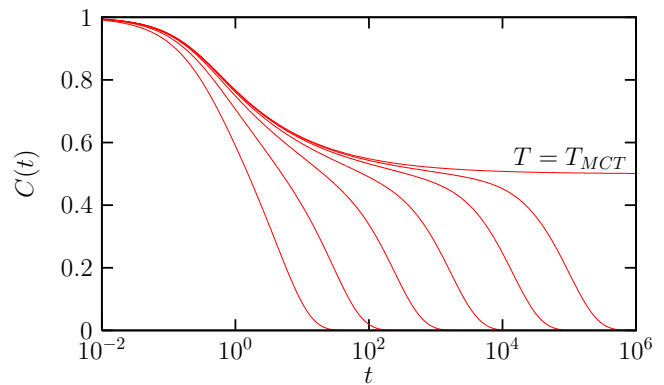,width=8.5cm}
\caption{\label{schematic:fig} Correlation $C(t)$ as a function of time 
for the $p$-spin model with $p=3$ 
for several temperatures approaching $T_{MCT}$, obtained 
from numerical solution of Eq.~(\ref{schematicp}). The curves show the 
appearance of a two-step decay characterized by several 
scaling laws discussed in the text.}
\end{figure}

Indeed the dynamics of many of these models can be
analyzed exactly~\cite{reviewLeticia}. In particular, mean-field $p$-spin 
models have been 
analyzed in great detail and provide a paradigm for mean field glassy 
dynamics. 
The equations of motion considered in the literature are Langevin equations, 
\be
\frac{\partial s_i(t)}{\partial t} = -\mu(t) s_i(t)
-\frac{\partial H}{\partial s_i(t)} + \eta_i(t),
\label{langevin}
\ee
where $\eta_i(t)$ is a Gaussian thermal noise of zero mean and variance 
$2T$ given by the fluctuation-dissipation theorem.
 
\begin{figure}
\psfig{file=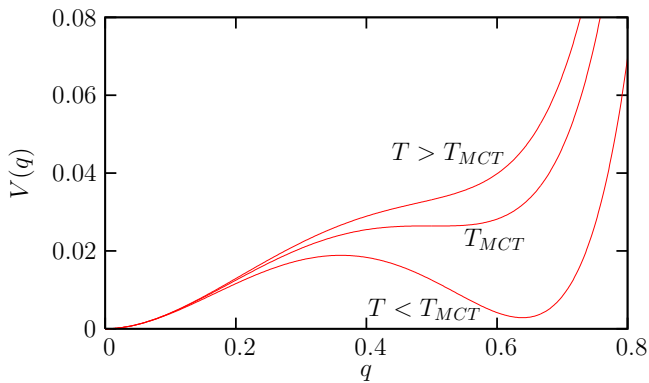,width=8.5cm}
\caption{\label{potential:fig} Evolution of $V(q)/T=-\frac{1}{2T^2}
q^{p}-\log (1-q)-q$ as a function of $q$ for different temperatures
across the dynamical transition. The solution with $q>0$
appears discontinuously at $T_{MCT}$.}  
\end{figure}

We focus on the spherical version of the model, and on the 
time autocorrelation function $C(t) =\frac 1 N\sum_i 
\langle s_i(t) s_i(0) \rangle$.
Note that $\mu(t)$ is the Lagrange multiplier
enforcing the constraint $C(0)=1$.
The equation of motion for $C(t)$ 
at thermal equilibrium reads:
\be
\frac{dC(t)}{dt} = -T C(t) - \frac{p}{2T} \int_0^t dt' 
C^{p-1}(t-t') \frac{d C(t')}{dt'}.
\label{schematicp}
\ee
We will meet this equation again in the next section about mode-coupling 
theory. 
We shall then postpone a detailed study and just anticipate some 
results that will
be derived later.  At high temperature, the correlation function 
decays quickly to zero.
Decreasing the temperature, the relaxation timescale increases and 
a two-step relaxation emerges,
see Fig \ref{schematic:fig} where we have plotted the numerical 
solution of the previous equation. 
At $T_{MCT}$ the timescale $\tau_\alpha$ corresponding to the 
slow relaxation diverges algebraically, 
\be
\tau_{\alpha} \sim \frac{1}{(T-T_{MCT})^\gamma},
\label{gamma}
\ee
where $\gamma$ is a critical exponent. The value of the plateau 
$q=\lim_{t\rightarrow \infty}C(t)$, 
called Edwards-Anderson parameter in the spin glass literature, 
satisfies a simple equation that can be obtained taking the infinite 
time limit of Eq.~(\ref{schematicp}):
\be
\label{qeq}
\frac{q}{1-q}=\frac{p}{2T^2}q^{p-1}.
\ee
A graphical solution of this equation 
is presented in Fig.~\ref{potential:fig}, where we plot 
$V(q)/T=\int_0^q dq' \left[\frac{q'}{1-q'}-\frac{p}{2T^2}q'^{p-1}\right]$. 
The minima of $V$
are the solution of Eq.~(\ref{qeq}). Clearly the minimum at $q=0$ is always 
present. Another solution, $q_{EA}$, appears 
at $T_{MCT}=\sqrt{p(p-2)^{p-2}(p-1)^{1-p}/2}$  and it can be 
interpreted as the long time limit of the correlation function inside 
one typical state.
Since the states have an infinite lifetime (in mean-field theory) the 
system remains trapped forever into the one
it started from.  It is important to remark that $q_{EA}$ is 
discontinuous at transition, which leads to the two-step behavior shown in 
Fig.~\ref{schematic:fig}. By contrast, $q_{EA}$ is 
continuously growing from 0 at the transition in spin glasses. 

Note from Fig.~\ref{potential:fig} that at 
$T_{MCT}^{+}$ the Edwards-Anderson parameter is zero and, concomitantly, 
$V(q)$ has a vanishing second derivative 
at $q_{EA}$. It is possible to show \cite{franzparisipotential}, that 
this is indeed related to the fact that the 
free energy Hessian of the states below $T_{MCT}$ develops zero modes at 
$T_{MCT}$. This behavior resembles very much 
the one of a spinodal transition. 
In fact, this analogy has been fruitfully explored, 
for instance to describe real space features of  
dynamic heterogeneity near $T_{MCT}$
\cite{KTW2,StevensonSchmalianWolynes,cavagna-review}.

Let us recap the overall picture arising from a mean-field 
analysis of the properties of the free energy 
landscape of glasses. 
At high temperature the dynamics is fast and the system is 
in the liquid state. 
Approaching $T_{MCT}$ the dynamics slows down because of the appearance 
of incipient stable states. 
Decreasing the temperature to $T_{MCT}$, it takes a  longer time 
to find an unstable direction, and thus to relax. 
Below $T_{MCT}$ there is an 
exponential (in the system size) number 
of states. The partition function and the thermodynamics are obtained by 
summing over all of them their corresponding
Boltzmann weight. This procedure is justified by the fact that, in a 
real finite dimensional system, the barriers between states 
should actually become finite. In the regime below $T_{MCT}$,
there is a competition between single state free energies that would 
favor the lowest free energy states, and 
configurational entropy that would favor the highest ones that are more 
numerous. Lowering the temperature
disfavors the entropy term and at $T_K$ the system 
undergoes a phase transition where  the sum in Eq.~(\ref{Zsum}) is 
again dominated by only few terms corresponding to states with 
free energy density $f_K$ given by $s_c(f_K,T)=0$. This transition 
corresponds to a {\it bona fide}  `entropy crisis' mechanism.

\subsubsection{Mode-coupling theory (MCT)}

\label{secmct}

The dynamical transition 
appearing upon approaching $T_{MCT}$ in RFOT 
landscapes is mathematically analogous 
to the one predicted to occur in supercooled liquids
by the Mode-Coupling Theory (MCT) 
of the glass transition,
although the latter has {\it a priori} no direct interpretation
in terms of a free energy landscape. 
This theory was introduced separately
by Leuthesser \cite{Leuthesser} and  Bengtzelius, 
G{\"o}tze, Sj{\"o}lander and 
coworkers~\cite{BGS}. It has been used to describe and predict 
the average dynamics, in particular the dynamical structure factor and 
the self-diffusion, for moderately supercooled 
liquids and colloids. Recently, it has been generalized to describe
dynamical correlations and some aspects of dynamic heterogeneity,
as described in Sec.~\ref{imct:sec}. 
In Sec.~\ref{current-MCT}, we discuss successes and limitations of MCT.

Originally, MCT was developed using the projection operator 
formalism~\cite{Leuthesser,BGS}. A good introduction to this 
method can be found in the book by Zwanzig \cite{Zwanzigbook}.
The starting point of the method is the derivation of the 
following equation for the dynamical structure factor $F(k,t)$
of a single component atomic liquid:
\begin{eqnarray}
\label{eq:exactmemory}
& & \frac{d^2F(k,t)}{dt^2}+\frac{k^2k_BT}{mS(k)}F(k,t) + \nonumber \\
& & \int_0^t d\tau M(k,\tau)
\frac{d}{dt}F(k,t-\tau) = 0\mathrm{.}
\end{eqnarray}
Generalizations to mixtures
and non-atomic liquid are also available.
This is an exact equation whose inputs are the static structure 
factor $S(k)=F(k,0)$ and the memory kernel $ M(k,\tau)$ for a given
particle mass $m$, and temperature, $T$. 
In a second, crucial step MCT 
suggests a self-consistent approximation for the 
memory kernel $ M(k,\tau)$. It is possible to 
show that the memory kernel 
corresponds to 
the variance of the random force acting on the density field,
see the review  \cite{CharbonneauReichmanreviewMCT}.
Thus, $ M(k,\tau)$ captures the effect of all 
degrees of freedom other than the density field on the density field itself. 
The physical idea motivating MCT is to focus on the slow part of the 
random force. Technically, 
the path is in principle straightforward: 
one should identify the dominant slow modes, project 
the random force onto them,
and derive the dynamical equations for their correlation functions. 
Of course, in practice this remains difficult 
because the number of slow modes is infinite.

Within MCT, only the bilinear density products
contribute to the slow part of $M$. After projection, 
the memory kernel is expressed in terms of a four-point function. 
In a final approximation, this function is  factorized
as the product of two-point density functions $F(k,t)$. 
This leads to the MCT self-consistent equations:
\begin{eqnarray}
\label{eq:MCTeq}
& & 0 = \frac{d^2F(k,t)}{dt^2}+ \nu(k)\frac{dF(k,t)}{dt}+
\frac{k^2k_BT}{mS(k)}F(k,t)+ \nonumber \\ 
& & \int_0^t
d\tau M_{MCT}(k,t-\tau)\frac{\partial F(k,\tau)}{\partial \tau} ,
\label{vertex1}\\
& & M_{MCT}(k,t)
=\frac{\rho k_BT}{16\pi^3 m}\int d\mathbf{k'}|\tilde{V}_{\mathbf{k}-
\mathbf{k'},\mathbf{k'}}|^2 \times \nonumber \\
& &
\quad \quad \times F(k',t)F(|\mathbf{k'}-\mathbf{k}|,t), \nonumber \\ 
& &  \tilde{V}_{\mathbf{k}-\mathbf{k'},\mathbf{k'}} \equiv 
\left\{(\mathbf{\hat{k}}\cdot\mathbf{k'})c(k')+
\mathbf{\hat{k}}
\cdot(\mathbf{k}-\mathbf{k'})c(|\mathbf{k}-\mathbf{k'}|)\right\} ,
\label{vertex2}
\end{eqnarray}
where we have rewritten the result using the direct correlation function 
$c(k)\equiv \left( 1- 1 / S(k) \right) /\rho$. 
The effective friction term represents the effect 
of the fast degrees of freedom.

This final expression clearly
shows that MCT is a particular closure of the equations on 
dynamical correlation functions. It is similar in spirit to 
several other closure schemes used in physics, 
such as Kraichnan's `Direct Interaction 
Approximation' for turbulence, or various large-$N$ field theoretical 
methods~\cite{MCTdiagram1}. Indeed, field theoretical 
derivations of MCT have long been sought, but this 
is in fact still a very active area.
The first pioneering works were published 
shortly after the original MCT derivation~\cite{DasMazenkoprl,Das1990}. 
The authors started from stochastic equations for the slow 
degrees of freedom of a liquid, the so-called non-linear fluctuating 
hydrodynamics, and rederived the MCT equations as a self-consistent, 
one-loop approximation. 
Motivations for the field-theoretical approach are that
it provides a complementary way to derive MCT which is in 
principle more suitable to nonequilibrium generalizations, 
and perhaps to systematic improvement.
Unfortunately this approach is plagued by difficulties related
to the preservation of time-reversal symmetry in self-consistent loop 
expansions \cite{MiyazakiReichman,ABL}. Very recent work aimed at 
getting fully consistent field-theoretical derivations of MCT equations
\cite{KimKawasaki,Hayakawa}, but this is technically
more involved than could naively be anticipated.

Within MCT, dynamical correlation functions are obtained by 
numerical integration, once the static structure factor of the liquid is 
known. This kind of analysis was performed on a
large number of different glassy liquids such 
as Lennard-Jones \cite{Bengtzelius,walternauroth} 
hard spheres systems \cite{barraths,barratlatz,foffi}, or 
silica melts \cite{sciortinokob}, 
and it has revealed a common behavior. It is found that
small changes in $S(k)$ lead, at high density or 
low temperature, to a great variation in $F(k,t)$ that resembles very much 
the one shown in Fig.~\ref{schematic:fig} for the correlation function of 
the $p$-spin model. The MCT equations display 
a dynamical glass transition to a phase where 
the average density of the liquid remains frozen in an amorphous profile.
 
The similarity with the $p$-spin model is not casual,  and 
there is indeed a deep connection between the two. 
A first, somewhat technical, way to unveil 
it consists in simplifying the wavevector 
dependence of the equations assuming that the
integral over $k$ is dominated by values close to $k_0$ where 
the structure factor has a strong peak.
This so-called `schematic' approximation
\cite{Leuthesser,BGS,gotze}, yields a 
simplified equation of motion for $F(k_0,t)$ that reads:
\ba
\nu(k_0)\frac{dF(k_0,t)}{dt} & = & 
-\frac{k_0^2k_BT}{mS(k_0)}F(k_0,t) \\ & - &  \frac{k_0 A^2}{8\pi^2 \rho} 
\int_0^t dt' 
F^{2}(k_0,t-t') \frac{d F(k_0,t')}{dt'}. \nonumber
\label{schematic}
\ea
where $A$ is the area under the peak of $S(k)$ at $k_0$. 
A simple change of variables maps this equation to 
that of the $3$-spin model, Eq.~(\ref{schematicp}).
This relation with fully-connected models 
suggests that MCT should be interpreted as a mean-field 
approximation. Note that this does not imply that
MCT becomes exact in the limit of large spatial dimensionality, 
as shown by recent calculations \cite{KUNI,schilling}.

The solution of MCT equations displays a 
very rich phenomenology as seen in Fig.~\ref{schematic:fig}.
There are three time-regimes. A fast relaxation toward a plateau, 
whose value depends on $k$, a slow relaxation close to the plateau, 
called $\beta$-regime, 
and finally the structural relaxation on the timescale of the 
$\alpha$-regime; see \cite{MCTreviewkob,MCTreviewdas} for more details.
In the following we shall denote $\epsilon$ the relative distance from 
the transition. For molecular liquids the control parameter is the 
temperature and, hence,
$\epsilon=(T-T_c)/T_c$; for colloids the control parameter is the
density and so $\epsilon=(\phi_c-\phi)/\phi_c$. 

Keeping track of the wavevector dependence, 
the detailed properties of the dynamics in the three regimes are 
as follows:

(1) {\it Fast relaxation --} Some degrees of freedom relax on timescales
of the order of $\tau_0$, even close to the transition. This regime is 
identified taking
the limit $\epsilon \rightarrow 0$ and keeping $t$ finite. In this case
$F(k,t)$ approaches a plateau at long times whose value is denoted $f_kS(k)$.
The non-ergodic parameter, $f_k$, is the fraction of density fluctuations
that become frozen at the transition. At large times the behavior of $F(k,t)$
is: 
\begin{equation} 
F(k,t) \approx  f_k + \frac{h(k)}{t^a}, \qquad t \gg \tau_0.  
\end{equation}
The exponent $a$ satisfies the equation
\begin{equation}
\label{gt-gen-a}
\frac{\Gamma^2(1-a)}{\Gamma(1-2a)}=\lambda,
\end{equation} 
where $\lambda$ is a number that can be computed using 
the structure factor only. Its expression is complicated,
see \cite{lambda}. The previous equation implies $0\leq a<1/2$. 

(2) {\it $\beta$-regime -- } In this sector, the timescale diverges as 
$\tau_\beta \sim \epsilon^{-1/2a}$ 
and the dynamical structure factor scales as:
\begin{equation} 
F(k,t) \approx f_k+\sqrt{\epsilon}h(k)g(t/\tau_{\beta}),
\end{equation}
where $g(x)\propto x^{-a}$ for $x \ll 1$ and $g(x)\propto x^b$ for $x 
\gg 1$. Note that the previous expression implies that all the $k$ dependence 
factorizes and is contained in $h(k)$ only, the so-called `factorization 
property'.  
The exponent $b$ satisfies the equation 
\begin{equation}
\label{gt-gen-b}
\frac{\Gamma^2(1+b)}{\Gamma(1+2b)}=\lambda,
\end{equation}
which implies that $0 \leq b \leq 1$. 

(3) {\it $\alpha$-regime --} In this sector the timescale diverges as 
$\tau_\alpha \sim \epsilon^{-\gamma}$, 
where $\gamma=1/2a+1/2b$. The factorization property 
does not hold anymore except for small $t/\tau_{\alpha}$ because 
the solution has to match the one found in the $\beta$-regime.

We refer again the reader to Fig.~\ref{schematic:fig} 
for a visual illustration of the different time regimes
predicted by MCT for dynamic structure factors in 
supercooled liquids.

Mode-coupling theory provides predictions also for other correlators such 
as the self-intermediate scattering function from which the 
mean-squared displacements and thus the self-diffusion
coefficient can be obtained. The previous properties remain essentially 
unaltered and all correlators display quite similar scalings.

\begin{figure}
\psfig{file=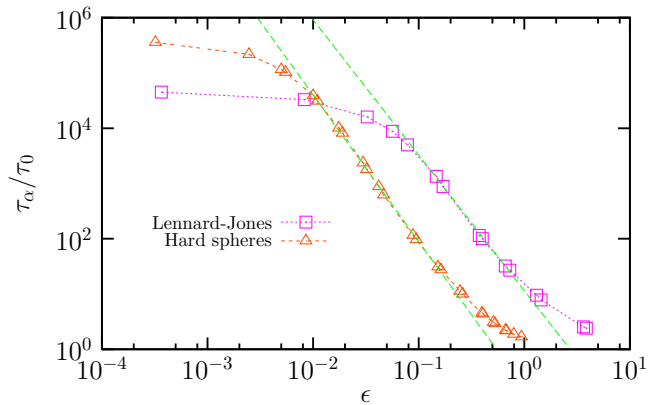,width=8.5cm}
\caption{\label{mct:fig} Fit of the evolution 
of the equilibrium structural relaxation time of a Lennard-Jones 
liquid (temperature is varied) and 
a hard sphere fluid (density is varied)
with the predicted MCT algebraic divergence, 
$\tau_\alpha \sim \epsilon^{-\gamma}$, where
$\epsilon$ is the reduced distance to the transition.
The plateau at low $\epsilon$ shows that
$\tau_\alpha$ remains finite at $\epsilon=0$, and 
the data directly indicate that the 
transition can actually be crossed at thermal 
equilibrium and is thus avoided.}
\end{figure}

All these predictions have been tested in great detail in numerical 
simulations and in experiments both on molecular liquids and in colloids.
It has also been shown that adding corrections to MCT does not spoil 
the main predictions and the universality of MCT has been established 
\cite{universalityMCT}. 
Different reviews \cite{gotze,MCTreviewkob,MCTreviewdas}
have already appeared on these tests. When fitting data 
using MCT, a central difficulty arises from the fact that
the actual transition is not present, as expected
from its mean-field nature. The absence of a 
genuine mode-coupling singularity is undisputed 
for molecular liquids both in simulations and 
experiments~\cite{gotze2}. 
Recent numerical and experimental works suggest that 
the same situation holds in hard sphere 
systems~\cite{krauth,LucaaboveMCT,szamelmct}. 
This is illustrated in 
Fig.~\ref{mct:fig} where the predicted MCT algebraic divergence
of the structural relaxation time for a Lennard-Jones liquid 
and a hard sphere fluid are superimposed on numerical data. While 
the fit is accurate over a window of 2 to 3 decades, it clearly 
fails to capture the low temperature or large density regimes of these
systems. 
Clearly, therefore, mean-field concepts 
cannot directly be applied to understand the glass transition, 
and a more refined analysis is needed.   

In conclusion, MCT predicts a transition where the system has a dynamical
arrest, particles stop to diffuse and the density becomes frozen around
an amorphous profile. Additionally, MCT yields non-trivial predictions for the
behavior of dynamical correlators that serve as a guideline in the study
of moderately supercooled liquids.  

\subsubsection{Dynamical correlations within Mode Coupling Theory}

\label{imct:sec}

We introduced MCT as a dynamical theory for 
two-time correlation functions. However,
the recent surge of interest on high-order correlation 
functions as probe of dynamic correlations and dynamic
heterogeneity suggests that it could be interesting 
to develop an MCT approach also for multi-point correlation
functions.

The MCT transition was originally described as a local phenomenon,  
the self-consistent blocking of the particles in their 
`cages'~\cite{gotze}. On general grounds, 
a diverging relaxation time is expected to arise from processes
involving an infinite number of particles (leaving aside the case of
quenched obstacles), as discussed in Sec.~\ref{Multi-point} and 
established by rigorous results \cite{MS0}. 
Actually, even the cage mechanism requires some kind 
of collective behaviour: in order to be blocked by one's 
neighbors, the neighbors themselves
must be blocked by their neighbors and so on 
until a certain scale that, intuitively, should 
set the relaxation timescale of the system. 
So one expects that even within MCT, `cages' should in fact 
be described as spatially correlated objects \cite{BBMR}.
It has indeed been shown that  
all susceptibilities and correlations 
defined in Sec.~\ref{Multi-point},
such as $\chi_4$ and $\chi_T$, diverge at the MCT dynamic
singularity.

Historically, the `local-cage' point of view was challenged in 
the context of 
mean-field disordered systems in \cite{franzparisi} 
(see \cite{KiTh,KTW2} for early results), since
these models are analogous to schematic MCT  equations.
Franz and Parisi argued that a dynamical susceptibility
similar to $\chi_4(t)$ diverges at the 
dynamical mode-coupling
transition. The first full MCT analysis was
developed in \cite{BB},  using a field-theoretical approach.
This clearly showed the existence of a diverging lengthscale 
within MCT. Later, a different susceptibility, $\chi_{\bq_0}(\bq_1,t)$
was introduced~\cite{BBMR}, which quantifies 
the response of the dynamical structure factor to a static oscillatory 
external perturbation with wavevector ${\bq_0}$. 
For a perturbation localized at the origin, $U(\bx)=U_0\delta(\bx)$, one finds 
$\delta F(\bq_1,\by,t)=U_0\int d\bq_0e^{i\bq_0\cdot\by}\chi_{\bq_0}(\bq_1,t)$. 
This susceptibility is akin (although not exactly related) to a 
three-point density correlation 
function in the absence of the perturbation. Although quite different from 
the four-point functions considered previously in the literature, 
$\chi_{\bq_0}(\bq_1,t)$ is expected to reveal the 
existence of a dynamical correlation length 
of the homogeneous liquid as well.
Physically indeed, $\chi_{\bq_0}(\bq_1,t)$ measures the influence of a density 
fluctuation at a given point in 
space on the dynamics elsewhere. Additionally
its scaling form is not affected by complications
due to conservation laws, as is the case for four-point correlators, 
see Sec.~\ref{Multi-point-current}. Recent work on Kac glassy models
also found a similar diverging length at the MCT transition 
\cite{FranzMontanari}
using replica techniques. 

Let us summarize the critical properties found for 
$\chi_{\bq_0}(\bq_1,t)$ in \cite{BBMR}. As
for the dynamical structure factor there is a different critical 
behavior in the $\beta$- and $\alpha$-regimes, although there is a 
unique diverging correlation lengthscale $\xi\propto \epsilon ^{-1/4}$.

(1)  In the $\beta$-regime, one has:
\be
\chi_{\bq_0}(\bq_1,t)=\frac{1}{\sqrt{\epsilon}+\Gamma q_0^2}S(q_1)h(q_1) 
\, g_{\beta}\left(\frac{q_0^2}{\sqrt{\epsilon}},
\frac{t}{\tau_{\beta}}\right)
\ee
where $g_{\beta}$  is a scaling function and $\Gamma$ a positive 
constant, which are both obtained 
quantitatively \cite{BBMR}. In particular, $g_\beta$ 
behaves as a $\sim t^a$ for small $t / \tau_{\beta}$, and as $\sim t^b$ 
for large $t / \tau_{\beta}$.

(2) In the $\alpha$-regime the critical behavior of 
$\chi_{\bq_0}(\bq_1,t)$ is:
\be 
\label{latebeta}
\chi_{\bq_0}(\bq_1,t) =\frac{\Xi(\Gamma q_0^2/
\sqrt{\epsilon})}{\sqrt{\epsilon} (\sqrt{\epsilon} + \Gamma q_0^2)} 
\, g_{\alpha,q_1}\left(\frac{t}{\tau_\alpha}\right),
\ee
with $\Xi$ a certain regular function with $\Xi(0) \neq 0$ and 
$\Xi(v \gg 1) \sim 1/v$ such that $\chi_{\bq_0}$ behaves as 
$q_0^{-4}$ for large $q_0$, independently of $\epsilon$. 
Also, $g_{\alpha,q_1}(u \ll 1) = S(q_1)h(q_1)u^b $, to match the 
$\beta$-regime, and  $g_{\alpha}(u \gg 1,q_1) \to 0$. 

The behavior of four-point quantities, such as $G_4$ and 
 $\chi_4$, is more complicated because of 
ensemble dependencies and the influence of conservation laws, 
see \cite{jcpI} for a detailed discussion. 
It is found that $G_4$ and $\chi_4$ should have a 
similar critical behavior, but 
not too close to the MCT transition, crossing
over to a distinct behaviour in its vicinity,
due to conservation laws. 
Therefore, for decisive tests of MCT predictions regarding
multi-point functions, the quantity 
$\chi_{\bq_0}(\bq_1,t)$ should be preferred, but no such study has been
reported yet. 
 
The study of critical properties of dynamical correlations for 
glassy liquids and, hence, the comparison 
with MCT predictions is still in its 
infancy \cite{TWBBB,Szamelnumchi4,jcpI,jcpII,AndersenStein,szamelmct}. 
At the time of this writing many questions 
remain open and deserve further studies. The 
determination of the dynamic correlation length is subtle and 
because of the complication brought
about by the existence of conserved quantities $G_4$ could be 
difficult to analyze. Certainly a numerical
study of $\chi_{\bq_0}(\bq_1,t)$ would be very important. 
Furthermore the role 
of finite size effects~\cite{berthierfss,dasguptafss}, the dependence of 
the results on the observable used to 
probe dynamical correlations, the universality of the results, 
the possible anisotropic character of 
dynamic heterogeneity \cite{Szamelanisotropy} all
figure among open questions needing further investigation.

\subsubsection{Current status of MCT}

\label{current-MCT}

The first and most important drawback of mean-field dynamics
and MCT is that the MCT transition it describes is not observed
in real materials. Additionally, 
a comparison between Eq.~(\ref{vft}) and (\ref{gamma}) makes it clear 
that MCT cannot be used to describe viscosity data close 
to the experimental glass transition $T_g$, 
since it does not even predict thermally activated behaviour.
Worse, MCT predicts a transition at which the system freezes completely: 
not only a fraction of the density fluctuations get frozen but also 
self-diffusion gets arrested \cite{mayr}. This is a theoretical 
artifact as it can be rigorously proven \cite{osada}
that the self-diffusion coefficient cannot go continuously
to zero at thermal equilibrium and finite temperature and pressure 
(this excludes the case of hard particles at close packing).

With all these major failures, then, 
why on earth should one continue to study and talk about MCT?
We provide several reasons.
 
It is now recognized that the MCT transition must be interpreted as an 
approximate theory of a crossover taking place in the dynamics. 
Actually, there are many other physical 
examples, such as  the Kondo model \cite{bookonkondo} or spinodal points 
\cite{debenedetti}, where crossovers become transition in 
approximate self-consistent theory. 
The fact that the transition is sharp in the 
theoretical treatment allowed the derivation 
of a variety of scaling laws which are as many predictions 
that can be tested in real materials, or, more prosaically, 
simple formula that can be used to `fit the data'. Part of 
the success of MCT has been its ability to propose
such clear-cut predictions (and fits), 
and graphical representations along which data could be analyzed.  

Indeed, one finds in the literature scores of 
papers where an MCT analysis of data is performed. 
Since the transition is (at best) 
only `avoided', its `crossover' nature offers 
quite a lot of flexibility for fitting, and judging 
the quality of the fits is often a difficult (and subjective) 
exercise, while `negative' results can always be attributed 
to `preasymptotic' corrections rather than 
deficiencies of the theory itself.
This has led to many controversies in the literature
which seem to persist to this day. 

A major achievement of MCT is the possibility 
to apply the same formalism to different materials 
and theoretical models, basically starting from the microscopic
interactions between atoms or molecules. This is again 
a reason for the success of MCT: each time a new model
with a different kind of interactions or chemical composition
is defined, MCT can be used to analyze its behaviour, 
and possibly predict new qualitative trends for the 
time dependence of correlation functions and their
evolution, even in cases where several control parameters
are relevant (e.g. mixtures, attractive vs. repulsive 
interactions, complex fluids, etc.). 
Striking and successful results and predictions, later 
confirmed by simulations and 
experiments, have been obtained in several cases: for example for the 
dynamical phase diagram of attractive colloids, 
for the behavior of the non-ergodic parameter as a function of the wave-vector 
for glassy liquids, for the treatment of molecules or particles
with non-spherical shapes, see the reviews 
\cite{MCTreviewkob,MCTreviewdas,gotze2}.
Thus, MCT is always a useful starting point when 
a new system with unknown behaviour arises.

Although the MCT predictions are limited to a 
modest time window of 2-3 decades corresponding 
to the onset of glassy dynamics in molecular liquids, 
this time window is actually the most experimentally 
relevant in colloidal materials \cite{lucalaurence}, since typical 
microscopic timescales for colloidal particles are in the 
millisecond range (instead of the nanosecond range for atoms).
Thus, MCT performs much better in soft matter systems, to the point 
that actually observing deviations from MCT predictions  
in an experiment can represent a challenge. 
Even for the canonical and well studied system of 
colloidal hard spheres, it was only recently suggested 
experimentally
that the MCT transition is avoided in the same manner 
as in molecular liquids \cite{LucaaboveMCT}. 
Thus we believe MCT will continue to be a useful tool, given 
the current rapid development in the synthesis of new colloidal particles.

On the theoretical side, it is now clear that MCT has a status of 
a mean-field theory \cite{universalityMCT}. As such, one expects 
major changes once fluctuations will be taken into account. 
As we pointed out already some kinds of fluctuations wipe out 
the sharp MCT transition and makes it a cross-over. Moreover, exactly
as for critical phenomena, one expects critical fluctuations to change 
the value of the MCT exponents below an upper critical dimension that 
was determined to be equal to $8$ \cite{BBdu,Franz2010}. The role of fluctuations 
on the MCT transition is a current topic of research.   

As a conclusion, MCT has clear and well-understood limitations 
and it will be never possible to test its predictions in a very sharp
way because it is not related to a true phase transition but, likely, 
just to a crossover. Still, its overall   
efficiency and flexibility, its ability to 
delivering actual predictions makes it useful. For 
this reason it continues to be developed, 
applied and generalized to study many different 
physical systems and situations, including aging systems and 
non-linear rheology of glassy materials,
see Sec.~\ref{aging}.

\subsubsection{Quantitative computations using  replica}

\label{monassonerie}

In Sec.~\ref{meanfield}, we presented the theoretical picture 
emerging from solving mean-field models (or geometries)
of the glass transition, and we found the resulting scenario rich and
encouraging: it generically supports
the existence of a configurational entropy vanishing transition 
associated, at a higher temperature, 
with a dynamical transition, {\it \`a la} MCT, which corresponds 
to the appearance of incipient metastable states. 

Still, to make a connection with experimental
results at least two main issues need to be addressed. First, one has 
to transform this set of mean-field ideas into a working tool 
able to produce quantitative calculations for the case of 
supercooled liquids. Second, dynamics within a rugged 
landscape must be revisited in order to explain the crossover
nature of the MCT transition, and the existence of a 
regime where dynamics is thermally activated and the viscosity
increases in a super-Arrhenius manner which is 
incompatible with the algebraic 
divergence predicted by mode-coupling theory.
In the present section and the next,  
we shall review briefly these two lines of research.

First, we shall focus on the approaches that have been devised to obtain 
quantitative microscopic predictions. In fact, MCT can be seen already 
as such a tool, but it is limited to the regime
above $T_{MCT}$. Below this temperature it cannot be applied anymore. At 
the time of this writing, a quantitative
microscopic theory of the dynamics valid below $T_{MCT}$ is still 
lacking, see \cite{bagchi} for a possible attempt. 
An alternative strategy is to leave dynamics for a while, 
and to turn to thermodynamics. The idea is to compute approximately the 
configurational entropy, the Kauzmann temperature, and the plateau value 
of dynamical correlation functions in the glass phase
\cite{monasson,Cardenas,MezardParisi,ParisiZamponi,reviewRFOTMP}. 

All the approaches developed to compute static properties 
of systems characterized with a rugged landscape 
make use of replica in one way or another~\cite{giorgio}. 
Physically, the reason
is that one aims at describing (or at least counting) metastable 
states which all have amorphous density profiles. This is 
similar to the spin glass case where 
the amorphous order is not revealed by looking at magnetization
profiles, as discussed in Sec.~\ref{spinglass}.
Inspired once more by the physics of 
disordered systems, the idea is again to let the system itself
indicate what are these states, and project distinct
copies of the system to `recognize' the metastable 
states. In the absence of quenched disorder, however, it is not enough 
to replicate the system, one is also forced 
to physically `couple' the different copies of the system
using an appropriate field.   

It is useful to first implement this idea for 
two copies \cite{monasson}, 1 and 2, 
of the system, characterized by density profiles $\rho_1({\bf x})$ 
and $\rho_2({\bf x})$. 
Using notations from Sec.~\ref{meanfield}, we write 
the free energy density of a single copy of the system as: 
\be
f = - \frac{T}{V} \log \int {\cal D} \rho_1 e^{-\beta H[\rho_1]},
\ee
where $H[\rho_1]$ is the microscopic Hamiltonian.
Let us now use the second copy of the 
system to scan the locally stable configurations
of the first one. To do so, we introduce a 
quadratic coupling of strength $g>0$
between the two 
copies and compute the new free energy 
\be 
f_2 [\rho_1] = - \frac{T}{V} \log \left[ \int {\cal D} \rho_2 
e^{ - \beta H[\rho_2] - g \int d^dx [\rho_1 - \rho_2]^2} \right].
\label{remi}
\ee
The free energy $f_2[\rho_1]$ will be small when $\rho_1$ is a configuration
which corresponds to a metastable state. Therefore, 
sampling all configurations of $\rho_1$ weighted with
$\exp (-\beta f_2[\rho_1])$ is a procedure to scan all metastable 
states, so that
\be
f_{\rm meta} = \lim_{g \to 0} \frac{1}{Z} \int {\cal D} \rho_1 f_2[\rho_1]
e^{-\beta f_2 [\rho_1 ]}
\label{limitg}
\ee 
is the average free energy of all metastable states; 
here $Z = \int {\cal D} \rho_1 \exp (-\beta f_2 [\rho_1])$ is a normalization.
When it exists, the difference between the total free energy density
of the system, $f$, and the free energy density 
of the minima, $f_{\rm meta}$, is related to the number of
metastable state available to the system, see also
Eq.~(\ref{Zeq}). As a consequence, this free energy cost 
is in fact equal to $-T s_c$.   
The lesson to be learnt from this example is that the introduction of 
identical copies of the system allows one to compute directly  
properties of the free energy landscape \cite{monasson,franzparisipotential}. 
It is important to remember, however, that the limit 
$g \to 0$ in Eq.~(\ref{limitg}) has a strong mean-field flavour, 
since genuine metastable states only exist in this 
limit, see \cite{giorgio} for a more precise discussion.
 
Several quantitative approaches have been developed and are based, in one 
way or another, on procedures similar to the one outlined above, see 
\cite{monasson,Cardenas,MezardParisi,ParisiZamponi}.
In the scheme of \cite{monasson}, one introduces 
$m$ copies of the system constrained to be in the same free energy 
minimum. Technically, this corresponds to take $m$ 
copies of the density configuration $\rho_1$.
This generalizes the partition function in Eq.~(\ref{Zeq}) to 
\[
Z_m =  \lim_{g \to 0} \frac{1}{Z} \int {\cal D} \rho_1 
e^{-\beta mf_2 [\rho_1 ]}=\]
\[
\int df \exp \left[ -\frac{N f m}{T} - N s_c(f,T) \right],
\]
associated to the replicated free energy $\psi(m,t) = -T \log Z_m$.
Note that $m$ only enters the first term, as all systems are identical
and are characterized by the same metastable states. 
Repeating the saddle point calculation yields 
\be 
f  = \frac{\partial \psi(m,T)}{\partial m}, \,\,\,\, 
s_c = \frac{m^2}{T} \frac{\partial (\psi(m,T)/m)}{\partial m}, 
\label{m}
\ee
which shows that the configurational entropy $s_c$ can be 
accessed by computing the thermodynamic properties 
of a system of $m$ coupled replicas, provided, as is
usual within replicas calculations, that the number 
of copies of the system is analytically continued to non-integer
values, as implicitly assumed in Eq.~(\ref{m}).

A remarkable achievement \cite{MezardParisi} is that not only 
the properties of the liquid state above $T_K$ can be 
computed analytically from Eq.~(\ref{m}), but also the ones of the `ideal'
glass below $T_K$. Recall from Fig.~\ref{sc:fig} that the Kauzmann
temperature is defined by $T_K = (\partial s_c / \partial f)^{-1}$
for the value of $f$ corresponding to $s_c=0$.
For the replicated system, one gets $T_K^{(m)} =  m (\partial s_c / 
\partial f)^{-1}$ and the location of the transition 
depends on $m$. For values of $m<1$, one typically finds 
that $T_K^{(m)} < T_K$. This means that the properties 
of the replicated liquid with $m \neq 1$ 
are deeply connected to the ones of the non-replicated glass
with $m=1$. Indeed, assuming that the transition has a nature similar
to the one found in mean-field models, one can use the fact that 
the free energy is continuous at $T_K$ and glass states
below $T_K$ have $s_c(T<T_K)=0$ to obtain the free energy of 
the (non-replicated) glass as 
\be
f_{\rm glass} (T) = \psi( m^*(T),T) /  m^*(T) ,
\ee   
where $m^*(T<T_K) < 1$ is self-consistently defined 
by $T_K^{(m^*)} = T$ and $\psi( m,T)$ is the energy of 
the replicated liquid defined above.

To summarize, starting from the 
hypothesis that the free energy landscape of 
supercooled liquids resembles the picture gained from 
mean-field models and geometries, one introduces
replicas as a useful mathematical tool to probe the 
thermodynamics of systems both above and below $T_K$. 
One ends up with an additional variational
parameter, $m^*(T)$, to describe the low temperature phase, 
which is formally strictly equivalent to the additional
parameter $m$ entering the one-step replica symmetry scheme
needed to solve mean-field models \cite{monasson}. 
Note that in this approach, the nature of the broken 
symmetry is a starting hypothesis
rather than a natural outcome of the calculations. 

In practice, of course, some approximations must be made 
to compute the free energy of the replicated liquid at low
temperature, which make heavy use of liquid
state theory and which might well depend on the studied system.
The main outcomes are the calculation of the location of the 
Kauzmann transition, the thermodynamic properties of the 
liquid (in particular the configurational entropy), 
and the glass (ground state energy, specific heat, 
structure).
These microscopic computations have been developed for a variety of 
glass-forming liquids, such as monoatomic or mixtures of
Lennard-Jones particles \cite{coluzziparisiverrocchio}, 
soft-spheres \cite{MezardParisi,coluzzimezardparisiverrocchio}, hard spheres 
\cite{Cardenas,ParisiZamponi}, sticky hard spheres \cite{reatto},
silica \cite{coluzzi}.
Perhaps the most impressive achievement is the detailed 
description of the large volume fraction behaviour of 
hard spheres \cite{ParisiZamponi,ParisiZamponireview}: 
the glass transition was located, 
the equation of state and structure of the glass obtained up to 
the `Glass Close Packing' density, which can be seen as a firm 
definition of the notion of random close packing \cite{bernal}, 
whose location can then be predicted accurately in any spatial
dimension \cite{francesco1} or for binary mixtures \cite{francesco2},
with excellent agreement with experimental results and simulations.

Although this quantitative side of RFOT theory is a most 
desirable feature, assessing quantitatively 
the quality of the results is not easy as experiments
and simulations typically fail to approach the Kauzmann 
transition. Even then, cases are known where a transition
is predicted in a regime where none is expected
\cite{thalmann,coluzzi}, suggesting 
that the quality of the approximations used 
to obtain quantitative results plays an important role \cite{KUNI}. 
Thus,
when accurate results are sought, the problem might
well become technically quite involved
\cite{ParisiZamponireview}.

\subsubsection{Scaling arguments beyond mean-field theory and point-to-set
lengthscale}

\label{scalingrfot}

The quantitative calculations described in the previous section
remain mean-field in nature
because they compute the properties of the supercooled liquid state, as if it 
were formed by a collection of states with infinite lifetime, as is the case 
in mean-field models. This is clearly incorrect: two thermodynamically 
stable states cannot coexist and have different intensive free energy 
at finite temperature, otherwise the system would nucleate from the one 
with highest free energy to the one with lowest, showing that the 
highest is in fact not a stable state. 
Furthermore, an exponential number (in the 
system size)
of stable states seems impossible: for large sizes there would not be 
enough  boundary conditions to select one state from the other 
\cite{NewmanStein}.

Additionally these calculations cannot address
the connection to dynamical properties, a crucial 
missing ingredient for a theory of the glass transition.
Presently, there only 
exist phenomenological arguments \cite{KTW2,rfotwolynes,BB2}, 
backed by microscopic computations \cite{Schmalian,Franz,FranzMontanari}, 
that yield a possible scenario for the dynamics, 
dubbed `mosaic state' in \cite{KTW2}. Since this aspect of 
RFOT theory was reviewed recently \cite{reviewRFOT,bbreview},
we shall be brief.
Schematically, the mosaic picture states that, in the regime $T_K<T<
T_{MCT}$, the liquid is composed of domains of linear 
size $\xi$. Physically, the lengthscale $\xi$ represents
the lengthscale above which it does not make sense to talk 
about metastable states anymore. 

The way to measure, or even to define precisely, the mosaic lengthscale
$\xi$ was not clear from the way it was initially introduced.
Recently, the so-called `point-to-set' correlation lengthscale 
was defined both in the context of RFOT theory
as a practical measure of the mosaic lengthscale~\cite{BB2}, and 
in more general settings~\cite{MS0,MM06}. 
The point-to-set length is a measure of the 
spatial extent over which the effect of equilibrium 
amorphous boundary conditions propagate.    

To understand the origin of this lengthscale, we consider 
the following `experiment' where, starting from an equilibrium
configuration of the system, we freeze the position of 
all particles outside a cavity of radius $R$. We then let 
the system evolve in the presence of this constraint, which acts
as a pinning field. The point-to-set lengthscale
is defined as the minimal cavity radius such that 
the pinning field has no influence at the center of the cavity.
As such, it is a measure of the many-body
correlations between a point (the center of the cavity) and 
a set of points (the pinned boundary). It is also important to emphasize
the similarity between this cavity procedure, and the 
coupling between copies applied in Eq.~(\ref{remi}). While the latter
was homogeneous, the former is spatially inhomogeneous,
and quantifies how far in space the coupling between states 
can propagate \cite{jorgelength}.

The constraint on the boundary of the cavity in fact acts as a `template'
for the particles inside the cavity, whose effect can be evaluated as follows.
By selecting a different state and deforming it only close to the 
boundary to satisfy the constraint, 
one would get a free energy cost of the order of 
$\Upsilon R^{\theta}$ with $\theta \le 2$. However, doing so, one would 
also gain entropy as the system could explore 
a multiplicity of different states, giving rise
to a free-energy contribution $-Ts_c R^3$. 
Entropy obviously gains on large lengthscales, $R \gg 1$, while
interface cost dominates at small $R$. Therefore,  
the crossover in the competition between these two terms takes place
for a lengthscale $R=\xi$ obtained by equating the two terms, 
\begin{equation}
\label{eqxi}
\xi = \left(\frac{\Upsilon}
{Ts_c(T)}\right)^{1/(3-\theta)}.
\end{equation}
In a real liquid, where there is no cavity, one can 
conjecture that there is a self-generated dynamical boundary 
condition acting on each patch of lengthscale $\xi$. 

The dynamical counterpart to this argument 
is as follows. Dynamically, the configurational entropy 
on scales smaller than $\xi$ is too small to stir the
configurations efficiently and loses against  the dynamically generated
pinning field due to the environment. By contrast, ergodicity 
is restored at larger lengthscale. Hence, the relaxation time of the 
system is the relaxation time, $\tau(\xi)$, of a finite size 
regions of the system. It is only after this long, but plausible series 
of arguments, that barriers encountered during 
relaxation finally become finite and involve a finite number of particles, 
unlike in the original mean-field treatment of the landscape
where barriers diverge with system size. 

Now, assuming thermal activation over 
energy barriers which are supposed to grow 
with size as $\xi^\psi$, with $\psi\ge\theta$, one predicts finally using 
Eq.~(\ref{eqxi}) that~\cite{BB2} 
\be
\log \left( \frac{\tau_\alpha}{\tau_0} \right) =
c \frac{\Upsilon}{k_B T}
\left(\frac{\Upsilon}
{Ts_c(T)}\right)^{\psi/ (3-\theta)},
\label{ag}
\ee
where $c$ is a constant. 
This argument is rather generic and therefore not very predictive.
Recent microscopic computations
\cite{Schmalian,Franz,MontanariSemerjian,FranzMontanari} 
attempted the computation 
of the exponents $\theta$ and $\psi$, directly addressing 
analytically the problem of the cavity described above.
Equations for the overlap profile 
between the initial `template' configuration and configurations
thermalized in the presence of the pinning boundaries were obtained.
These calculations confirmed the existence of a non-trivial crossover
lengthscale above which the overlap inside the cavity vanishes, 
indicating that order does not propagate on lengthscales much larger
than $\xi$. 

The results are unfortunately not yet conclusive. Although part of the 
computations
can be justified and controlled by using for instance
Kac models \cite{Franz}, other parts involve uncontrolled replica 
symmetry breaking schemes. The calculations provided 
the estimate $\theta = 2$. Note that some other phenomenological 
arguments suggest the  value of $\theta=3/2$ \cite{KTW2}.
There are no detailed computation available for $\psi$, 
only the suggestion that $\psi=\theta$ \cite{KTW2}. 

Note that using the value $\theta=3/2$ with $\theta=\psi$ simplifies
Eq.~(\ref{ag}) into a form that is well-known experimentally
and relates $\log \tau_\alpha$ directly 
to $1/S_c$, which is the 
celebrated Adam-Gibbs relation~\cite{ag} between 
relaxation time and configurational entropy. As discussed in 
Sec.~\ref{phenomenology}, such a relation  
is in rather good quantitative agreement with many
experimental results~\cite{Angell,Hodge,Johari}.
RFOT theory, therefore, reformulates and generalizes 
the mechanism suggested by Adam-Gibbs \cite{rfotwolynes}. 
Furthermore, using the fact that 
the configurational entropy vanishes linearly at $T_K$,
Eq.~(\ref{ag}) becomes similar to the VFT 
divergence of Eq.~(\ref{vft}), with the 
identification between two important temperatures, 
\be
T_0 = T_K,
\label{totk}
\ee  
which embodies the deep connection between thermodynamics
and dynamics characterizing RFOT theory.
The above equality between two temperatures that are 
commonly used in the description of experimental data
constitutes a central achievement of RFOT theory, since 
it accounts for the empirical relation found between the kinetics and the 
thermodynamics of supercooled liquids. It should be kept in mind,
however, that experiments have not established its validity
beyond any doubts, as discussed in detail in Sec.~\ref{phenomenology}. 

Wolynes and co-workers have obtained several other results using  
phenomenological arguments based on RFOT. Two remarkable ones are: 
the relation between fragility and specific heat jump at the glass transition
and stretching exponent $\beta$ of time-dependent correlation functions,  and 
the speeding up of the dynamics close to a free surface (recently observed in 
\cite{Gruebel}). These predictions and several others are discussed in detail in 
\cite{reviewRFOT}.

\subsubsection{Current status of RFOT theory}

We described RFOT theory as a `patchwork' (not to say `mosaic') of 
apparently distinct theoretical approaches
to the glass problem:  Adam-Gibbs theory, mode-coupling 
theory, mean-field spin glass theory and replica approaches, 
and the mosaic state
scaling picture. As such, RFOT theory is clearly an
impressive theoretical piece of work, which gives
a very consistent overall scenario for the glass transition
and nonequilibrium phenomena related to the glassy state, 
based on peculiar features of the free energy landscape
as well as tools to perform microscopic calculations.

Coming from high temperatures, dynamics primarily slows down because
there appear incipient metastable states, in a restricted 
temperature window described in full microscopic
detail by mode-coupling theory. Decreasing further the temperature, 
the dynamics becomes dominated by the thermally activated barrier 
to be crossed from one metastable state to another, in a way consistent with 
the deep relation between dynamical correlation length and timescale 
discussed in Sec.~\ref{dh}. In this regime, the thermodynamic
behaviour can also be described at the thermodynamic
level using replica calculations which predict the 
existence of a finite temperature thermodynamic 
transition towards a genuine `ideal' glass state. 
A description of the dynamics near the transition exists,
but contains several steps that heavily rely on empirical 
scaling concepts.

There are of course several weaknesses in this construction.
First, although we attempted here to give a unified view,
the theory is still pretty much made of distinct
pieces that do not necessarily smoothly fit together
and have perhaps no strict boundaries.
For instance, the details of the crossover between 
MCT and activated regimes are not well understood.
In early works, attempts were made to include 
`hopping effects' in mode-coupling theory, deriving expressions 
for the memory kernels which transform the sharp 
MCT transition into a crossover~\cite{DM,GS87}. 
However, the status of these `extended' MCT 
is debated~\cite{ABL,CR06}. 
Moreover, there are also strong indications that thermal activation
is in fact already at play in the temperature regime
usually described by MCT \cite{denny,heuer,crossover}. 
Finally, recent works attempted to include nonperturbative 
processes in the MCT description~\cite{mayerdave,bagchi}, but no 
treatment generically applicable to liquids is yet available. 

Second, when applied to three-dimensional liquids, 
RFOT theory relies on several, sometimes distinct, 
types of approximations.  For finite dimensional 
systems, a complete and solid version of RFOT theory remains
to be worked out. This is especially true for the `mosaic' part of RFOT theory
which yield dynamic predictions, namely a Vogel-Fulcher
divergence of the relaxation time.
This implies that opponents can criticize RFOT theory
because it contains too many uncontrolled assumptions, 
while supporters can always hide a weak point as resulting
from some approximation rather than from the approach itself.
We do not see how this issue can be resolved, as data
themselves do not allow clear-cut conclusions. Although the 
ultimate consequences of the theory
are sometimes in very good agreement with experiments, as
Eq.~(\ref{totk}), one should not conclude too fast that 
RFOT theory is correct. In this context, a pertinent line of investigation,
allowed by numerical simulations, is to more directly
test the microscopic mechanisms underpinning the derivation of the mosaic
picture \cite{rob2,cavagna}.
In particular, recent works have shown that the static `point-to-set' 
correlation lengthscale described in Eq.~(\ref{eqxi}) does increase
upon supercooling \cite{cavity}. 
Furthermore, first measurements of the exponent $\theta$ and $\psi$
of the mosaic theory have been obtained with 
the somewhat surprising results $\theta\simeq 2 > 
\psi\simeq 1$ \cite{exponents,exponents2}. 
At present, direct tests of the mosaic picture are quite 
involved (even conceptually),
and hence are rare and not yet conclusive. This line of research 
appears nevertheless very promising
to establish, disprove, or develop further the mosaic picture, 
see \cite{cavagna-review} for more details.   

The fact that these approximations become exact
in several mean-field settings (fully connected models,
Bethe lattices) suggests that RFOT theory might have a
status similar to the Curie-Weiss theory for the ferromagnetic
systems which does contain correct elements of the
real theory. Current research can thus be seen as an attempt
to understand and describe better non mean-field
effects. Going beyond mean-field theory is not only technically
but also conceptually difficult. For instance, Stillinger claimed 
that the configurational entropy vanishes only at zero temperature, thus
suggesting that no entropy crisis can take place \cite{stillinger}.
This was related to his identification
of metastable states with energy minima, which is now recognized to be
an incorrect approximation, 
even in mean-field models \cite{birolimonasson}, where 
the well-defined metastable states are in general made of a
large number of energy minima, in the spirit of the 
`metabasins', sometimes described in numerical works
as a large assembly of inherent stuctures~\cite{heuer}.
The same criticism applies to recent work 
who similarly claimed having demonstrated 
the absence of a glass transition in bidimensional 
binary mixtures~\cite{donevmix}.
However, a correct definition of metastable
states beyond mean-field theory is still lacking, see 
\cite{birolikurchan} for a discussion and a
first attempt.
Thus, going beyond mean-field theory is clearly a difficult, but
quite exciting and important task,
from which new results can be expected in the future.

A further source of concern for RFOT theory is the fact that
quite often the mean-field models from which it largely
originates seem to behave quite differently when 
studied in finite dimensions \cite{pspin3d,potts,moore,moore06a,moore06b}. 
In fact, there does not 
yet exist a theoretical model in finite
dimension, for which the RFOT theory scenario can be shown to 
apply, see \cite{sarlat,ricci,hrem} for recent efforts.
From a theoretical perspective, such a discovery
would be a highly decisive step, even if the model 
were very abstract and not obviously connected to 
experimental systems. 

Thus, there is hope that
in the next few years, joint theoretical and experimental
efforts will drive RFOT theory into a corner, 
where its status can be made precise. At the time of writing, 
one can state that RFOT theory is still imperfect, but
the broadness of its scope and predictions, the 
number of distinct approaches that systematically give back at least 
some piece of it, the generality of the concepts it uses, 
makes one believe that it contains at least some
useful seeds to construct a `final' theory of the 
glass transition--if such a thing exists.  

\subsection{Kinetically constrained models and dynamic facilitation}

\label{theorykcm}

\subsubsection{The physical picture}

Another approach to the glass transition problem is based
on the concept of dynamical facilitation, see 
\cite{GCannualreviewchemistry} for a review.
In short, `facilitation' captures the physical idea
that since viscous liquids are almost solid, mobility
is so sparse at any given time that any local relaxation
event is likely to trigger, or `facilitate' the relaxation
of nearby molecules after a time which is short compared to 
the macroscopic relaxation time but large compared to the microscopic one,
so that mobility can propagate throughout the sample. Thus,
the focus is less on molecules than on their mobility \cite{FA}. 

There is no doubt that at least some degree of 
facilitation is present 
in nearly jammed materials, but the theoretical
approach described in this section goes well beyond this 
simple observation and posits that the entire dynamics
of the system is mainly due to facilitation effects.
This means that typically a mobile region of the sample can become 
unjammed and thus  mobile only if
it is adjacent to a region which is already unjammed \cite{gcpnas}. 
This is equivalent to postulating that, except for very rare events, 
mobility in a viscous liquid cannot spontaneously arise in an 
immobile region in space, 
nor can it spontaneously decay. This is obviously a very strong 
assumption.

This constraint is conjectured to become effective roughly below $T_{MCT}$ 
and to dominate the dynamical evolution close to $T_g$ \cite{gcpnas}. 
This is far from a trivial assumption since it implicitly
uses the fact that there exists a temperature below 
which the material is nearly jammed, so that a description
in terms of sparse mobility is valid. This approach can thus
not self-consistently capture the microscopic origin 
of the dynamical slowing down in supercooled liquids. Instead, 
it can possibly become a very effective and useful description   
of structural relaxation occurring near the glass 
transition \cite{palmer}.

At present, it is still unclear whether this main 
assumption of mobility conservation is correct, only approximate, or 
whether it plays the central role suggested by the facilitation
approach. An important point is the fact 
that a large number of theoretical models have been 
defined based solely on the idea of kinetic constraints. They
are called `Kinetically Constrained Models' (KCMs) and they
all display a phenomenology which is strikingly consistent to the one 
described in Sec.~\ref{phenomenology} for molecular 
glass-formers \cite{solrit}. 
In the last decade or so, several KCMs have been defined 
and studied in quite some detail. In turn, this large
body of theoretical work has produced new results and predictions, and 
has thus also triggered more research on competing theories. 
 
\subsubsection{Kinetically constrained models (KCMs)}

\label{KCMsection}

Over the last 25 years, KCMs have been central 
to the development of the facilitation approach to the glass transition.

The first example we shall present is the Kob-Andersen 
model \cite{KAgas}. Interestingly, this model 
is quite similar to the lattice glass models described in 
Sec.~\ref{giuliotheory}. The model is again an attempt to capture 
the physics of a hard sphere system, and the fact that
dynamics becomes slow at high density because the environment of each 
particle is very crowded. The Kob-Andersen model is a lattice gas, 
with occupation number on each site $i$ of a regular lattice
$n_i$. There is no interaction apart from 
the hard-core constraint, and the Hamiltonian is thus trivial: 
\begin{equation}
H[\{ n_i \}] = 0, \quad n_i = 0, \, 1.
\label{hamka}
\end{equation}
Geometric frustration is introduced at the level 
of the kinetic rules, that are defined as constrained 
local moves. Namely, a particle can jump to a nearest neighbor site 
only if (i) that site is empty, to satisfy the hard-core constraint;
(ii) the sites occupied before and after the move
have less than $m$ neighbors, $m$ being an adjustable parameter.
Kob and Andersen studied the case $m=4$ for a cubic lattice in 
$d=3$ ($m=6$ corresponds to the unconstrained lattice gas), and the model
displays glassy dynamics at large density \cite{KAgas}.
Such kinetically constrained lattice gases
have been studied in various spatial dimensions, for different values
of $m$, for different constraints, or even different 
lattice geometries~\cite{solrit}.
They can be thought of as models capturing the idea of a `cage' effect
in a strict sense, since a particle with a
dense neighbor shell cannot diffuse. Dynamic facilitation is thus 
a direct consequence of steric effects in this model \cite{pinaki-ka}.

The crucial difference with lattice glass models is that 
here all configurations are allowed by the Hamiltonian, 
but their kinetics depends on geometry, while in lattice glass
models kinetic rules ignore the geometry but not all
configurations are allowed \cite{BiroliMezard}. 
This is actually a deep difference:
KCMs assume that geometrical constraints act at the level 
of kinetic rules with no thermodynamic counterpart and no reference
to the interaction which are responsible for them.

In this lattice gas picture, the connection with glass-formers
is not obvious because density, rather than temperature
controls the dynamics. Thermal models with similar features
can in fact be defined by focusing on holes rather than particles.
This points toward a model with a small concentration of 
mobile regions, which move by creation and annihilation.
This is actually very reminiscent of Glarum `defects' theory where
the relaxation proceeds via simple diffusion
of a low concentration of independent defects \cite{glarum}.
Using the conjugated ideas of kinetic constraints, facilitation 
and rare defects, Fredrickson and Andersen
defined and studied a family of kinetic Ising models for the glass 
transition in \cite{FA}. This last article is a seminal 
paper that opened a whole new research avenue, making 
possible the study of phenomena associated to the glass transition
via simple kinetic models.
The Fredrickson-Andersen models
consist in an assembly of non-interacting `defects', or `spins', 
\begin{equation}
H[\{ n_i \}] = J \sum_{i=1}^N n_i , \quad n_i = 0, \, 1,
\label{hamfa}
\end{equation}
where $J$ is an energy scale for creation of mobility, and 
$n_i=1$ represent the mobility `state' at site $i$,
whose averaged concentration becomes exponentially small at low temperature, 
$\langle n_i \rangle \approx \exp(-J/T)$. As for the Kob-Andersen 
lattice gas, the non-trivial ingredient lies in the chosen rates 
for the transition between states. The kinetic rules stipulate
that a transition at site $i$ can happen with a 
usual Glauber rate, but only if site $i$ is surrounded by
at least $k$ defects ($k=0$ corresponds to the unconstrained
limit). 

As for kinetically constrained lattice gases, these models 
have been studied
in different spatial dimensions, on different lattices, and using 
a number of distinct definitions of the kinetic rules, 
yielding a large number of possible glassy behaviours 
\cite{solrit,leonard}.
The similarity between those spin facilitated models
and the kinetically constrained lattice gases is striking.
Altogether, they now form a large family of models generically called 
KCMs.  

These models can be divided into several classes. 
`Non-cooperative' models, such as the Fredrickson-Andersen 
model for $k=1$ (the 
least constrained model) display Arrhenius dynamic slowing down
and are thus reminiscent of strong glass-formers. They 
are well-described by simple diffusion of point 
mobility defects. 
`Cooperative' models display a super-Arrhenius dependence of the 
structural relaxation time. This is the case for 
the Fredrickson-Andersen model with a stronger 
kinetic constraint, $k>1$. Another example is the `East' 
model where the kinetic constraint with $k=1$ has a directional
character: only excited sites to the left in each space dimension can 
facilitate the dynamics \cite{jackle,nef}, which can 
be rationalized on the basis that displacements in liquids
have a vectorial character that could extend to facilitation \cite{gcpnas}. 
For an overwhelming majority of KCMs,
relaxation times diverge only in the limit of zero temperature, 
as even one single defect can diffuse and relax the entire system
(at least in the simplest models), and the defect concentration
only vanishes at $T=0$. However, KCMs can also be defined with 
kinetic rules and geometries for which the existence 
of a finite temperature glass transition can be established
\cite{TBF2}. 

Let us remark that the facilitation approach, 
and in particular KCMs, encode in a new
and more microscopic way the older, but well-known, concept of free volume.
Free volume models are among the most widely used models to 
analyze experimental data, especially in polymeric 
systems.
They have been thoroughly reviewed before~\cite{freevol,debenedetti}.
Here, the main idea is that dynamic slowing down occurs because 
the `free volume to move' available to each particle, $v_f$, vanishes at some 
temperature $T_0$ as $v_f \approx \alpha(T-T_0)$, a relation 
which connects volume to temperature.
Statistical arguments then relate relaxation timescales to free volume 
assuming that movement is possible if locally there is `enough' 
free volume available, more than a typical value 
$v_0$. A VFT divergence is then predicted:  
\be
\frac{\tau_\alpha}{\tau_0} \sim \exp\left(\gamma \frac{v_0}{v_f}\right) \sim 
\exp \left( \frac{\gamma v_0/\alpha}{[T-T_0]^{\mu}} \right), 
\label{vftfree} 
\ee
where $\gamma$ is a numerical factor and $\mu=1$. 
Predictions such as Eq.~(\ref{vftfree}) are used to justify the wide use
of free volume approaches, despite the many (justified) criticisms
that have been raised. The physics at work is 
obviously strongly reminiscent of the above description of the 
Kob-Andersen kinetically constrained lattice gas.  
In fact, the analogy is even semi-quantitative in some cases,
since cooperative KCMs with a finite temperature glass transition have been 
shown rigorously to be characterized by divergence of the relaxation
time as in Eq. (\ref{vftfree}), albeit
with a value of $\mu$ different from one \cite{TB}. 

As with free volume approaches, it is not exactly 
clear what is meant by `mobility defects' within KCMs in terms of 
the original interacting system they seek to describe, nor how 
kinetic constraint can truly emerge from the unconstrained 
dynamics of a many-body system. 
A good news on this front is that at least 
a proof of principle that kinetic rules can emerge 
has been obtained \cite{juanpe}. Several examples are available 
but here we only mention the simple case of the bidimensional 
plaquette model defined by 
a Hamiltonian of a $p$-spin type, but in two dimensions on a square lattice
of linear size $L$,  
\be
H = -J \sum_{i=1}^{L-1} \sum_{j=1}^{L-1} S_{i,j} S_{i+1,j} 
S_{i,j+1} S_{i+1,j+1},
\label{spm}
\ee
where $S_{i,j}=\pm 1$ is an Ising variable lying at node $(i,j)$
of the lattice. Contrary to KCMs, the Hamiltonian in Eq.~(\ref{spm}) 
contains genuine interactions,
which are no less (or no more) physical than $p$-spin models
discussed in Sec.~\ref{giuliotheory} to which 
it actually strongly resembles. Interestingly
the dynamics of this system is (trivially) mapped onto 
that of a KCM by analyzing its behaviour in terms of plaquette 
variables, $p_{i,j} \equiv S_{i,j} S_{i+1,j} 
S_{i,j+1} S_{i+1,j+1}$, such that the Hamiltonian 
becomes a non-interacting one, $H = -J \sum_{i,j} p_{i,j}$, 
as in Eq.~(\ref{hamfa}).  
More interestingly, the analogy also applies to the dynamics
\cite{juanpe}. 
The fundamental moves are spin-flips, but when a single spin is
flipped the states of the four plaquettes surrounding that spin
change. Considering the different types of moves, one quickly realizes
that excited plaquettes, $p_{i,j} = + 1$, act as
sources of mobility, since the energetic barriers to spin flips
are smaller in those regions. This observation allows to
identify the excited plaquettes as defects,  
by analogy with KCMs. Similarly to KCMs, also,
spatially heterogeneous dynamics, 
diverging lengthscales accompanying diverging timescales
and scaling behaviour sufficiently close to 
$T=0$ (see below) can be established by further analysis \cite{spm}, 
providing a simple, but concrete example, of how an interacting 
many-body system might effectively behave as a model with 
kinetic constraints. This type of plaquette models, and
other similar spin models, were originally introduced \cite{Sethna,Lipowsky} 
to show how ultra-slow glassy dynamics can emerge because 
of growing free energy barriers.

\subsubsection{Diffusing defects, excitation lines and space time bubbles}

On the quantitative side, research on 
the facilitation approach to glassy dynamics 
has mainly consisted in the analysis of the physical behavior of KCMs. 
Motivation for this work largely stems from 
the observation that KCMs behave, at the phenomenological level,
very much as real glass-formers, as noted long ago both
for thermal models \cite{FA,FAb,FA2,FA3}
and constrained lattice gases \cite{tlg,KAgas,sellitto}.
Actually, theoretical research on dynamical 
heterogeneity was partly sparked by the numerical
observations that KCMs display strong spatial
fluctuations of the local relaxation rate 
\cite{harroa,harrob,harrofirst}. 
Therefore, KCMs provide a simplified context to understand 
glassy phenomena in detail, or at least study one of their 
possible explanations. 

In the following we shall not review all the different models
and their physical behavior, as this was done 
thoroughly in \cite{solrit}. Neither shall we review in great detail
the dynamic facilitation approach as a theory of the glass transition,
as this is the object of another recent review 
\cite{GCannualreviewchemistry}. 
Instead, we shall present a simple physical picture, 
common to all KCMs, which is helpful to grasp the main 
physical behavior of KCMs and, therefore, the main predictions 
and limitations of the facilitation approach. 

A common feature of all KCMs is that their relaxation can be 
accurately described in terms of the motion of sparse 
defects. In the simplest cases like the 
non-cooperative Fredrickson-Andersen model, these defects 
diffuse, but they might also undergo sub-diffusive motion in some cases, like in the 
directional East model. In cooperative models,
such as the Kob-Andersen lattice gas or the Fredrickson-Andersen model
with $k>1$, defects do not coincide with mobile sites, but can be formed
by extended clusters moving in a cooperative manner. 

\begin{figure}
\psfig{file=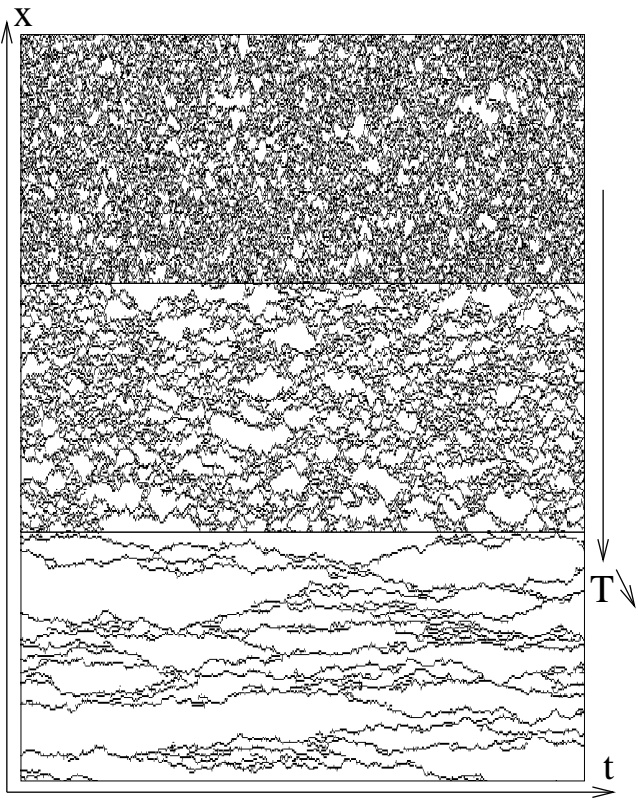,width=8.5cm,clip}
\caption{\label{fig:bubble1d} 
Space-time representation of the dynamics of mobility defects
in the non-cooperative Fredrickson-Andersen model with $k=1$ in
$d=1$. Dynamic facilitation, 
imposed by the kinetic constraints, implies the existence
of excitation lines which can branch and coalesce.
The physics changes from a homogeneous description 
at high temperatures to a strongly spatially and temporally 
heterogeneous dynamics at low temperatures when mobility
is sparse. After \cite{crossover}.}
\end{figure}

Since local relaxation at a given site occurs when it is visited 
by one such `defect', explaining structural
relaxation is equivalent to explaining the defect dynamics.
To make the discussion more concrete, let us explain how this defect 
description emerges for the simple case of the non-cooperative
Fredrickson-Andersen model with $k=1$. In that case, 
a spin $n_i$ can flip only if it has at least one spin $n_j=1$
among its nearest neighbors. At low temperature, defects 
diffuse in an activated manner. Typically an isolated defect 
facilitates the excitation of one of the neighboring site
with rate $\exp(-J/T)$, and moves there with 
probability $1/2$. Schematically, one gets:
\be
0100 \to 0110 \to 0010 \to \cdots
\ee
Overall,
defects perform random walks with a diffusion coefficient 
$D \sim \exp(-J/T)$. Defects can also annihilate when they meet and 
be created from an existing defect. A representation
of this diffusion, branching and annihilation process is shown 
in Fig.~\ref{fig:bubble1d} for the one-dimensional case. 
In particular, the central feature of the 
facilitation approach, namely that a defect cannot 
be created out of a completely immobile region, is obvious
from this figure, in particular at low temperature where defects
are sparse.  

An approximate but quite good description of KCMs consists in 
focusing on sparse, possibly extended, defects of density 
$\rho_d$, which  diffuse anomalously with an exponent $z$ 
and a generalized diffusion coefficient $D$ which
is temperature (or density) dependent \cite{TWBBB}.  The relaxation time 
of the system can then easily be obtained by computing the time over 
which a finite fraction of the system, say $1/2$, has been visited by 
at least one defect. The number of distinct sites visited by a given 
defect after a time $t$ is by definition
equal to $(Dt)^{d_F/z}$, where $d_F$ 
is the fractal dimension of the walk (or the space dimensionality $d$
when $d_F > d$). For 
a random walk  in $d=3$, one has $d_F=z=2$, 
and therefore a number of sites visited per defect is 
simply equal to $Dt$. 
The relaxation time $\tau_\alpha$ is then given by the equation 
$\rho_d (D \tau_\alpha)^{d_F/z} \simeq 1/2$. Both the defect density $\rho_d$ 
and the diffusion coefficient vanish when decreasing the temperature 
with expressions that are model-dependent. The relaxation times thus
generically scales as
\be
\tau_\alpha  \propto \frac 1 D \rho_d^{z/d_F}.
\ee
In the case of the non-cooperative Fredrickson-Andersen 
model discussed before ($k=1$), one gets 
$\tau_\alpha  \propto e^{2J /T}$ for $d>1$ and 
$\tau_\alpha  \propto e^{3J /T}$ in $d=1$, i.e. 
an Arrhenius dependence in all dimensions, similar to the one 
describing the viscosity of strong glass-formers. 

As expected, stronger temperature dependencies are obtained when 
kinetic constraints become more and more severe. For the 
directional East model, for instance, the energy
barrier to relax a domain of the form $1 0 \cdots 0 1$ 
containing $\ell$ unexcited sites increases with $\ell$
since the leftmost defect must in principle 
facilitate all sites to its right. The minimal 
energy barrier needed for this process 
to occur scales as $E(\ell) \sim \log (\ell)$, and corresponds to a 
`hierarchical' path \cite{sollichevans}. 
Since the typical distance between defects 
at equilibrium is $\ell_{\rm eq} \sim \exp(J/T)$, one finds 
that $\log \tau_\alpha \sim E(\ell_{\rm eq}) / T \sim 1/T^2$, which is 
reminiscent of the B\"assler law in Eq.~(\ref{basslerlaw}).
Even more fragile behaviour is obtained 
for cooperative Fredrickson-Andersen models and Kob-Andersen 
lattice gases. For instance, for the well-studied Fredrickson-Andersen
model in $d=2$ with $k=2$ which is the original 
model studied in \cite{FA}, the relaxation time increases 
as $\tau_\alpha \sim \exp \left[ \exp( c/ T) \right]$, with $c$ a numerical 
constant \cite{solrit,TBF}. It is interesting to note that 
both dependencies found for the East and 
$k=d=2$ Fredrickson-Andersen fragile 
models seem to describe the viscosity data   
obtained in real molecular liquids quite well \cite{dyrenature,elmatad}.
Actually, it has been argued that in order to apply 
the East model results to real liquids, one should use a scaling 
$\ell_{\rm eq} \sim \exp(J/T-J/T_{\rm on})$. This leads to a modified version 
of the B\"assler law, which has been discussed and tested 
in \cite{elmatad,elmatad2010}.
 
The space-time representation of the defect diffusion 
dynamics in Fig.~\ref{fig:bubble1d} can be understood as
an alternative, very illustrative way to represent the 
dynamical behaviour of the system \cite{gc,GCannualreviewchemistry}. 
To the extent that dynamics is facilitated, mobility cannot disappear 
except if it is close to another mobile region.
Furthermore, the defect dynamics implies that mobile regions organize 
along `excitation lines', when represented in a space-time 
phase diagram. Hence, in a system with facilitated dynamics 
space-time is structured with strings of excitations that are directed in time,
can coalesce or branch but never die, nor appear spontaneously.
In between these lines lie large inactive regions, also called `bubbles'. 
The larger the spatial extension of the bubbles, the longer it takes
an excitation line to relax the corresponding domain, and the slower
is the structural relaxation. Thus, all dynamical properties 
can in fact be understood characterizing the space-time structure 
of bubbles and excitations lines. One should keep in mind, 
however, that this is fully equivalent to the defect
dynamics laid out above.  

To summarize, KCMs can be thought of as `defect' models, 
where defects can be non-trivial, extended objects
possibly with non-trivial sub-diffusive or cooperative dynamics. 
When temperature decreases both the density of defects
become small and their dynamics slows down dramatically, 
implying that the overall relaxation slows down in an activated,
possibly super-Arrhenius manner. Additionally, space-time 
representations, as shown in Fig.~\ref{fig:bubble1d}, suggest
the appearance when temperature decreases 
of large immobility domains with broadly distributed spatial and 
temporal extensions. When temperature decreases, these 
fluctuations are amplified  and the corresponding 
timescales and lengthscales increase rapidly and diverge 
(for most KCMs) in the limit of zero-temperature (or
unit density for lattice gases), which can thus
be considered as a dynamic critical point \cite{steve2},
as we now discuss.  

\subsubsection{Main predictions and results}

Within the facilitation approach and, hence, KCMs,
thermodynamic properties are trivial, and the interesting
physics is in the dynamics. As a consequence, non-trivial 
predictions and results concern the dynamical behaviour, 
and, more precisely, dynamic heterogeneity, 
as was realized soon after their introduction \cite{harrofirst,silvioKA,gc}. 
Remarkably, virtually all the aspects related to dynamic heterogeneity 
mentioned in Sec.~\ref{dh} can be investigated, and qualitatively
or quantitatively rationalized in the language of KCMs. 

Detailed numerical and analytical studies have indeed shown that in 
these systems, non-exponential relaxation patterns do stem 
from a spatial, heterogeneous distribution of timescales, 
directly connected to a distribution of dynamic 
lengthscales~\cite{gc,solrit,steve,TBF,panetal,mayerjack}. 

A decoupling phenomenon between the structural
relaxation time and the self-diffusion coefficient naturally
appears in KCMs and can be shown 
to be a very direct, quantifiable, consequence of 
the dynamic heterogeneity \cite{jung}. In fact, two distinct 
timescales emerge in the facilitation/KCMs 
description which both characterize the dynamics. 
One is the persistence time, $t_p$, which corresponds to the typical time a 
tracer particle has to wait to start moving, given an arbitrary 
start time for observation. The second timescale is the exchange time,
$t_x$, which is the typical time between elementary moves. 
The average value of $t_p$ and $a^2/\langle t_{x}\rangle$
are respectively estimates of the relaxation 
time $\tau_\alpha$ and the self-diffusion coefficient 
$D$, with $a$ the typical size of particle jumps. As a consequence one finds
$D \tau_\alpha \simeq \langle t_{p}\rangle/\langle t_{x}\rangle$. In 
some KCMs, this ratio increases substantially when decreasing the 
temperature, thus these models exhibit a substantial decoupling phenomenon. 

This is due to the fact 
that bubbles in space-time typically persist for the time scale of 
structural relaxation, while 
particles can diffuse faster by `surfing' on the excitation lines 
that surround these bubbles. Another way to 
understand the decoupling is to express $\langle t_p \rangle$
and $\langle t_x \rangle$ as distinct moments of the distribution 
of waiting time between jumps performed by a tracer particle,
which do not coincide when the distribution broadens
with decreasing temperatures \cite{Gilles,hedges,heuer}. 

The above description of self-diffusion in 
fluctuating mobility fields such as shown in Fig.~\ref{fig:bubble1d} 
has several non-trivial consequences such as for instance
a strongly non-Fickian relation between timescales and 
lengthscales at distances
smaller than a crossover lengthscale which grows when 
temperature decreases and decoupling increases \cite{berthierepl},
as observed in numerical simulations of supercooled liquids
\cite{berthier}.

Another useful aspect of KCMs is that
multi-point susceptibilities, multi-point spatial 
correlation functions such as the ones defined 
in Eqs.~(\ref{chi4def}) and (\ref{g4def}) can be studied 
in much greater detail than in molecular systems, to the point
that scaling relations between timescales, lengthscales, and 
dynamic susceptibilities can be 
established~\cite{silvioKA,steve2,TWBBB,panetal,Ck,jcpII}.

\begin{figure}
\psfig{file=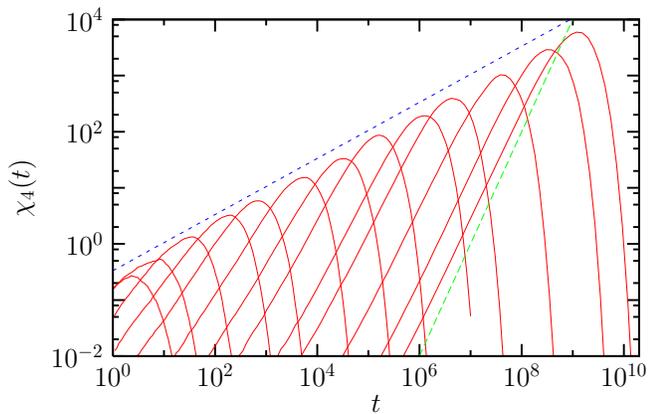,width=8.5cm}
\caption{\label{fig:chi4fa} Time evolution of 
$\chi_4(t)$ for different temperatures for the non-cooperative ($k=1$)
Fredrickson-Andersen model in three dimensions. Temperature
decreases from left to right. 
The peak height increases at low
$T$ as $\chi_4(\tau_\alpha) \sim 
1/\rho_d \sim \exp(J/T)$, while the relaxation time
scales as $\tau_\alpha \sim \exp(2J/T)$, such that
$\chi_4(\tau_\alpha) \sim \tau_\alpha^{1/2}$, as shown with a dotted line,
while the approach to the peak is diffusive, 
$\chi_4(t) \sim t^2$ (dashed line).}
 
\end{figure}

The type of scaling behaviour with the temperature or the density
depends on the details of the model at hand, in particular whether it has
or does not have a transition. However, a unified qualitative 
description is still possible in terms of the defect motion description 
given in the previous section \cite{TWBBB}.
On timescales small compared to $\tau_\alpha$, all sites 
that have been visited by the same defect are
dynamically correlated.
As a consequence, the dynamical correlation length $\xi_4(t)$ increases 
as $\xi_4(t) \sim 
(Dt)^{1/z}$ and the corresponding four-point 
susceptibility, $\chi_4(t) \simeq \rho_d (Dt)^{2 d_F /z}$, which
is the square of the volume of the regions visited by the same defect times
the defect density. When $t$ is comparable to 
the relaxation time $\tau_\alpha$, one finds
$\chi_4(\tau_\alpha) 
\simeq 1/\rho_d$ which, as expected, increases when decreasing
the temperature and increasing the density. 
On longer timescales, $\chi_4( \gg \tau_\alpha)$ decreases with $t$,
because a site can now be visited by more than one defect, and 
dynamical correlations are progressively erased.
As a direct example of this behaviour, we show in Fig. \ref{fig:chi4fa}
the time and temperature  evolution of $\chi_4(t)$ for the non-cooperative
($k=1$) Fredrickson-Andersen model in three dimensions \cite{steve},
where the above scaling relations hold \cite{TWBBB}.
 
In summary, relaxation timescales and dynamic 
lengthscales are found to diverge with well-defined
critical laws \cite{steve2,mayerjack}, which however are 
model dependent. 
The discovery of such `dynamic criticality' 
has proven useful, since it actually triggered several works aimed at 
computing similar laws within competing theories. It also implies 
the possibility that some  universal behaviour 
might truly emerge in the physics of supercooled liquids, 
precisely of the type observed 
numerically in Fig.~\ref{chi4ludo}, and 
experimentally in Fig.~\ref{cecile}.

\newcommand{\tobs}{t_{\rm obs}}

As suggested by the space-time representations shown in
Fig.~\ref{fig:bubble1d}, dynamical heterogeneity studies
are mostly concerned with distributions of dynamical quantities, 
most often through their variance, recall the definition of $\chi_4$
in Eq.~(\ref{chi4def}). However, there is potentially
useful information also in the overall shape of the distributions.
It seems reasonable that regions where the correlation function
is large possess rather stable structure at the molecular level,
while regions where it is small correspond to relatively
unstable local structure. 
To identify such trajectories over an observation time $t_{\rm obs}$, 
it is useful to define a measure
of dynamical activity, for example \cite{gcscience}
\be
K(N,t_{\rm obs}) = \sum_{i=1}^N \sum_{j=1}^{t_{\rm obs}/\Delta t} 
| {\bf r}_i(t_j) - {\bf r}_i (t_j-\Delta t)|^2,
\label{equ:def_K}
\ee
where ${\bf r}_i(t)$ is the position of particle $i$ at time
$t$ and the $t_j=j\Delta t$ are equally spaced times.
For large $N$ and $t_{\rm obs}$, the distribution of $K$
becomes sharply peaked about its average,
$\langle K \rangle$. In general,
for large $N$ and $t_{\rm obs}$, one expects the 
distribution of $K$ to have the form
\be
P(K) \simeq \exp[-N t_{\rm obs} f(K/Nt_{\rm obs})],
\ee
where the function $f(k)$ resembles a free energy density.

In some KCM \cite{merolle}, 
the distribution $P(K)$
has a characteristic shape, skewed towards small activity,
with an apparently exponential tail.
Further, on estimating $f(k)$ from this plot, there is a
range of $K$ over which $f(k)$ is non-convex (that is,
$f''(K)<0$).  
Within such a framework, non-convexity of $f(K)$ has
a direct interpretation as 
a `dynamical phase transition' 
in the system \cite{orderdisorder}.
The existence of
these phase transitions has been proven in simple
models~\cite{orderdisorder}
and numerical results for Lennard-Jones model
liquids are also consistent with the existence
of such a transition~\cite{gcscience}, which also 
exists in mean-field models where RFOT theory of Sec.~\ref{giuliotheory} 
applies \cite{jack-rom}. This leads to the hypothesis that the nature of the
dynamically heterogeneous fluid state could be generically 
interpreted in terms of coexistence between an active liquid and
inactive `ideal glass' states, with seems to suggest 
intriguing connections to the mean-field picture  based on the existence 
of numerous metastable states. 

\subsubsection{Current status of the facilitation approach}

We have conveyed the idea that despite the large number of distinct
KCMs capturing the idea of dynamic facilitation, these 
models are characterized by a common physics, based on  
the propagation of some rare `defects'. Quantitatively, however,
different models might behave quite differently. 
The variety of physical outcomes obtained from the variety of models
and kinetic rules has proven useful, as it has widened the spectrum 
of all the possible behaviours  that real material could display.
Thus, in the recent effort aiming at devising tools to
quantify and understand dynamic heterogeneity in glass-formers,
as described in Sec.~\ref{dh}, KCMs have played a leading role. 
 
As far as building a unique theory for the glass transition
is concerned, this variety is also a source of worry since   
it is difficult to obtain a unified description: many models
do not have a glass transition at finite temperature, but some have; 
some models seem to describe strong glass-formers, other are better for 
fragile materials; quantitative results about the behaviour 
of high order correlation functions or decoupling phenomena quite strongly
depend on the chosen model, the value of 
an upper critical dimension qualifying the strength
of fluctuations is also different from one model to another. 
At the moment, it is not clear whether or why one of these models
should be preferred to the others, since there exists
no systematic coarse-graining procedure to represent  
a molecular model with continuous degrees  
of freedom in terms of a lattice model with no interactions 
and kinetic constraints \cite{glotzersilica,kennett}.
Thus, the choice of a `reference' model to fit 
real data is an ambiguous issue, which would clearly deserve 
more work. 

It is interesting to remark that the theoretical status of the dynamic
facilitation approach is almost opposite to that of the RFOT theory, 
since there are several well-defined finite dimensional theoretical models 
capturing the idea of dynamic facilitation which can 
be thoroughly understood, but none of them can be 
derived from (even approximate) microscopic calculations. 
By contrast, RFOT theory is supported by several microscopic approaches,
but theory is not confirmed within the framework
of finite-dimensional models. This comparison raises the 
following question: is it possible to obtain more microscopic
theoretical insights from the dynamic facilitation approach beyond
qualitative or scaling considerations?
This absence of a microscopic derivation, or at least of some empirical 
procedure to back these ideas makes the approach prone to
criticisms. 

The idea that all thermodynamic
aspects are unimportant implies that 
the facilitation approach has little to say about 
the thermodynamic behavior of glass-formers close to $T_g$.
A possible coincidence between VFT and Kauzmann temperatures, $T_0$ 
and $T_K$, is not expected (except if they are both zero),
nor can the dynamics be deeply connected to thermodynamics,
as in Adam-Gibbs relations. Some view this as a deep flaw \cite{BBT}, 
while within the KCM approach, one is forced to disregard the significance 
of thermodynamics \cite{GCreply,GCannualreviewchemistry}.  
This is certainly the point where KCMs and RFOT theory 
differ most evidently.  
Even though the dynamics of KCMs shares similarities 
with systems characterized with a complex energy 
landscape~\cite{nontopo,steve3} and KCMs even 
show a MCT-like transition on Bethe lattices \cite{TBF,SBT}, 
their thermodynamical behaviours are widely different from RFOT theory, 
as was recently highlighted in \cite{rob2}. 
It will be interesting to follow how the facilitation approach
will handle the recent surge of interest in the definition and measurements 
of {\it static} `point-to-set' correlations, 
which are nonexistent in KCMs in the defect formulation.

A second central assumption made in KCMs is that mobility can neither
be spontaneously created nor destroyed unless adjacent regions 
are mobile, or, at least, that events violating this constraint
are rare and become rarer when approaching the glass 
transition \cite{gcpnas,elmatad2010}.
Without these assumptions, KCMs
become trivial models where death and birth rates for mobility
ultimately control the dynamics, and the features 
described in previous sections become irrelevant. 
The fact that some degree of facilitation is present in the dynamics is 
very reasonable and partially proven by numerical 
simulations \cite{glotzersilica,kennett,candelier1,GCannualreviewchemistry}.
However, the facilitation assumption goes much further and states that
almost nothing else is possible and that rare events, 
violating the kinetic constraints,
become rarer approaching $T_g$, so that the mobility is effectively nearly
conserved. There is at present no data supporting this assumption.
A recent experimental analysis of a granular system close to its glass 
transition actually suggests the opposite behaviour that 
mobility becomes less conserved and facilitation 
plays a decreasing role when the transition is approached \cite{candelier2}.
Certainly, this is an important issue to be addressed in the future 
in order to validate or disprove
this theoretical approach to the glass transition. It would be interesting
in particular  to 
repeat the analysis of \cite{candelier2} in a model of supercooled liquid. 

\subsection{Geometric frustration, 
avoided criticality, and Coulomb frustrated theories}

\label{sectiongilles}

\subsubsection{Physical picture and simple models}

In all of the above models, 
`real space' was present in the sense that special attention was paid to
different lengthscales characterizing the physics of the models
that were discussed. However, apart from the `packing models' with 
hard-core interactions, 
no or very little attention was paid to the geometric structure 
of local arrangements in molecular liquids close to a glass transition. 
This is generally justified using concepts such as 
`universality' or `simplicity', meaning that 
one studies complex phenomena using simple models, a typical 
statistical mechanics perspective.
However, important questions remain: what is the liquid structure
within mosaic states? How do different states differ? What is the 
geometric origin of the defects invoked in KCMs? Are they similar
to defects (disclinations, dislocations, vacancies, etc.) 
found in crystalline materials? 

There exists a line of research 
which attempts to provide answers
to these questions~\cite{gillesreview}. It 
makes heavy use of the concept of geometric frustration.
Broadly speaking, frustration refers to the
impossibility of simultaneously minimizing all 
the interaction terms in the energy function of
the system~\cite{toulouse}. Frustration might arise from quenched disorder
(as in the spin glass models described above), but liquids have 
no quenched randomness. In that case, frustration has a purely 
{\it geometric} origin. It is 
attributed to a competition between a short-range 
tendency to extend spatially a `locally preferred order', and 
global constraints that prevent the periodic tiling of space
with this local structure.

This can be illustrated 
by considering once more the packing problem of spheres 
in three dimensions~\cite{sadoc}. 
In that case, the locally preferred cluster of 
spheres is an icosahedron. However, the five-fold rotational symmetry 
characteristic of icosahedral order is not compatible
with translational symmetry, and formation of a periodic 
icosahedral crystal is impossible~\cite{frank}. 
By contrast, disks on a two-dimensional plane arrange 
locally as a regular hexagon, with one atom at the center and six neighbours
at the vertices. If periodically repeated, this structure can then form
a triangular lattice that can fill space with no influence 
of geometric frustration. 

The geometric frustration that 
affects spheres in three dimensional Euclidean space
can be relieved in curved space~\cite{Nelson}. This corresponds for instance 
to studying particles on a sphere, 
or on the hyperbolic plane (a surface of constant negative curvature). 
Indeed by changing 
the metrics and topology of the underlying space it may become possible 
for the local order to extend over larger lengthscales.     
One can fruitfully exploit this idea in two ways. 
A first possibility is to start from a curved space carefully chosen 
such that geometric frustration is entirely absent for the 
considered system. The structure of the system minimizing the energy 
can then be determined, and serves as a useful reference state. 
Changing the space curvature to go back to the physical Euclidean
space then generates topological defects that disturb the initially 
perfect order. Detailed analysis      
along those lines showed in particular that a sphere packing
possesses, in Euclidian space, topological
defects (mainly disclination lines), as the 
result of forcing the ideal icosahedral
ordering into a `flat' space. Thus the relevant topological defects
in that case are one-dimensional objects forming a disordered network,
presumably having a very complex dynamics. 
Nelson and coworkers have developed a solid theoretical framework based 
on this picture to suggest that the slowing down of supercooled liquids
is due to the slow wandering of these topological defects~\cite{Nelson}, 
but their treatment remains so complex that few quantitative, explicit results 
have been obtained, in particular concerning the dynamical behaviour of the
frustrated systems \cite{nelson3}.

The picture of sphere packing disrupted by geometric 
frustration is appealing and provides handles to attack the 
problem of glass formation from an atomistic perspective. 
Furthermore, the idea of uniform frustration can be incorporated
into simple models allowing for a more 
abstract approach in terms of simple statistical mechanics ideas and
scaling-type of approaches~\cite{gillesreview}.
To build such models, one must be able to identify, then
capture, the physics of geometric frustration. Using the 
concept of an ideal long-range ordering in a system of size $L$
in a curved space, which is then strained back in the three-dimensional
Euclidian space, Kivelson {\it et al.}~\cite{gillesphysica}
suggest that the corresponding free energy should scale as 
\be 
F(L,T) = \sigma(T) L^2 - \phi(T) L^3 +s(T) L^5. 
\label{fgilles}
\ee   
In this expression, 
the first two terms express the tendency of growing local preferred order 
and they represent respectively the energy cost of having an interface 
between two phases and a bulk free energy gain inside the domain. 
It is assumed that without the last term long range order would sets 
in at $T=T^*$.  
Geometric frustration is encoded in the third term which represents 
the strain free energy resulting from the frustration. 
This last term is responsible for the fact that 
the transition is avoided. 
The remarkable feature of Eq.~(\ref{fgilles}) is the 
super-extensive
scaling of the energy cost due to frustration which opposes
the growth of local order; in dimension $d$ it would scale as $L^{2+d}$.  
As a consequence, when $L$ is large, the last term becomes dominant
and prevents thus the extension of order to the entire space. 
Thus, the system is broken up in a `patchwork' of locally ordered 
domains separated by domains 
made of topological defects---hence the name of
frustration limited domains. 
Furthermore, minimizing the free energy per unit volume, $F/L^3$, 
one finds that the characteristic linear size $L^*$ of the patches 
scales as $(\sigma/s)^{1/3}$. 
Since $\sigma$ increases below $T^*$, one finds that the characteristic 
size of the patchwork increases below $T^*$ too. 

Turning to the dynamics, Kivelson {\it et al.} further argue 
that dynamics of the system involves restructuring of these domains 
in a thermally activated manner, using arguments similar to the ones
used within the mosaic picture of RFOT theory.
The typical energy barrier is given by 
a high temperature constant plus a second term $\sigma(L^*)^2$,
increases below $T^*$, which means that the assumption 
$\psi=2$ is made from the beginning,
see Eq.~(\ref{ag}).
Using more refined but still heuristic arguments, Kivelson {\it et al.} 
argued that $L^*$ grows as $(1-T/T^*)^\nu/K^{1/2}$ where $K$ is an 
adimensional parameter measuring the 
strength of frustration, and $\nu$ is the exponent governing the growth 
of the correlation length of the unfrustrated transition ($K=0$). 
The corresponding prediction for the energy barrier is 
\be\label{deltaT**}
\Delta(T < T^*) = \Delta_> + \frac{A T^*}{K} \left(1-\frac{T}{T^*}
\right)^{4\nu},
\ee
where $\Delta_>$ is the high temperature value of the barrier and $A$ 
is a positive constant.
Furthermore, it has also been argued that barrier fluctuations
lead to typical glassy features such as broad 
distribution of relaxation times or spatially heterogeneous 
dynamics, which can thus be discussed in a 
phenomenological manner \cite{Gilles}.  
We refer the reader to \cite{viot} for more details on these
aspects. An important prediction obtained from scaling considerations
is the variation of the glass fragility with frustration: in this 
approach, larger frustration means smaller domain sizes, and therefore 
less collective relaxation yielding smaller fragility.

A complementary route 
consists in the analysis of relatively simple, 
finite dimensional statistical models that supposedly retain the basic 
physical elements of Eq.~(\ref{fgilles}). 
We consider first an interacting spin model where the magnetization is meant
to represent the `preferred local order', nearest neighbour 
ferromagnetic interactions the
tendency to local ordering, and longer-ranged antiferromagnetic
interactions the opposite effect of the frustration. The 
following Hamiltonian possesses, in three dimensions, 
these minimal ingredients:
\be
H = - \sum_{\langle i,j \rangle} {\bf S}_i \cdot {\bf S}_j 
+ Q \sum_{i \neq j} \frac{{\bf S}_i \cdot {\bf S}_j }{|{\bf x}_i 
- {\bf x}_j|},
\label{coulomb}
\ee
where the spin ${\bf S}_i$ occupies the site $i$ at position 
${\bf x}_i$. It can be shown that the long-range 
Coulombic interaction plays a role analog to 
the super-extensive free energy in Eq.~(\ref{fgilles}).
Of course, such Coulomb frustrated spin models 
can be for Ising or multi-component spins, for hard or soft spins,
or generalized to different
spatial dimensions~\cite{gillesreview}. For theoretical studies
it is convenient to study field-theoretical versions of the Hamiltonian
(\ref{coulomb}):
\ba
H &=&  \frac{1}{2} \int d^d x \left[ 
r_o \phi({\bf x})^2 + [\nabla \phi ({\bf x})]^2 + \frac{u}{2} 
\phi({\bf x})^4 \right] \nonumber \\
& + & \frac{Q}{8 \pi} \int d^d x \int d^d x' 
\frac{\phi({\bf x})  \phi({\bf x}')}{|{\bf x}-{\bf x}'|},
\label{fieldtheo}
\ea
where the ferromagnetic and Coulombic terms are easily recognized, 
and the magnetization $\phi({\bf x})$ is now a field.

It is interesting to notice that the competition of interactions
acting at different lengthscales in fact describes a 
much larger body of problems, 
including the physics of diblock copolymers, magnetic multi-layer compounds,
Rayleigh-B\'enard convection, 
doped Mott insulators, etc. A well-studied, related field theory is given 
by:
\be
H = \int d^d x \left[ 
\frac{1}{2} \phi({\bf x}) \left( r_0 + k_0^{-2} ( \nabla^2 + k_0^2)^2
   \right) \phi({\bf x}) + 
\frac{u}{4} 
\phi({\bf x})^4
\right].
\label{braz}
\ee  
Although this field theory was mainly considered from the point of view
of diblock copolymers \cite{leibler} and Rayleigh-B\'enard 
convection \cite{swift}, it was explicitly considered in the 
context of the glass problem in \cite{comment,reply,phill}. 
Dimensional analysis shows that, in 
that case, the strength 
of frustration is related to $k_0$ through $k_0 \sim Q^{1/4}$.

This family of statistical models yields a rich physical behaviour, with 
a complex phase diagram and dynamical behaviour. 
In the absence of frustration, $Q=0$, they generally undergo a 
second order phase transition from a paramagnetic to a ferromagnetic phase, 
which should be viewed, in this context, as the analog 
of the ordering transition occurring in the curved space relieving 
the geometric frustration of the sphere packing. That transition
occurs at some finite temperature, $T_c(Q=0)=T^\star$. The presence
 of frustration, $Q >0$, generally has dramatic effects whose details
depend, however, on the studied model~\cite{chayes,zohar}. 
The transition can be entirely suppressed, $T_c(Q>0)=0$, or 
severely depressed as soon as $Q > 0$ yielding a genuine 
discontinuity when $Q \to 0$. By decreasing $T$ at small but finite 
$Q$, the system gets close to, but narrowly `avoids', the critical point
at $T^\star$.
This situation occurs for instance in the $O(N \to \infty)$ 
and spherical versions of the model.
In the more canonical Ising case, 
the situation is different since the second order transition 
occurring for $Q=0$ becomes first order at finite $Q$, but there is 
no discontinuity of the transition temperature 
as $Q \to 0$~\cite{braz}. At finite $Q$, the ordered 
phase is a spatially modulated phase (stripes in $d=2$, lamellae in 
$d=3$, etc.). Interestingly, the limit of stability of the 
paramagnetic phase (the spinodal) is depressed down to $T=0$, and it 
is thus (in principle) possible to supercool the first order 
transition and study the disordered phase down to low temperatures. 

In all cases, therefore, there exists a temperature regime below 
the `avoided critical point', $T<T^\star$, where 
the system is disordered and the dynamics is potentially affected 
by the presence of the frustration. Detailed Monte Carlo studies 
show that, for a given frustration strength $Q$, the 
dynamics slows down considerably when $T$ decreases 
below $T^\star$~\cite{grousson1,grousson2,gillesreview}. 
Grousson {\it et al.} quantified dynamics using spin-spin 
autocorrelation functions, $C(t) = \left\langle {\bf S}_i(t) \cdot
{\bf S}_i(0) \right\rangle$,
in a variety of models (Ising, 5-state spins, and XY models) 
in three-dimensions. They show that the time decay of $C(t)$ gets 
non-exponential at low $T$ and extract a relaxation timescale
that shows activated, super-Arrhenius behaviour. Remarkably, they also 
find that the fragility decreases when frustration increases, 
in agreement with the scaling approach of 
\cite{gillesphysica}, and 
simulations of Lennard-Jones models 
on the hyperbolic plane~\cite{Sausset2}, 
suggesting that Coulomb frustrated 
spin models do indeed capture an essential part of the physical behaviour
of molecular glass-formers.    

This optimistic view was contradicted in more recent work
\cite{phill}, 
where  numerical simulations of the Langevin dynamics 
of the field-theory of Eq.~(\ref{braz}) were reported. 
The numerical analysis yields
results for spin autocorrelation functions $C(t)$ in agreement with those 
of Grousson {\it et al.}. However, Geissler and Reichman 
also measured the time decay of 
fluctuations of the Fourier components of the field, 
$C_{\bf k}(t) = \left\langle \phi({\bf k},t) \phi({\bf -k},0) 
\right\rangle$, and found exponential decay 
for all Fourier modes, $C_{\bf k}(t) \sim \exp(-t/\tau(k))$. Therefore
they interpret the stretched exponential relaxation reported 
by Grousson {\it et al.} as being the result of a trivial 
superposition effect,
$C(t) \propto \int d^d k C_{\bf k}(t)$. Even more striking was their finding
that the relaxation times $\tau(k)$ were very accurately predicted 
by a simple dynamical Hartree approximation mirroring the classic 
static treatment of the field-theory (\ref{braz}) 
in \cite{braz}. This 
finding implies that the dynamical slowing down detected numerically 
results from the strong temperature dependence of static
correlations which will eventually yield the system to undergo 
a first-order transition towards a modulated phase. This scenario
bears little resemblance with the physics of supercooled liquids where
dynamics is not obviously driven by simple static correlations, as we emphasized 
several times in this paper.   

\subsubsection{Current status of the Frustration Limited Domains theory}

We believe that the generic idea 
of geometric frustration is a very appealing one, 
because it directly addresses the 
physics in terms of the `real space' at the molecular level.
Moreover, it seems to yield quite naturally the idea that 
the system organizes itself, at low temperature and finite frustration
level into some `mosaic' of domains corresponding to some 
local order whose size increases, but does not diverge, 
when $T$ decreases. 
Tarjus, Kivelson and co-workers clearly 
demonstrate that such an organization of supercooled liquids
into mesoscopic domains allows one to understand most of the fundamental
phenomena occurring in glass-formers~\cite{gillesreview}. 
Our use of the word `mosaic', as in
RFOT theory, is not accidental. It was demonstrated in several works 
\cite{schmalian1,schmalian2,nussinov}  that, 
when treated as systems with one-step of replica symmetry 
breaking as discussed in Sec.~\ref{scalingrfot}, Coulomb frustrated 
systems such as in Eq.~(\ref{fieldtheo}) undergo a Kauzmann 
transition very similar to the one found in mean-field spin glasses.  
This is remarkable and suggests that the two theories may not be so different 
as they appear at first sight.  
The presence of structured domains also 
connects to ideas such as cooperativity, dynamic heterogeneity 
and spatial fluctuations, that directly explains, at 
least qualitatively, non-exponential relaxation, decoupling phenomena 
or super-Arrhenius increase of the viscosity. The variation 
of fragility with the strength of the geometric frustration
is perhaps the most striking and original prediction stemming from 
this approach.

As for the RFOT mosaic picture, direct
confirmations of this scenario from molecular simulations 
are difficult to obtain. Two independent routes have been followed 
recently. On the one hand there have been studies focusing on 
the role of locally preferred structures. The identification of the 
locally preferred structure is in general quite difficult \cite{mossa}, 
except when it can be linked
to formation of local icosahedral order. Coslovich and Pastore
have found by numerical simulations of several models of glass-forming
liquids that the increase of icosahedral order upon super-cooling 
is more rapid and more pronounced for liquids characterized by a 
higher fragility~\cite{coslovich}. This success of the theory is tempered 
by recent simulations \cite{tanaka}, that  suggest instead using  
different model systems, 
including three dimensional polydisperse hard spheres, that
the increasing local order is the hexagonal bond 
orientational order
(characteristic of the crystal state) and not the icosahedral one. 
In yet another recent paper \cite{dyrepeter}, local order 
is again linked with slow structural relaxation but the interplay between 
local order and frustrated crystallization appears much 
more complicated than originally surmised in the frustration limited domain
approach. Thus, we are still far from understanding whether the basic 
content of this approach are wrong, correct, or too simplistic 
to describe supercooled liquids, and more work is needed to resolve this
issue.

The other route that has been followed consisted in analyzing a situation
very similar to the one FLD advocates to happen for glass-formers,
such as a two-dimensional monoatomic Lennard-Jones system on the hyperbolic
plane \cite{Sausset1,Sausset2,Sausset3}.  The hyperbolic plane
is a two-dimensional surface of constant {\it negative} curvature, 
and its relevance had been  suggested earlier 
by Nelson and co-workers~\cite{Nelson2}, but these numerical 
studies are technically not straightforward \cite{Sausset1}.
In the hyperbolic plane, the hexagonal crystal order is frustrated. 
In this setting the crystal order  
represents {\it mutatis mutandis} the locally preferred `glass' order that 
is conjectured to be frustrated for three dimensional glass-formers. 
This provides a concrete example of a particle model
characterized by an avoided phase transition. Furthermore, by changing the 
curvature of the space one can test the predictions on the fragility dependence
on frustration, which is related to the curvature of the plane.  
Sausset {\it et al.} confirmed  numerically that 
by curving space, they can prevent crystal ordering, as expected, and that
the low temperature dynamical behaviour of the liquid they obtain 
is very sluggish and has a behaviour very similar to glass-formers: 
time correlation functions decay in a two-step manner with a final 
decay which is not exponential, the dynamics is spatially heterogeneous, 
decoupling occurs~\cite{Sausset3}.
Interestingly, they also find that the fragility
of the glass-formers they obtain decreases when frustration 
grows~\cite{Sausset2}.  
The physical explanation is that increasing the frustration also increases
the density of topological defects, so that the putative collective behaviour
responsible for super-Arrhenius relaxation at small frustration 
becomes shorter-ranged when frustration increases. 

As a conclusion, this approach to the glass transition is based
on a very appealing real space description. As others, it is compatible 
with several experimental results, and it has passed non-trivial benchmark 
tests.
A possible criticism, which we also formulated in the case 
of the facilitation approach, 
is that no quantitative or microscopic results 
beyond scaling have been obtained,
and it is not even clear how they could actually be 
derived. In fact, approximate 
analytical approaches to FLD, that could have provided quantitative 
predictions, have found back results consistent with RFOT theory. 
In order to further test the FLD approach, it would be interesting to 
have more detailed, quantitative,
testable predictions regarding the behaviour of 
thermodynamic (e.g., specific heat) and dynamic (e.g. four-point 
dynamic susceptibility) observables from this perspective, see 
\cite{saussetchi4} for recent work in this direction.

\section{Off-equilibrium dynamics: Aging and rheology}
\label{aging}

\subsection{Why aging? Phenomenology and simple models}

\label{whyaging}

We have dedicated most of the above discussion to physics taking place 
when approaching the glass transition at thermal equilibrium. 
We discussed a rich phenomenology and serious
challenges for both our numerical and analytical capabilities to account
for these phenomena. Most people, however, focus on the properties
of glasses, i.e. below the glass transition, so deep in the glass 
phase that the material
seems to be frozen forever in an arrested amorphous state, 
endowed with enough mechanical stability for a glass to retain, say, the 
liquid it contains (preferentially a nice red wine either from 
France or Italy). One could naively think
that in this state of matter the dynamical evolution is arrested. Is it true?
The answer is clearly `no'. There is still life (and physics) below the glass 
transition. We recall that for molecular glasses, $T_g$ is defined
as the temperature below which relaxation is too slow to 
occur within a given experimental timescale. Much below $T_g$, therefore, 
the equilibrium relaxation timescale is so astronomically large 
that thermal equilibrium is out of reach. One enters therefore 
the realm of off-equilibrium dynamics. 
A full physical understanding of the non-equilibrium 
glassy state remains a central challenge~\cite{youngbook,leszouches}.

An immediate consequence of studying materials in a time window 
smaller than equilibrium relaxation timescales is that the system 
can, in principle, remember its complete history, a most 
unwanted experimental situation since all details of the experimental
protocol may then matter. The simplest protocol to study aging phenomena
in the glass phase is quite brutal~\cite{struik}: 
take a system equilibrated above the glass 
transition and suddenly quench it to a low temperature.
The system  then tries
to slowly reach thermal equilibrium, even though
it has no hope to ever get there. 
Aging  means  
that the system never forgets the time $\tw$ spent
in the low temperature phase, its `age', and that any measurement
started at time $\tw$ might have an outcome that 
explicitely depend on the value of $\tw$, unlike the situation
at thermal equilibrium.  

This implies in particular that any physical property of the glass becomes 
an age-dependent quantity in an aging protocol, and more 
generally dependent on how the glass was prepared. One can 
easily imagine using this property to tune mechanical or optical
characteristics of a material by simply changing the way it
is prepared, like how fast it is cooled to the glassy state.  
A very striking experimental realization of this idea 
was recently provided for organic glasses prepared 
in two different ways~\cite{edigerscience}. Ediger and collaborators
have compared the properties of glasses prepared in a canonical
way (slow cooling of bulk samples), to the ones 
of samples grown using slow vapor deposition at about 50~K
below the glass transition. They find that the latter samples
are much more `stable', in the sense that they behave as 
canonically prepared samples with a very large 
effective aging time (40 years for some sample).

Coming back to the simplest situation 
of a sudden quench to low temperature, it is found that one-time physical
observables such as density or energy evolve very slowly 
with the age of the sample. In polymer glasses for instance, 
the volume of the sample slowly decreases with $\tw$~\cite{struik}. 
Power laws with small exponents, $v(\tw) \sim 
v_\infty + (t_0/\tw)^{\alpha}$, or even logarithmic 
relaxations, $v(\tw) \sim  v_0 - v_1 \log(\tw/t_0)$, 
are frequently reported. In some cases, 
the time evolution of these observables is so slow that aging
behaviour is not obviously revealed by their study and the system
might superficially appear to have reached equilibrium. 

\begin{figure}
\psfig{file=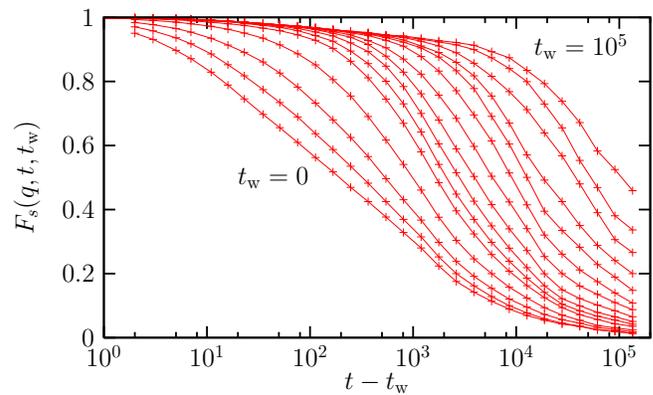,width=8.5cm}
\caption{\label{agingfig} Aging dynamics in a 
Monte Carlo study of a Lennard-Jones
glass-forming liquid at low temperature. The system is quenched at 
time $\tw=0$ to a low temperature which is then kept constant. 
Two-time self-intermediate scattering functions 
are then measured for 20 logarithmically spaced waiting times $\tw$
from $\tw=0$ to $\tw=10^5$ (from left to right). The relaxation
becomes slower when $\tw$ increases: the system `ages'.} 
\end{figure}

In order to show that the system never equilibrates, two-time quantities, 
such as density-density or dipole-dipole correlation functions, are much more
useful. A typical example is presented 
in Fig.~\ref{agingfig} where the self-part of the intermediate function
in Eq.~(\ref{isf}) is shown for a Lennard-Jones molecular liquid
at low temperature. Immediately after the quench, the system 
exhibits a relatively fast relaxation: particles still move
substantially. However, when the age of the system increases, 
dynamics slows down and relaxation becomes much slower. 
The relaxation then separates into two well-defined `time sectors'. 
For short time differences, $t-\tw \ll \tw$, which
corresponds to short-time dynamics of the 
amorphous structure, correlation functions at different
$\tw$ superpose. However, for large time differences, 
$t-\tw \gg \tw$, different curves relax at different rates, implying
that structural relaxation becomes slower when the system
gets older. Eventually,
when $\tw$ becomes very large, the relaxation becomes too slow to be followed
in the considered time window and the system seems frozen on that 
particular timescale. For practical purpose, it has now become a glass.

A striking feature conveyed by these data is that an aging system
not only remains out-of-equilibrium forever,
but its typical decorrelation or relaxation time\footnote{This is 
sometimes called relaxation time. It does not mean that 
the system relaxes after this time but just that its structure changes 
substantially, i.e. decorrelates, over
this time scale.} is in fact directly 
set by its own age $\tw$, which also separates 
short-time, equilibrated dynamics from long-time, aging relaxation. 
Therefore, although the system itself has no intrinsic characteristic 
relaxation timescales for relaxation, 
it remains able to relax during an aging experiment 
in a finite time which is set by the experimental protocol.
More quantitatively, it is found 
in the simplest cases that two-time correlators $C(t,\tw)$, 
or equivalently response functions, 
can be decomposed in the following way:
\be
C(t,\tw) \approx C_{\rm eq}(t-\tw) + C_{\rm aging}(t/\tw), 
\label{simpleaging}
\ee
in which case the relaxation time $\tau_\alpha(\tw)$
is directly proportional to $\tw$.  
This `time-aging time' superposition principle is 
reminiscent of the more standard `time-temperature' superposition
principle often found at thermal equilibrium.
It is found in many (but not all) polymeric liquids that
$\tau_\alpha$ grows sub-linearly with $\tw$,  
\be
\tau_\alpha(\tw) \sim \tw^\mu,
\label{muexp}
\ee
where $\mu$ is the so-called aging exponent~\cite{struik}, which 
usually takes value in the range $\mu \in [\frac{1}{2}, 1]$ ($\mu$ 
is equal to $0.8$ in Fig. \ref{agingfig}).
The simple description (\ref{simpleaging}) has been challenged by 
dielectric experiments of aging of molecular liquids 
\cite{NagelLeheny,Loidl,Richert}. These studies
have revealed that the aging part of the correlation function cannot be 
rescaled just by changing the relaxation time, i.e. the time-aging 
superposition is not always satisfied.
One has to stress however that in experiments one can access only the behavior
at frequency much higher than the inverse of the decorrelation time. 
This is due to the fact that 
the quench can be only performed at finite speed, of the order of 
$1 {\rm K}/{\rm min}$, in order to
avoid temperature gradients in the sample. As a consequence, when the 
final temperature
is reached, the decorrelation time has already grown so much that it is 
out of the experimental window. 
Thus, it is not really known 
whether the violation of the time-aging time superposition principle 
affects only the high
frequency tails of the dielectric susceptibility or whether 
Eq.~(\ref{simpleaging}) is incorrect even for $t$ of the order
of the decorrelation time for many glassy liquids.

Since the complete history of a sample in the glass phase matters,
there is no reason to restrain experimental
protocols to the simple aging experiment mentioned above. Indeed, 
experimentalists have investigated scores of more elaborated protocols
that have revealed an incredibly rich, and sometimes quite unexpected, 
physics \cite{youngbook}, a very old tradition in the field 
of polymer glasses \cite{kovacs,struik}.
In particular, striking `memory' and `rejuvenation' effects are 
observed during temperature cycling experiments~\cite{refregier} 
(one can also imagine applying 
an electric field or a mechanical constraint, be they constant 
in time or sinusoidal, etc.). 
These two effects were first observed in spin glasses, but the protocol
was then repeated in many different materials, from polymers
\cite{Bellon3,fukao,montes} 
and organic liquids \cite{NagelLeheny,yardim} to 
disordered ferroelectrics \cite{levelut}. 

On top of being elegant and quite intriguing, such protocols 
are relevant because they probe more deeply the dynamics 
of aging materials, allowing one to ask more precise 
questions beyond the simplistic observation that 
`this material displays aging'.  
Moreover, the observation of 
similar effects in many different glassy materials
implies that these effects are probably quite generic to systems 
with slow dynamics. Interesting also are the subtle 
differences possibly observed from one material to another.
A large body of experimental, numerical and theoretical papers have 
been devoted to this type of experiments, see \cite{hiking}
for a more extensive review on this topic. 

A popular interpretation of the aging phenomenon was obtained
by considering trap models~\cite{jptrap,jptrap2}. 
In this picture, reminiscent of the Goldstein 
view of the glass transition
mentioned above~\cite{Goldstein}, 
the system is described as a single particle evolving 
in a complex energy landscape with a broad distribution 
of trap depths. Thus, this is a paradigmatic mean-field approach. 
In particular, it is not easy to be very specific about what the `traps' of 
the `rugged landscape' represent in physical terms, although
much effort was recently dedicated to establish
connections between landscape and real space 
pictures~\cite{nontopo,bertin2,heuer}.

The simplest version of the trap model makes no reference 
to a spatial organization of the traps, and therefore
does not take into account  possible interactions between 
traps~\cite{jptrap2}. The dynamics of the model in written 
in terms of an evolution equation for $P(E,t)$, the probability
that the system is in a trap of depth $E$ at time $t$, and it 
is assumed that dynamics is thermally activated:
\be
\frac{\partial P(E,t)}{\partial t} = 
- \Gamma_0 e^{-\beta E} P(E,t) + \Gamma(t) \rho(E), 
\label{trap_eq}
\ee
where 
$\Gamma(t) = \Gamma_0 
\int dE \, e^{-\beta E} P(E,t)$ is the average hopping rate at
time $t$, $\Gamma_0$ being an attempt frequency. The 
complexity of the glassy material is now hidden into 
the probability distribution of the trap depth, $\rho(E)$.  
In Bouchaud's trap model, an exponential distribution 
of trap depth is assumed, 
\be
\rho(E) = \frac{1}{k_B T_0} \exp \left( - \frac{E}{k_B T_0} \right),
\ee
but the Gaussian trap model was also often considered in the context
of the equilibrium dynamics of supercooled 
liquids~\cite{bassler,dyre,jptrap2,heuer}.
In the perspective of aging studies, the exponential trap distribution
is more relevant because it induces a broader distribution 
of trapping times, $\tau$, 
\be
\varphi(\tau) \sim \frac{\Gamma_0}{(\Gamma_0 \tau )^{1+T/T_0}}. 
\label{fat}
\ee
Remarkably, the first moment of this distribution diverges 
at low temperatures, $T < T_0$, so that the system undergoes 
a true glass transition with a relaxation timescale
which diverges at $T_0$. 
By contrast, the Gaussian distribution does not yield 
to ergodicity breaking (i.e. aging is interrupted at long times) 
and it is thus mostly used in 
equilibrium studies, where it seems to account well for
properties of the potential energy landscape accessed 
in computer simulations~\cite{heuer}.  

When the system is suddenly quenched below 
$T_0$, aging behaviour arises because the system visits traps 
that are increasingly deep when $\tw$ increases, corresponding 
to more and more stable states. It takes therefore 
more and more time for the system to escape, and the dynamics slows 
down with time, in a manner reminiscent of Fig.~\ref{agingfig}.
More quantitatively, one can compute two-time correlation 
functions, such as the persistence function 
$C(t,\tw) = \int dE P(E,\tw) \exp[ -(\Gamma_0 e^{-\beta E})]$ 
which represents the probability the system has not changed trap
between times $\tw$ and $t+\tw$:
\ba
C(t,\tw) & = & \frac{\sin (\pi x)}{\pi} \int_{t/(t+\tw)}^1 
(1-x)^{T/T_0-1} x^{-T/T_0} \, dx \nonumber \\
& = & {\cal C}(t/\tw),
\label{trapaging}
\ea
which represents therefore an explicit example where the aging exponent 
defined in Eq.~(\ref{muexp}) can
be computed exactly, $\mu=1$, and correlation functions are
scaling functions of the rescaled time $t/\tw$, as in 
Eq.~(\ref{simpleaging}). 
 
In Eq.~(\ref{trapaging}), the correlation function is 
computed as an ensemble average. Important physical insights are also 
provided by trap models beyond averages at the level of the fluctuations, 
since the existence of waiting time distributions with 
fat tails indeed implies 
that time series are typically highly intermittent, and that
run-to-run fluctuations are also large, suggesting that it is 
interesting to consider also statistical distributions of the fluctuations. 
For instance, one can easily imagine that van-Hove functions 
in aging systems described by trap models are far from 
Gaussian~\cite{actrw}, and that non-ergodic effects can be quite 
strong in systems described by Eq.~(\ref{fat}), see \cite{ergodic}
for an example and \cite{barkai2} for a more formal approach. 

\subsection{Mean-field aging and effective temperatures}

\label{mfaging}

Theoretical studies of mean-field glassy models have provided important
insights into the aging dynamics of both structural and spin
glasses~\cite{CugKur1,CugKur2}. We have already encountered these
models in Sec.~\ref{meanfield}. They provide a simple setting to study glassy
models with a rugged free energy landscape. As a consequence it is natural 
to analyze their aging dynamics and use the corresponding results as a 
mean-field guide line for real systems. 

In Sec.~\ref{meanfield}, we described several alternative theoretical 
paths leading to essentially similar results for the
equilibrium properties of glasses, which could all
be described as `mean-field'. Because aging 
studies now directly deal with states that 
are non-stationary, protocol dependent, and far from 
equilibrium, not all of the mean-field equilibrium approaches 
are easily extended to low temperature aging studies. We expect 
for instance that mode-coupling theory should be able to treat 
such time-dependent phenomena, but this seems 
to be technically quite involved~\cite{latz} and a compete extension 
of MCT to the aging regime remains an open problem.
For completeness, we mention a recent work where 
the phenomenological RFOT mosaic scaling arguments
were used to describe also nonequilibrium relaxation below the glass 
transition~\cite{rfotaging}. This framework makes contact with the 
older theorical approach of 
Nayaranaswamy-Moynihan-Tool~\cite{tool,nara,mohican} but it also 
predicts important deviations from their findings.
  
Even by focusing on fully connected disordered models,
often described as `simple' models,  
we note that it took several years to derive 
a proper asymptotic solution of the long-time dynamics 
for a series of mean-field spin glasses~\cite{reviewLeticia}. These 
results have then triggered an enormous activity~\cite{crisrit}
encompassing theoretical, numerical and also experimental 
work trying to understand further these results, and to
check in more realistic 
systems whether they have some reasonable range of 
applicability beyond mean-field. 
This large activity, by itself, easily demonstrates
the broad interest of these results.  

Mathematically, understanding the aging dynamics
of mean-field glass models means solving a closed set of 
dynamical equations. For concreteness, let us consider 
the following spin glass Hamiltonian~\cite{theo}, 
\be
H = - \sum_{p=2}^\infty \sum_{j_1 < \cdots < j_p} 
J_{j_1 \cdots j_p} s_{i_1} \cdots s_{i_p}, 
\label{hamp}
\ee
where $s_i (i=1, \cdots, N)$ are spin variables interacting 
through coupling constants which are random Gaussian variables
of zero mean and variance $p! J_p^2/(2N^{p-1})$. 
This model is a straightforward generalization of the 
$p$-spin model of Eq.~(\ref{pspin}).
We consider soft spin
variables, which are real variables satisfying the spherical constraint, $\sum_i s_i^2 = N$.  
Due to the mean-field nature of the Hamiltonian (\ref{hamp}), 
a closed set of dynamical equations involving two-time correlation 
and response functions can be derived~\cite{reviewLeticia}. 
Defining
\ba
C(t,\tw) & = & \frac{1}{N} \sum_{i=1}^N 
\overline{\langle s_i(t) s_i(\tw) \rangle},  \nonumber \\ 
R(t,\tw) & = & \frac{1}{N} \sum_{i=1}^N \overline{\frac{\partial \langle
s_i(t) \rangle}{\partial h_i(\tw)}}\Bigg|_{h_i=0},  
\ea
where $h_i$ is a magnetic field that couples to spin $s_i$, and
\be
g(x) = \frac{1}{2} \sum_{p=2}^\infty J_p^2 x^p,
\label{defdeg}
\ee
one gets the time evolution of the two-time dynamic functions 
following a quench from a completely disordered state at $\tw=0$: 
\begin{widetext}
\begin{equation}
\begin{aligned}
\frac{\partial C(t,\tw)}{\partial t}  = & - \mu (t) C(t,\tw) + 2 T
R(t,\tw) 
+ \int_{0}^{\tw} d t' D(C(t,t'))R(\tw,t') +  
\int_{0}^{t} d t' \Sigma(t,t'') C(t'',\tw) ,\\
\frac{\partial R(t,\tw)}{\partial t}  = & 
-\mu (t) R(t,\tw) + \delta(t-\tw)  +  \int_{\tw}
^{t} d t' \Sigma (t,t') R(t',\tw), \\ 
\mu(t)  = & \,T + \int_{0}^{t} d t' 
\left[  D(C(t,t')) R(t,t') + \Sigma(t,t') C(t,t') \right],
\label{eqs}
\end{aligned}
\end{equation} 
\end{widetext}
where the kernels are defined as
\be
D(t,\tw) = g'[C(t,\tw)], \,\, \Sigma(t,\tw) = g''[C(t,\tw)]R(t,\tw).
\ee
The unique feature that makes the dynamics of mean-field
spin glass models soluble is that the 
dynamical equations (\ref{eqs}) are closed and only involve two-point 
functions. This great simplification stems from the 
fully-connected nature of the Hamiltonian (\ref{hamp}), and 
allows one to formulate an exact asymptotic solution for the 
dynamics of mean-field models~\cite{CugKur1}.

A comparison with Eq.~(\ref{schematicp}) immediately reveals 
why the aging regime is much harder to treat analytically, 
since one has to face two difficulties. 
First, two-time correlation functions now depend on both their
arguments, $C(t,\tw) \neq C(t-\tw)$.   
Second, the equations of motion (\ref{eqs}) 
in the aging regime not only involve time correlations, 
but also time-dependent response functions. At thermal equilibrium 
response and correlations are not independent, since the 
fluctuation-dissipation theorem (FDT) relates both quantities, 
\be
R(t,\tw) = \frac{1}{T} \frac{\partial C(t,\tw)}{\partial \tw}.
\label{fdt2}
\ee
Indeed, imposing both FDT and time translational invariance in Eqs.~(\ref{eqs})
yields the much simpler Eq.~(\ref{schematic}) for the $p$-spin model.

The solution to Eqs.~(\ref{eqs}) has been reviewed 
in great detail before~\cite{reviewLeticia}, so we only briefly describe
the most striking physical outcomes. 
As mentioned in Sec.~\ref{meanfield}, in these mean-field models, thermal
equilibrium is never reached when the quench is performed
below a critical temperature $T_c$, below which the relaxation 
time is infinite. It can be shown that aging proceeds 
forever by downhill motion in
an increasingly flat free energy landscape~\cite{laloux}, 
with subtle differences between spin glass and structural 
glass models. In both cases, however, time
translational invariance is broken, and two-time correlation and
response functions explicitly 
depend on both their time arguments. 
When $g(x)$ contains  a single term with $p>2$, 
as for instance in the spherical $p$-spin model, 
it can be shown that two-time correlation functions take
the form~\cite{reviewLeticia,alexandre}
\be
C(t,\tw) \approx C_{\rm eq}(t-\tw) + 
{\cal C} \left( \frac{h(t)}{h(\tw)} \right), \,\,
h(t)  = \exp(t^{1-\mu(T)}),   
\label{alex}
\ee
where $\mu(T)$ is an exponent that cannot be computed analytically. 
Numerical solutions show that $0< \mu(T) <1$ and $\mu(T\to 0) \to 1$.
The scaling form (\ref{alex}) is appealing since 
it is very similar to the empirical form in Eq.~(\ref{simpleaging}), 
and provides an explicit example where the use of the 
aging exponent $\mu$ defined in Eq.~(\ref{muexp}) and 
introduced by Struik~\cite{struik}
is analytically justified.  
From a broader perspective, it is interesting to note
that the exact dynamic solution of the equations 
of motion for time correlators~\cite{arnulf} displays a behaviour 
in strikingly good agreement with the numerical results 
reported for structural glasses, 
e.g., in Fig.~\ref{agingfig}.

Note that a more complex time dependence for $C(t,t')$ can be obtained 
for some mean-field glassy systems. These generally display a phenomenology
reminiscent of spin-glasses and not structural glasses 
(the transition is continuous,
the spin glass susceptibility diverges, etc). This is for 
example the case of the 
`$p=2+4$ model' for which $g(x)$ takes a more complex form. 
In these cases the scaling form with 
a single function $h(t)$ in Eq.~(\ref{alex}) does not hold anymore, 
but has to be generalized to include a continuous hierarchy
of such functions. Physically, this implies that 
the relaxation of correlation functions in the aging 
regime is associated to an infinite number of aging 
timescales, which all diverge with the age of the system.
 
The temporal behaviour of time correlation functions already
shows that mean-field spin glass models display a rich 
aging phenomenology.  
In aging systems, however, there is no reason to expect the FDT 
in Eq.~(\ref{fdt2}) to hold
and both correlation and response functions carry, 
at least in principle, distinct physical
information. Again, the asymptotic solution obtained for 
mean-field models quantitatively establishes that the FDT does not 
apply in the aging regime. The solution also 
shows that a generalized form of the FDT holds at large 
waiting times~\cite{CugKur1}. This 
generalized form of the FDT reads 
\be R(t,\tw) 
= \frac{X(t,\tw)}{T}
\frac{\partial   C(t,\tw) }{\partial \tw},
\label{fdr_def}
\ee 
which defines $X(t,\tw)$, the so-called fluctuation-dissipation ratio (FDR).
At equilibrium, correlation and response functions are time
translational invariant and
equilibrium FDT imposes that $X(t,\tw) = 1$ at all times. A parametric
fluctuation-dissipation (FD) plot of the step response or susceptibility
\be
\Chi(t,\tw)= T \int_{\tw}^t dt'\,R(t,t'),
\ee
against
\be
\Delta C(t,\tw)=C(t,t)-C(t,\tw),
\ee
is then a straight line with unit slope. These simplifications do
not necessarily occur in a non-equilibrium system. 
But the definition of an FDR
through Eq.~(\ref{fdr_def}) becomes significant for 
mean-field aging
systems~\cite{CugKur1,CugKur2}. In mean-field spin glass models the
dependence of the FDR on both time arguments is only through the
correlation function, 
\be
X(t, \tw ) \approx x (C (t, \tw )), 
\label{simple}
\ee
valid at large
wait times, $\tw \to \infty$. 

\begin{figure}
\psfig{file=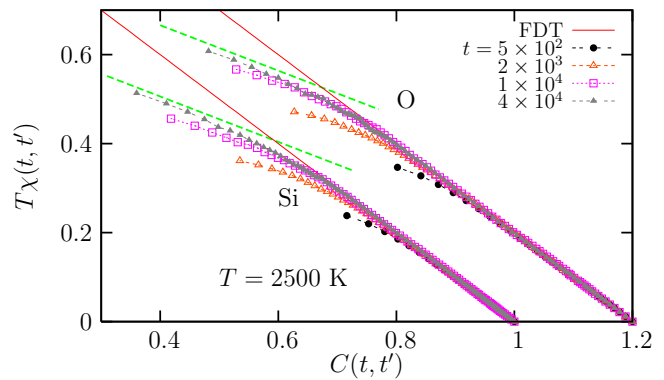,width=8.5cm}
\caption{\label{fdrfig} Parametric correlation-response plots 
measured in the aging regime of a numerical model for a silica
glass, SiO$_2$~\cite{prlsilica}. 
The plots for both species smoothly converges
towards a two-straight line plot of slope 1 at short times (large $C$ values), 
and of slope $X^\infty \approx 0.51$ at large times
(small values of $C$), yielding an effective temperature of about 
$T_{\rm eff} = T/X^\infty \approx 4900$~K. Note that $T_{\rm eff} > 
T_{MCT} \approx 3330 {\rm K} > T_g \approx 1450$~K.}
\end{figure}

For mean-field structural glass models (such as 
the $p$-spin model with $p>2$),
time correlation functions display a two-step relaxation 
process as in Eq.~(\ref{alex}). Correspondingly, 
the simplification (\ref{simple}) is even more spectacular since 
the FDR is shown to be characterized by only two numbers instead 
of a function, namely 
$X \sim 1$ at short time differences (large values of the correlator) 
corresponding to a quasi-equilibrium regime, with a crossover 
to a non-trivial number, $X \sim X^\infty$ for
large times (small values of the correlator). 
This implies that parametric FD plots are simply made of
two straight lines with slope $1$ and $X^\infty$, instead of the
single straight line of slope 1 obtained at equilibrium. Formally,
the infinite-time FDR, $X^\infty$, is defined as 
\be
X^\infty = \lim_{t\to\infty} \lim_{\tw \to \infty} X(t,\tw).
\ee
These theoretical predictions were tested with success 
in numerical simulations~\cite{barratkob,angelani}. 
In Fig.~\ref{fdrfig} we present more recent numerical data obtained 
in an aging silica glass~\cite{prlsilica}, presented in the form 
of a parametric response-correlation plot. 
The measured correlation functions are the self-part of the intermediate
scattering functions defined in Eq.~(\ref{isf}), while 
the conjugated response functions quantify the response of 
particle displacements to a spatially modulated 
field conjugated to the density.  
Plots for silicon and oxygen atoms at different
ages of the system are presented. They seem to smoothly converge 
towards a two-straight line plot, as obtained in mean-field 
models for structural glasses (note, however, 
that this could be a pre-asymptotic, finite ``$\tw$'', effect). 
Note also that the value of $T_{\rm eff}$ is much larger than
other relevant temperature scales, in particular than $T_{MCT}$, 
while both quantities nearly coincide in mean-field models. 

The behaviour of {\it spin glass} mean-field models, such as the  
$p=2+4$ model, is again more involved. It is found that 
to each timescale of the continuous hierarchy 
corresponds to a given value of the FDR. This means that the 
FDR has become a continuous function, $X \sim x(C)$ in the aging regime, 
instead of the single number found for the $p$-spin model.
For the spherical $p=2+4$ model, for instance, one finds
\be
x(C) = \frac{T}{2} \frac{g'''(C)}{[g''(C)]^{3/2}}, 
\label{eq_fdr}
\ee
for $C$ values corresponding to the aging regime;
$g(x)$ was defined in Eq.~(\ref{defdeg}).
Thus, the parametric FD plot is now made of a straight line 
of slope 1 for large value of the correlation function
(quasi-equilibrium regime), followed by a continuous curve
with slope $<1$ for smaller values of the correlation, that differs 
from the FDT even in the infinite time limit, $\tw \to \infty$.  

Since any kind of behaviour
is in principle allowed in non-equilibrium situations, 
getting such a simple, equilibrium-like structure for the 
FD relations is a remarkable result.  
This immediately led to the idea that aging systems might be characterized
by an effective thermodynamic behaviour and the idea
of `effective equilibration' at different timescales~\cite{CugKurPel97}.  
In particular, generalized FD relations in 
Eq.~(\ref{fdr_def}) suggest to define an 
effective temperature, as 
\be
T_{\rm eff} = \frac{T}{X(t,\tw)}, 
\label{teffdef}
\ee  
such that mean-field structural glasses are characterized by a unique 
effective temperature, $T_{\rm eff}=T/X^\infty$, 
in their aging regime.
It can be interpreted as the temperature at which 
slow modes are quasi-equilibrated~\cite{CugKurPel97,silvio}. 
One finds in general that
$0<X^\infty <1$, such that $T_{\rm eff} > T$, 
as if the system had kept some memory of its 
high temperature initial state. It was then proposed that 
FDRs, or equivalently effective temperatures, can be measured
by focusing on any type of physical observables~\cite{CugKurPel97}.
The FDR is then defined in terms of the two-time connected correlation
function for generic physical observables $A(t)$ and $B(t)$, 
\be C_{AB}(t,\tw) = \langle A(t) B(\tw)
\rangle - \langle A(t)\rangle \langle B(\tw) \rangle, 
\label{corr}
\ee 
with $t \ge
\tw$, and the corresponding two-time (impulse) response function 
\be R_{AB}(t,\tw) =
\frac{\delta \langle A(t) \rangle}{\delta h_B (\tw)}\Bigg|_{h_B=0}.  
\label{resp}
\ee
Here $h$ denotes the thermodynamically conjugate field to the
observable $A$ so that the perturbation to the Hamiltonian
is $\delta H = - h_B \cdot B$. A practical measurement
of the FDT is then performed by building a parametric FD plot
of the integrated response function, 
\be
\chi_{AB} = T \int_{\tw}^t dt' R_{AB}(t,t'),
\ee
versus the correlation function $[C_{AB}(t,t)-C_{AB}(t,\tw)]$.

The name `temperature' for the quantity defined in Eq.~(\ref{teffdef})
is not simply the result of a dimensional analysis but has a deeper, 
physically appealing meaning that is revealed by asking the 
following questions. How does one measure temperatures in
a many-body system whose relaxation involves well-separated 
timescales? What is a thermometer (and a temperature) in a 
far from equilibrium aging material? 
Answers are provided in \cite{CugKurPel97,Kurchan} both 
for mean-field models and for additional toy models 
with multiple relaxation timescales. The idea is to couple 
an additional degree of freedom, such as a harmonic oscillator, $x(t)$,  
which plays the role of the thermometer operating at frequency 
$\omega$, to an observable of interest $A(t)$ via a linear coupling,
$-\lambda x(t)A(t)$. Simple calculations show then that the 
thermometer `reads' the following temperature, 
\begin{equation}
\frac{1}{2} k_B T_{\rm meas}^2 \equiv \frac{1}{2} \omega^2 
\langle x^2 \rangle
= \frac{\omega C_{AA}'(\omega,\tw)}{2 R_{AA}''(\omega,\tw)},
\label{meas}
\end{equation}  
where $C_{AA}'(\omega,\tw)$ is the real part of the 
Fourier transform of Eq.~(\ref{corr}), and 
$R_{AA}''(\omega,\tw)$ the imaginary part of the 
Fourier transform of Eq.~(\ref{resp}), with 
$h=\lambda x$. The relation (\ref{meas})
indicates that the bath temperature is measured, 
$T_{\rm meas} = T$, if frequency is high and FDT is satisfied, while
$T_{\rm meas} = T_{\rm eff} > T$ if frequency is slow enough to be
tuned to that of the slow relaxation in the aging material.
More generally, relaxation in mean-field glassy
systems may occur in several, well-separated time 
sectors~\cite{CugKur2}. It is
then easy to imagine that each sector could be associated with 
its own effective temperature~\cite{Kurchan}.  

The direct
link between the FDR in Eq.~(\ref{fdr_def}) and the effective temperature
measured in Eq.~(\ref{meas}) was numerically confirmed in computer
simulations of glassy molecular liquids. In \cite{jl}, a tracer 
particle was used as a thermometer, its frequency being tuned 
by modifying its mass. It was verified that its kinetic 
energy was controlled by the temperature of the heat bath for light (high
frequency) tracers, while it was related to the effective 
temperature of the slow degrees of freedom for heavy (low frequency)
tracers. In the same spirit, a two-level system with adjustable
frequency was studied in \cite{ilg}, and the activation rate 
changed from being proportional to $\exp(-E/T)$ to $\exp(-E/T_{\rm eff})$
when decreasing the frequency. These two examples show that 
$T_{\rm eff}$ truly deserves the name of a temperature
in a fundamental sense. We note that
these modern concepts are related to, but make much more 
precise, older ideas of quasi-equilibrium and fictive 
temperatures in aging glasses~\cite{tool,nara,mohican}. 
Attempts were even made to develop a thermodynamic
of the glass state, heavily relying on the idea
of effective temperatures. This has become the subject 
of a book~\cite{theobook}. 

The final piece of information extracted from the behaviour of mean-field
spin glass models is the existence of a connection between
a non-equilibrium dynamic quantity, namely the 
asymptotic FDR $x(C)$ defined in Eq.~(\ref{simple}), and 
the thermodynamic behaviour of the system in the low temperature phase.
It turns out that the thermodynamics of 
mean-field models for structural glasses 
is characterized by one-step replica symmetry breaking,
while a full-step solution is needed to solve the 
thermodynamics of models for spin glasses~\cite{giorgio}.
Remarkably, the structure of the spin glass order parameter, 
the Parisi function $P_{\rm eq}(q)$ describing the probability 
distribution of overlaps between equilibrium states, 
is directly related to the structure of the function $x(C)$. 
For the $p=2+4$ model, for instance, static
calculations yields an explicit expression for 
$P_{\rm eq}$~\cite{theo}. Comparing with 
Eq.~(\ref{eq_fdr}), it turns out that the following
equality holds:
\be
x(C) = \int_0^C dq P_{\rm eq}(q). 
\label{theorem}
\ee
The situation for mean-field structural glasses is more complicated, 
since Eq.~(\ref{theorem}) does not hold, but 
the integrated Parisi function (the right hand side in (\ref{theorem}))
has the same structure as the FDR (the left hand side)~\cite{CugKur1}. 
Therefore, the full-step or one-step replica symmetry 
breaking schemes needed to solve the static problem in these models 
have a direct dynamical counterpart, the 
FDR being a function or a number, respectively, in 
the aging regime. It was further argued that Eq.~(\ref{theorem})
might hold for finite dimensional glassy systems as 
well~\cite{FraMezParPel98}, raising the 
exciting possibility that 
the Parisi function might become experimentally accessible through 
aging experiments, which triggered a large research 
activity~\cite{crisrit}. However, it is not easy to 
determine the conditions under which Eq.~(\ref{theorem})
might hold~\cite{giorgio}. 
Moreover, aging studies are very often performed
very far from any asymptotic regime, and  
very little is know about how Eq.~(\ref{theorem}) is modified
when the $\tw \to \infty$ limit cannot be taken~\cite{alain,malcom}, 
making the thermodynamic interpretation of the outcome of 
aging measurements delicate.

Taken together, these results make the mean-field description of aging
very appealing, and they nicely complement the mode-coupling/RFOT
description of the equilibrium glass transition described above.
Moreover, they have set the agenda for a very large body of numerical
and experimental work, as reviewed in~\cite{crisrit}.
It should be clear, however, that these results are strictly valid in the 
mean-field limit in the sense discussed in Sec.~\ref{meanfield}. 
Nonequilibrium aging dynamics in mean-field spin glasses turns
out to describe the slow descent of the system in an energy landscape 
which becomes more and more flat as the age increases~\cite{laloux}, 
with no access to the deeper metastable states that are supposed 
to play an important role in glasses near the 
experimental glass transition. This view is thus in striking contrast 
with the purely activated description given for instance
by trap models. Additionally, it is important 
to understand the role of spatial fluctuations which are
not naturally included in the mean-field description, see
\cite{malcom,chamon} for recent work in this direction.
Thus it is important to test further the mean-field 
concepts to understand how they apply to the three-dimensional world.  
 
\subsection{Beyond mean-field: Experiments, critical points and kinetically
constrained models}

Despite successes such as shown in Fig.~\ref{fdrfig}, the broader
applicability of the mean-field scenario of aging dynamics remains
unclear, however. 
While some experiments and simulations indeed seem to support the
existence of well-behaved effective
temperatures~\cite{Grigera99,Abou04,Wang06,israeloff2010}, 
other studies also reveal
the limits of the mean-field scenario.  Experiments have for instance
reported anomalously large FDT violations associated with intermittent
dynamics~\cite{Bellon1,Bellon2,Buisson1,Buisson2,bartlett}, while theoretical
studies of model systems have also found non-monotonic or even
negative response functions~\cite{Viot03,nicodemi,kr,DepSti}, and
ill-defined or observable-dependent FDRs~\cite{FieSol02}.  
Some experiments even reported an absence of violations to 
the fluctuation-dissipation theorem~\cite{jabbari,jop}. In
principle, these discrepancies with mean-field predictions are to be
expected, since there are many systems of physical interest in which
the dynamics are not of mean-field type, in particular displaying activated
processes. However, it is not possible to draw any consistent picture from 
the experiments at this stage. As a consequence it is not quite clear yet
how and to what extent mean field results are violated in real systems. 

It is thus an important task to
understand from the theoretical point of view when the mean-field
concept of an FDR-related effective temperature remains viable.
However, studying theoretically the interplay between relevant dynamic 
lengthscales and thermally
activated dynamics in the non-equilibrium regime of disordered 
materials is clearly a
challenging task. As a consequence, much work has been devoted to analyze
simple effective models. A lot of attention has been focused on spin models.

A first class of system that we discuss are coarsening systems.
Although not directly
related to the glass problem, they provide a simple, 
yet non-trivial, theoretical
framework to study situations where both aging and spatial 
heterogeneity are present, and where time correlation
and response functions display interesting scaling behaviour. 
The paradigmatic situation  is that of an Ising ferromagnetic 
model (with a transition at $T_c$) 
suddenly quenched in the ferromagnetic phase at time 
$\tw=0$. For $\tw > 0$ domains of positive and negative magnetizations
appear and slowly coarsen with time. The appearance of 
domains that grow with time proves the presence 
of both aging and heterogeneity in this situation. 

The case where the quench is performed down to $T<T_c$ is 
well understood. 
The only relevant lengthscale is the growing domain size, $\ell(\tw)$
and the physical behavior can be understood by scaling theory~\cite{reviewbray}.
Correlation functions display aging, and scale invariance 
implies that $C(t,\tw) \sim f(\ell(t)/\ell(\tw))$. Response 
functions can be decomposed into two 
contributions~\cite{barrat,BBK2}: one part stems from 
the bulk of the domains and behaves as the equilibrium response, 
and a second one from the domain walls and becomes vanishingly small
in the long time limit where $\ell(\tw) \to \infty$ and the density
of domain walls vanishes. This implies that for coarsening systems
in $d \geq 2$, one has $X^\infty = 0$, or equivalently an infinite 
effective temperature, $T_{\rm eff} = \infty$. The case $d=1$ is special
because $T_c=0$ and the response function remains dominated by the
domain walls, which yields 
the non-trivial value $X^\infty = 1/2$~\cite{ising1d,ising1d2}. 

Another special case has retained attention. When the quench 
is performed at $T=T_c$, there is no more distinction
between walls and domains and the above argument yielding 
$X^{\infty}=0$ does not hold. 
Instead one studies the growth
with time of critical fluctuations, with $\xi(\tw) \sim \tw^{1/z_c}$
the correlation length at time $\tw$, 
where $z_c$ is the dynamic critical exponent.
Both correlation and response functions become non-trivial
at the critical point~\cite{GL}. 
It proves useful in that case to consider the dynamics 
of the Fourier components of the magnetization fluctuations, 
\be
C_q(t,\tw) = \langle m_q(t) m_{-q}(\tw) \rangle,
\ee
and the conjugated response 
\be
R_q(t,\tw) = \frac{\delta \langle m_q(t) \rangle}{\delta 
h_{-q}(\tw)}\Bigg|_{h_{-q}=0},
\ee
where $h_q(t)$ is the Fourier component of 
the magnetic field at time $t$.
From Eq.~(\ref{fdr_def}) a wavevector dependent 
FDR follows, $X_q(t,\tw)$, which has 
interesting properties that can be computed
by a number of means, including dynamical 
renormalization techniques, see \cite{pasquale} for a review.
One of the main outcomes of these studies is that the effective
temperature for quenches at the critical point might in some cases
depend on the observable~\cite{gambassi1} 
and on the initial condition \cite{gambassi2}, thus weakening 
the interpretation of $X_q$ in terms of effective temperature.

In dimension $d=1$, it is possible to compute 
$X_q(t,\tw)$ exactly in the aging regime at $T=T_c=0$. 
An interesting scaling form is found, and numerical simulations 
performed for $d>1$ confirm its validity \cite{pre}: 
\be
X_q(t,\tw) \approx {\cal X}[ q \xi(\tw) ],
\label{scaling_xq}
\ee
where the scaling function ${\cal X}(x)$ is  
${\cal X}(x \to \infty) \to 1$ at small lengthscale, $q \xi \gg 1$, 
and ${\cal X}(x \to 0) \to X^\infty = 1/2$ (in $d=1$)
at large distance, 
$q \xi \ll 1$; recall that $z_c=2$ in that case.
 
Contrary to mean-field systems where geometry played no role, 
here the presence of a growing correlation lengthscale 
plays a crucial role in the off-equilibrium regime since 
$\xi(\tw)$ allows one to discriminate between fluctuations 
that satisfy the FDT at small lengthscale, $X_q \sim 1$, 
and those at large lengthscale which are still far from equilibrium, 
$0< X_q \sim X^\infty < 1$. These studies suggest therefore that 
generalized fluctuation-dissipation relations in fact 
have a strong lengthscale dependence---a result which is
not predicted using mean-field approaches.  

Another interesting result is that the FDT violation
for global observables (i.e. those at $q=0$) takes a particularly 
simple form, since the introduction of a single number is sufficient, 
the FDR at zero wavevector, 
$X_{q=0}(t,\tw) \equiv X^{\infty} = 1/2$ (in $d=1$). 
This universal quantity takes non-trivial values 
in higher dimension~\cite{GL}, e.g. $X^\infty \approx 0.34$ is measured 
in $d=2$~\cite{pre}. This shows that the study of global rather than local 
quantities makes the measurement of $X^\infty$ much easier. 
Finally, having a non-trivial value of $X^\infty$
for global observables suggests that the possibility to 
define an effective temperature remains valid, 
but it has become a more complicated object, related to global
fluctuations on large lengthscale.  
The first experimental determination of the value of $X^\infty$
near a critical point was reported only very recently
in a system of liquid crystals, where the director is the relevant
fluctuating observable~\cite{joubaud}. 
An intriguing value $X^\infty \simeq 0.31$ was 
reported, which has received, to our knowledge, 
no theoretical justification \cite{pasquale}. 
 
Kinetically constrained spin models represent a second 
class of non-mean-field 
systems whose off-equilibrium has been 
thoroughly studied recently~\cite{leonard}.
This is quite a natural thing to do
since these systems have 
local, finite ranged
interactions, and they 
combine the interesting features of being defined in 
terms of (effective)
microscopic degrees of freedom, having local dynamical rules, and 
displaying thermally activated and heterogeneous dynamics.

\begin{figure}
\psfig{file=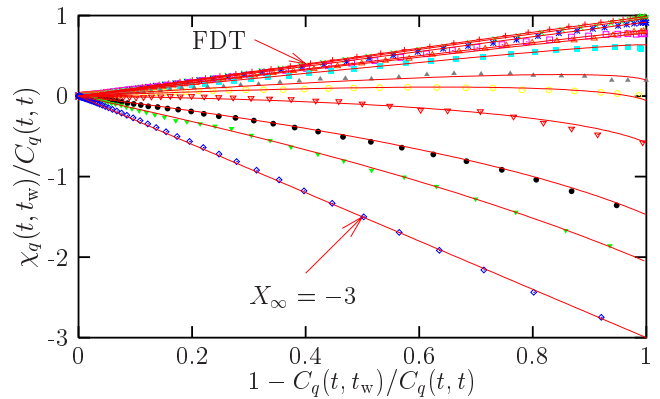,width=8.5cm}
\caption{\label{fdrfig2} 
Parametric response-correlation plots for the Fourier components 
of the mobility field in the $d=3$ Fredrickson-Andersen model.
Symbols are from simulations, lines from analytic calculations, 
and wavevectors decrease from top to bottom.
The FDT is close to being satisfied at large $q$ 
corresponding to local equilibrium. At larger distance
deviations from the FDT are seen, with an asymptotic 
FDR which becomes negative. Finally, for energy fluctuations
at $q=0$ (bottom curve), the plot becomes a pure straight line
of (negative!) slope $-3$, as a result of thermally activated dynamics.} 
\end{figure}

The case of the Fredrickson-Andersen model, described in Sec.~\ref{theory},
has been studied in great detail~\cite{leonard}, and we summarize the main 
results. Here, 
the relevant dynamic variables are the Fourier components 
of the mobility field, which also correspond in that case 
to the fluctuations of the energy density.
Surprisingly, the structure of the generalized fluctuation-dissipation
relation remains once more very simple. In particular, in dimension 
$d>2$, one finds a scaling form similar to Eq.~(\ref{scaling_xq}), 
$X_q(t,\tw) = {\cal X}[ q \xi(\tw)]$, with a well-defined 
limit at large distances, $X_{q=0}(t,\tw) \equiv X^\infty$. 
The deep analogy with critical Ising models stems from the fact
that mobility defects in KCMs diffuse in a way similar to 
domain walls in coarsening Ising models. It is in fact
by exploiting this analogy that analytic results are 
obtained in the aging regime of the 
Fredrickson-Andersen model~\cite{peters}.

There is however a major qualitative difference between 
the two families of model. The (big!) surprise lies in the sign 
of the asymptotic FDR, since calculations show that~\cite{nveX} 
\be
X^\infty = -3, \quad \,\, d>2.
\ee
In dimension $d=1$, one finds 
$X_{q=0}(t,\tw) = f(t/\tw)$ with 
$X_{q=0}(t\to\infty,\tw) = \frac{3\pi}{16-6\pi} 
\approx -3.307$.
Numerical simulations in various spatial dimensions 
nicely confirm these calculations. 
In Fig.~\ref{fdrfig2}, we show such a comparison between 
simulations (symbols) and theory (lines) in the case
of the $d=3$ Fredrickson-Andersen model~\cite{nveX}. 
Fourier components
of the mobility field yield parametric FD plots that follow
scaling with the variable $q \xi(\tw)$, as a direct result
of the presence of a growing lengthscale for 
dynamic heterogeneity, with a simple diffusive behaviour 
in that case, $\xi(\tw) \sim \sqrt{\tw}$.  
Again, generalized fluctuation-dissipation relations
explicitly depend on the spatial lengthscale considered, 
unlike in mean-field studies.  
In Fig.~\ref{fdrfig2}, the limit $q =0$ 
corresponding to global observables is also 
very interesting since the plot is a pure straight line, 
as in equilibrium. Unlike equilibrium, however, the slope is not 1 but
$-3$. A negative slope in this plot means a negative FDR, and therefore
suggests a negative effective temperature, 
a very non-intuitive result at first sight. 

Negative response functions in fact directly follow
from the thermally activated nature of the dynamics 
of these models~\cite{nveX}. First, one should note that 
the global observable shown in Fig.~\ref{fdrfig2} corresponds 
to fluctuations of the energy, $e(\tw)$, whose conjugated field is 
temperature. 
In the aging regime the system slowly drifts towards equilibrium.
Microscopic moves result from thermally activated 
processes, corresponding to the local crossing of energy barriers.
An infinitesimal change in temperature, $T \to T+\delta T$ 
with $\delta T >0$, accelerates these barrier crossings and 
makes the relaxation dynamics faster.
The energy response to a positive temperature pulse
is therefore negative,  $\delta e < 0$, which directly
yields $\delta e / \delta T <0$, and explains 
the negative sign of the FDR. This line of reasoning does not apply to 
mean-field glasses, where thermal activation plays no role.  
To our knowledge, negative effective temperatures in realistic aging 
materials has not yet been observed.

Finally, another scenario holds for local observables in some KCMs when 
kinetic constraints are stronger, such as the 
East model~\cite{leonard} 
or a bidimensional triangular plaquette model~\cite{robfdt}. 
Here, relaxation is governed by a hierarchy of energy 
barriers that endow the systems with specific dynamic
properties. In the aging regime following a quench, 
in particular, the hierarchy yields an energy relaxation that 
arises in discrete steps which take place 
on very different timescales, reminiscent of the `time sectors' 
encountered in mean-field spin glasses (but not in mean-field structural
glass models!). 
Surprisingly, it is found that to each of these 
discrete relaxations one can associate a well-defined (positive)
value of the fluctuation-dissipation ratio, again 
reminiscent of the dynamics of mean-field spin 
glass models.  Therefore, even in models that are very far from 
the mean-field limit the physical picture 
of a slow relaxation taking place on multiple timescales 
with each timescale characterized by an effective temperature 
seems to have some validity. 

In conclusion, we have described multiple non mean-field 
situations where collective behaviour (critical point
in standard ferromagnets, dynamic criticality emerging
in kinetically constrained models) produce
non-trivial aging dynamics characterized by the emergence
of universal fluctuation-dissipation properties. 
These results are instructive for 
understanding how and to what extent mean-field results 
can change in finite dimensional systems.  However, 
it is not clear that aging in structural glasses may be understood
in terms of domain coarsening (domains between what?), nor that 
kinetically constrained models are faithful effective models
of glasses. As a consequence,
it is difficult to reach for theoretical models
a consistent picture of effective temperature 
and aging valid for finite dimensional structural glasses.  
However these results suggest that, most certainly, 
nonequilibrium effective temperatures have a 
validity much beyond the realm of mean-field disordered 
spin models where they were first discovered \cite{Kurchan}.   
 
\subsection{Driven dynamics of glassy materials}

We have introduced aging phenomena with the argument that
in a glass phase, the timescale to equilibrate becomes 
so long that the system always remembers its complete history. This is
true in general, but one can wonder whether it is possible to invent 
a protocol where the material history could be erased, and the system
`rejuvenated'~\cite{mckenna}. 
This concept has  been known for decades in the field of polymer 
glasses, where complex thermo-mechanical histories are often 
considered, for obvious practical reasons. 

Let us consider an aging protocol where a supercooled liquid is quenched 
to a low temperature at time $\tw=0$, but is simultaneously
forced by an external mechanical constraint.
Experimentally, one often finds
that a stationary state can be reached, which explicitly depends
on the strength of the forcing: a system which is forced more strongly 
relaxes faster than a material which is less perturbed, a phenomenon
called `shear-thinning'. The material
has therefore entered a driven steady state, where 
memory of its age is no longer present and dynamics has become 
stationary: aging is stopped~\cite{jorgeaging}. 

Many studies of these driven glassy states have been performed 
in recent years. Note that the drive is in general mechanical and, hence,
these are relevant for the rheology of supercooled liquids
and glasses, and the $T \ll T_g$ limit corresponds to studies 
of the plasticity of amorphous solids, which is a broad field in 
itself, see \cite{falk,maloney,tanguy,schall,bocquet2009} 
for a few modern perspectives in this direction.
In the colloidal world, such studies are also relevant 
for the newly-defined field of the rheology of `soft glassy
materials'. These materials are (somewhat tautologically) 
defined as those for which the non-linear 
rheological behaviour is believed to result precisely
from the competition between intrinsically slow relaxation processes
of glassy origin, 
and an external forcing~\cite{prlsollich}. It is believed that 
the rheology of dense colloidal suspensions, 
foams, emulsions, pastes,
or even biophysical systems are ruled by such a competition, 
quite a broad field of application 
indeed~\cite{cates}.

From the point of view of statistical mechanics modeling, the rheology of  
soft glassy materials can be naturally studied
from the very same angles as the equilibrium glass transition 
and aging phenomena itself. 
As such, trap models~\cite{prlsollich,presollich,suzanne}, 
mean-field spin glasses~\cite{ledoussal,BBK} and the related
mode-coupling theory 
approaches~\cite{dave,dave1,dave2,fuchs,fuchs1,fuchs2,fuchs3} 
have been explicitly extended
to include an external mechanical forcing. In all these cases, one 
generically finds that a driven steady state can be reached and aging is
indeed expected to stop at a level that depends on the strength of the 
forcing.  A most interesting aspect is that 
the broad relaxation spectra predicted to occur 
in glassy materials close to a glass transition 
directly translate into `anomalous' laws both 
for the linear rheological behaviour (seen experimentally 
in the broad spectrum of elastic, $G'(\omega)$, 
and loss, $G''(\omega)$, moduli)\footnote{$G(\omega)$ relates the stress to an imposed sinusoidal strain 
oscillating at frequency $\omega$, see \cite{barnes}.}, and the non-linear 
rheological behaviour (such as a strong dependence of the 
steady state viscosity $\eta$ upon an imposed shear rate $\dot\gamma$),
while aging and rheology would compete deep in the glass phase 
to produce interesting phenomena such as slow creep behaviour. 
However, all these different theoretical approaches have their strengths
and weaknesses, as we now briefly describe.

The trap model described in Sec.~\ref{whyaging} was extended into 
the `soft glassy rheology' (SGR) 
model~\cite{prlsollich,presollich}. Its simplicity makes it 
a nice tool to investigate complex thermo-mechanical
histories where aging and mechanical forcing compete~\cite{suzanne}, 
leading to several interesting predicted behaviours. Thus, 
the model is often used by experimentalists to rationalize 
non-trivial rheological behaviours commonly encountered in complex
materials, see e.g. \cite{rouyer}. 
 
The SGR model is a direct extension of the trap model, where 
each trap is characterized not only by its depth, $E$, but additionally
by a `strain' variable, $\ell$. The evolution equation now involves
$P(E,\ell,t)$, which generalizes $P(E,t)$ 
to include the statistical fluctuations of the 
strain variable. 
Using notations similar
to the ones used in Eq.~(\ref{trap_eq}), the dynamics of the 
SGR model is defined as:
\ba
\frac{\partial P(E,\ell,t)}{\partial t} & = & - \dot{\gamma} 
\frac{\partial P(E,\ell,t)}{\partial \ell}
- \Gamma_0 e^{- \beta (E - \frac{1}{2}k \ell^2)} P(E,\ell,t) \nonumber \\ 
& + & \Gamma(t) \rho(E) \delta(\ell), 
\label{sgr_eq}
\ea
where 
$\Gamma(t) = \Gamma_0 
\int d\ell \int dE \, e^{-\beta (E- \frac{1}{2}k \ell^2)} P(E,\ell, t)$. 
The first term represents the effect of the global shear rate,
$\dot{\gamma}$, in the absence of hopping between traps, namely
affine deformation of the traps. The second term 
describes the probability to leave the trap occupied at 
time $t$, and takes into account in a linear manner the fact that
shearing promotes hopping between traps by lowering the 
barrier heights, thereby 
defining an elastic constant, $k$. Note that activated dynamics
with an `effective temperature factor', 
$T_{\rm eff} = 1/\beta$, is assumed, even though its meaning in the context
of a driven colloidal system or emulsion is not clear~\cite{cates}. 
The last term now includes the new factor $\delta(\ell)$,
implying that after a hop, the newly found trap is unstrained.
Of course, for such a mean-field description (a single particle
hopping between traps), `strain', `shear rates', or `shear
stress' are just names and are not intended to carry 
physical information about a real three-dimensional flow.
As in the original trap model, 
an exponential form is adopted for the distribution 
of trap depths, $\rho(E)$~\cite{prlsollich}.
The global shear stress is defined by 
\be 
\sigma(t) = k \langle \ell  \rangle_P = k \int d\ell \int dE \, 
P(E,\ell,t) \ell ,
\ee
so that knowledge of $P(E,\ell,t)$ from solving 
Eq.~(\ref{sgr_eq}) allows one to predict any 
needed rheological quantity~\cite{presollich}.
The success of the SGR model stems partly from the
fact that depending on the value of the effective temperature, 
a broad variety of non-trivial, but experimentally realistic,
rheological behaviour can be predicted both for linear
and non-linear response. The model can therefore easily be used 
to fit a set of experimental data
by adjusting a few quantities, such as 
$T_{\rm eff}$ or $\Gamma_0$, see \cite{livingcell} for a 
biophysical example. Moreover, it is simple 
enough that extremely complex thermo-mechanical histories
can be easily implemented, and compared to experiments, see 
\cite{virgile} for an elegant illustration.
The weaknesses of the approach are the same as for the original 
trap model, as far as the interpretation of the traps 
in real space is concerned.  
Moreover, the lack of a spatial representation of the physics
implies that the different traps are completely independent from one another,
and the model cannot describe shear and kinetic heterogeneities. 
It would be interesting to develop spatial variations 
of the original SGR trap model to include
mechanically realistic interactions between traps \cite{hebraud}, 
as is also done in theoretical modelling of elasto-plastic 
response of amorphous solids~\cite{picard,bocquet2009}.   

Mean-field glass models can also be modified to include the physical
effect of an external mechanical forcing in the 
dynamics~\cite{horner,fabrice,ledoussal,BBK}. 
Since these models contain fully-connected Hamiltonians 
defined with no reference
to geometry, the modeling of an external flow is necessarily crude. 
In the case of a sheared glassy material, it is argued that 
the main effect of the imposed shear flow in the equations of motion 
is to inject energy into the system~\cite{jorgeaging}. 
Taking again the example of the 
$p$-spin model in Eq.~(\ref{pspin}), one now considers the driven dynamics
\be
\frac{\partial s_i(t)}{\partial t} = -\mu(t) s_i(t)
-\frac{\partial H}{\partial s_i(t)} + f_i^{\rm drive} (t) + \eta_i(t),
\label{langevindrive}
\ee
where $f_i^{\rm drive}(t)$ stands for an external driving force. 
A natural choice in this context is to consider a driving force which 
has a functional form similar to the $p$-spin interaction 
but involves coupling constants which contain an asymmetric part, so that
the resulting force cannot be derived from an energy function. The 
specific choice made in \cite{BBK} is 
\be
 f_i^{\text{drive}}(t) =  \sigma(t)
\sum_{\substack{j_1<\cdots<j_{k-1} \\
j_1,\cdots,j_{k-1} \neq i}} 
\tilde{J}_i^{j_1 \cdots \,j_{k-1}} s_{j_1} \cdots s_{j_{k-1}},
\label{force}
\ee  
with coupling constants that are random Gaussian variables of variance
$k!/(2N^{k-1})$, which are symmetrical about 
permutations of $(j_1, \cdots, j_{k-1})$, but are uncorrelated 
about permutations of  $i$ with any element of 
$(j_1, \cdots, j_{k-1})$.
With this particular choice of asymmetrical couplings and a constant
amplitude of the driving force, $\sigma(t) = \sigma$, a numerical
solution of the two-time dynamical equations of the form 
similar to Eq.~(\ref{eqs}) shows that the dynamics becomes stationary
for any $\sigma > 0$~\cite{ledoussal} and any temperature $T>0$ (even 
for $T<T_c$). 
Therefore, in the stationary state following a quench 
at time $\tw=-\infty$ in the presence
of a constant driving force, the 
dynamical equations become:
\begin{widetext}
\begin{equation}
\begin{aligned}
\frac{d C(t)}{d t}  = & -\mu  C(t) + \int_{0}^{t} d t' 
\Sigma (t -t')  C(t')  +  \int_{0}^{\infty} d t' 
\left[ \Sigma (t+t') C(t')  + D(t + t') 
R (t') \right], \\
\frac{d R(t)}{d t}  = & -\mu  R (t) + \int_{0}^{t} d t' 
\Sigma (t-t') R(t') , \quad \quad 
\mu  =   \, T +  \int_{0}^{\infty} d t' \left[  D(t') 
R (t') + \Sigma (t') C(t') \right] ,
\label{system}
\end{aligned}
\end{equation}
\end{widetext}
with kernels given by: 
\ba
D(t)  & = & \, \frac{p}{2} C(t)^{p-1} + 
\sigma^2 \frac{k}{2} C(t)^{k-1}, \\ 
\Sigma (t) & =  & \frac{p(p-1)}{2} C(t)^{p-2} R(t).
\ea

Mathematically, the asymmetry in the coupling constants of the driving 
force shows up in the expressions of the kernels, since only 
$D(t)$ contains the driving term proportional to $\sigma$, 
with no counterpart in the 
expression for $\Sigma(t)$. It can be shown 
formally~\cite{superjorge} that
detailed balance imposes specific symmetries of the kernels 
$D$ and $\Sigma$, which is indeed explicitly broken in 
Eqs.~(\ref{system}) by the term proportional to $\sigma$.
As in the aging regime, Eq.~(\ref{eqs}), these equations involve
both correlation and response functions, and are thus more 
difficult to solve than the equilibrium, high
temperature dynamics in Eq.~(\ref{schematic}).
However, the solution proceeds very much as in the aging 
regime~\cite{dynultra}.

The connection to rheological quantities is done using energetic 
considerations. Using Green-Kubo type of arguments, the viscosity
$\eta$ is related to the relaxation timescale obtained 
from the time decay of $C(t)$ from Eq.~(\ref{system}),
so that $\eta(T,\sigma) \sim \tau_\alpha$. The power dissipated by the 
driving force is then shown to be $\propto \sigma^2/\tau_\alpha
\sim \sigma^2/ \eta$, while
it is $\propto \sigma \dot\gamma = \sigma^2/\eta$ 
for a sheared fluid. Thus one 
identifies $\sigma$ as the `shear stress' for the present system,
and steady state constitutive rheological relations can 
readily be studied. As for the SGR model, the results 
show that the interplay between the glass transition physics
and external forcing induces strong shear-thinning behaviour
which depends quantitatively on the temperature, with 
experimentally relevant scaling laws for the 
viscosity across the $(\sigma, T)$ plane~\cite{BBK}.
For $T \le T_c$ for instance, a shear-thinning 
behaviour is predicted at low shear rates,
\be
\eta \sim \dot\gamma^{-\nu(T)},
\ee
where $\nu(T)$ is a temperature dependent exponent, 
with $\nu(T_c) \equiv \nu_c = 2/3$ and $\nu \to 1$ as $T \to 0$. This 
behaviour is typical of a `power-law fluid' in the 
rheological literature~\cite{larson}. 
The non-trivial shear-thinning
exponent $\nu_c = - 2/3$ at $T_c$ reveals a complex
interplay between thermal processes and mechanical forcing, while
in the low-$T$ limit the `natural' exponent $\nu = -1$ is recovered, 
as expected on dimensional grounds.
Above $T_c$, the following scaling form is obtained:
\be 
\eta(\dot\gamma,T) \simeq 
\eta_0(T) [ 1 + \dot\gamma / \dot\gamma_0]^{-\nu_c},
\label{carreau}
\ee 
where $\eta_0(T)$ is the equilibrium value of the viscosity
and $\dot\gamma_0$ is a constant. This form of scaling 
is well-known and commonly found in the rheology  literature~\cite{larson}, 
and  these predictions 
compare rather well with computer simulations~\cite{onuki,jllong1,jllong2},
and experiments~\cite{crassous,weeks_shear}.

As compared to the SGR model, the equations of motion 
(\ref{system}) are mathematically more involved, even 
in the steady state, so that it is technically more difficult to 
solve for the response of the system to more complicated histories, 
although this is a technical rather than a fundamental limitation.
A more fundamental difference, as already discussed in equilibrium
and aging situations, is the absence, at the mean-field level, of thermally
activated processes which would allow the system to visit 
the numerous deep metastable states that contribute to 
its configurational entropy between the mode-coupling and
Kauzmann transitions.  

Just as these activated processes explain why 
mode-coupling dynamic singularities are avoided in 
finite dimensional systems, 
activated processes could in principle affect the rheological behaviour
of glassy materials. As discussed above, including these processes 
in the context of mean-field glass theory is a challenging task. 
A phenomenological description in the framework 
of RFOT theory was recently proposed \cite{rheolub}, extending
in particular the validity of Eq.~(\ref{carreau}) to lower temperatures.
Note that distinct predictions were recently obtained 
in \cite{bbreview}.  
Staying at the mean-field level, however,  
it can be argued~\cite{pisa} that thermal activation should allow 
the system to visit metastable states with lower free energy.
In such a state, it can be  shown that 
the system must be forced above a finite level of stress 
in order to flow. In rheological terms, this means that the presence
of metastable states leads to materials
with a static finite yield stress (although dynamically there no
finite yield stress).
Very recent calculations in the replica framework
exposed in Sec.~\ref{scalingrfot} have confirmed the existence 
of a finite shear modulus originating from the existence 
of metastable states \cite{yoshino}.  
In a realistic description, yet to be developed, one should describe 
within a single 
framework the effect of shear and thermal activation within 
a complex energy landscape. 
 
Another difference with the SGR approach concerns the study 
of effective temperatures, which again mirrors the mean-field results
obtained in aging situations (see Sec.~\ref{mfaging}).
Indeed, many of the results obtained in aging systems  
about deviations from fluctuation-dissipation relations
can be shown to apply to the driven case as well. 
In particular, a fluctuation-dissipation ratio
can still be defined from Eq.~(\ref{eq_fdr}) leading to the notion
of effective temperatures for driven systems. The solution of
Eqs.~(\ref{system}) yields the time dependence of both 
correlation and response functions, from which a parametric FD
plot can be built. As for aging materials, it is predicted
that these FD plots conserve an equilibrium shape, being
piecewise linear, with 
each relaxation timescale being associated to its own 
value of the fluctuation-dissipation ratio and being interpreted 
in terms of an `effective equilibrium' for slow degrees
of freedom. 
An interesting feature is that effective temperatures
are predicted to occur even above the glass transition, provided
the driving force is large enough that non-linear rheological effects
are observed. Thus, a deep relation between anomalous response
functions and deviations from thermal equilibrium 
is established~\cite{BBK}.

These predictions were confirmed in a number of numerical 
studies of sheared supercooled liquids~\cite{jl,ilg,jllong1,jllong2,liuteff}, 
and the main physical result, the existence of an effective temperature
describing the relaxation of sheared glassy systems, 
now forms the basis of several recent phenomenological description
of the plastic deformation of amorphous 
solids~\cite{liuteff2,langerteff,falk3,eran}. 
We are aware of no experimental attempt to quantify 
violations of the fluctuation-dissipation theorem under 
stationary conditions created by a shear flow, although this 
should in principle be much easier than in non-stationary, 
aging  situations where several experiments have already 
been performed. 

We mentioned in Sec.~\ref{mfaging}
that mode-coupling theory had not been fully extended to deal
with nonequilibrium aging situations. By contrast, in recent years,
a large research activity has been dedicated to the 
derivation of mode-coupling approximations to deal with the 
rheology of glassy liquids and colloidal suspensions. 
A first derivation is obtained starting from generalized 
fluctuating hydrodynamic equations, as in field-theoretic derivations
of the equilibrium MCT~\cite{DasMazenkoprl}. Among the several 
approximations involved, the standard mode-coupling decomposition 
of four-point static correlations as product of two-point 
functions is performed, yielding closed dynamical equations  
for the time evolution of the intermediate scattering function under shear. 
For the case of a stationary simple shear flow with an imposed strain 
in the $x$-direction $\gamma(t)$, one gets the 
following dynamical equations~\cite{dave}:
\begin{widetext}
\ba
\frac{dF({\bf q},t)}{dt} &  = & - \frac{D {q}(-t)^2}{S(q(-t))} 
F({\bf q},t)   -  
\int_0^t dt' M({\bf q}(-t),t-t')\frac{dF({\bf q},t')}{dt'}, 
\label{mctdave} \\
M({\bf q},t) & = & \frac{\rho D}{2} \frac{q}{q(t)} 
\int d {\bf q'} V({\bf q}, {\bf q}') V({\bf q}(t), {\bf q}'(t))
F({\bf q}(t)-{\bf q}'(t),t ) F({\bf q}'(t),t), \nonumber
\ea
\end{widetext}
where ${\bf q}(t) = (q_x, q_y + \gamma(t) q_x, q_z)$ is 
a time dependent wavevector, the vertex function 
has the standard mode-coupling expression as in Eq.~(\ref{vertex2}), 
and $F(q,t) = \langle \rho_{{\bf q}(-t)} (t) \rho_{-{\bf q}} (0)  \rangle$ 
is the intermediate scattering function modified to 
take into account the global advection by the shear flow. Formally, 
these equations are very similar to the ones derived at
equilibrium, see Eqs.~(\ref{vertex1}, \ref{vertex2}).  
This implies that the physics captured by this approximation 
stems from the advection, and thus the distortion,   
of density fluctuations along the $x$-direction by the shear flow. Due to the
mode-coupling mechanism in Eq.~(\ref{mctdave}), relaxation of 
density fluctuations in the $x$-direction triggers the relaxation of all
the modes. In a parallel effort~\cite{fuchs,fuchs1}, very similar MCT dynamic 
equations for glasses under shear flow were derived employing the
technique of projection operators also used at thermal equilibrium
to derive mode-coupling equations. Here also, a similar
decoupling of four-point correlations into products of two-point quantities
is performed, and in its latest version~\cite{fuchs3} the final   
dynamical equations are very similar to the above expressions, 
Eqs.~(\ref{mctdave}), see \cite{fuchs3} for a detailed 
discussion of the technical differences between the two approaches. 
 
The rheological
behaviour in steady shear flow above the mode-coupling transition 
resembles the description in Eq.~(\ref{carreau}), although
the `simple' shear-thinning exponent $\nu=-1$ is found.
In the glass~\cite{fuchs}, the system develops a 
finite yield stress, which jumps discontinuously from zero 
when the mode-coupling transition is crossed. 
Below $T_c$, the rheological behaviour resembles that of a 
Bingham fluid~\cite{larson}.
We note that although fully connected glass models and schematic 
mode-coupling models are fully equivalent at thermal equilibrium, 
both approaches seem to predict different behaviours 
out of equilibrium. It is not yet clear whether these differences
are due to the set of approximations involved in the derivation
of Eqs.~(\ref{mctdave}), or to the very peculiar form of the non equilibrium 
drive (\ref{force}) chosen for mean field glass models, which 
may be unrealistic.
Both approaches suffer from the same fundamental
limitation that emerging `critical' or `universal' properties
near the mode-coupling transition will be drastically modified
in realistic numerical simulations or real experiments, 
as the mode-coupling singularity is not present in real 
materials, even in colloidal hard sphere systems 
where the theory was often applied~\cite{crassous}. 
This implies that all the problems and ambiguities
due to the `crossover' nature of the dynamic transition 
encountered by MCT at thermal equilibrium (see Sec.~\ref{current-MCT})
will be again present under shear, and the theory should fare 
with experimental data no better or no worse than at equilibrium.  

Although initially developed to study stationary shear flows, 
$\gamma(t) = \dot\gamma t$, MCT-based rheological equations 
have been now derived for arbitrary flows and 
shear histories~\cite{fuchs2,fuchsprl2008,fuchspnas2009}.
The resulting equations are more complicated than Eq.~(\ref{mctdave}),
involving in particular three-time memory functions
in the general case. The case of oscillatory 
simple shear flows was also studied~\cite{dave2}. 
The possibility to study time dependent shear flows makes the
MCT approach quite appealing, as it can thus compete 
with the flexibility offered by the SGR model described above, 
with the advantage that one is  
dealing with a microscopically realistic description 
of the liquid. However, it is possible to study the 
interplay of shear flow and aging in the SGR model~\cite{suzanne}, 
which is not yet feasible within MCT approaches. 

\begin{figure}
\psfig{file=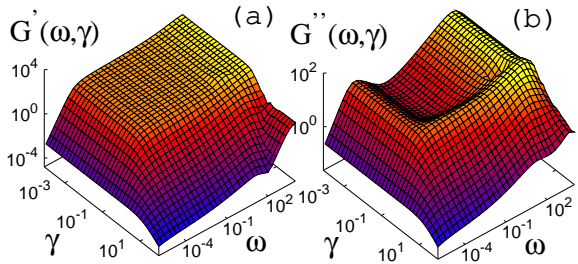,width=8.5cm,clip}
\caption{\label{kuni} Frequency and strain dependence of (a) the storage
modulus and (b) the loss modulus for a hard sphere suspension near the 
mode-coupling glass transition. 
The broad viscoelastic spectra emerging in fluids near the 
glass transition is affected in a non-linear manner 
by a large external shear flow which accelerates the dynamics.
From \cite{dave2}}
\end{figure}

As concrete applications of the theoretical framework, 
step strain protocols~\cite{fuchs2}
or sinusoidal shear flows of arbitrary amplitudes were
studied~\cite{dave2}.  In Fig.~\ref{kuni}, we show the 
behaviour obtained for the storage and loss moduli of
a dense suspension of hard spheres close
to the colloidal glass transition as a function of frequency
(as in linear viscoelasticity) and strain amplitude
(as in nonlinear steady shear flows). These results illustrate
both the broad viscoelastic spectra emerging in fluids near the 
glass transition, and the competition between intrinsic 
slow dynamics and the external perturbation imposed by the shear flow
which accelerates the dynamics.

The issue of a nonequilibrium extension to the fluctuation-dissipation
theorem in shear flows was also discussed within 
MCT. A formal derivation of the Einstein relation between self-diffusion
and mobility under shear flow is presented in  \cite{szamelfdt}, while
FDT violations are acknowledged~\cite{dave}, but do not play a role 
during the derivation of Eqs.~(\ref{mctdave}) to arrive at
dynamical equations involving correlation functions only, as opposed
to coupled equations for response and correlations
in Eqs.~(\ref{system}). Very recently, an approximate expression for 
time dependent susceptibilities was obtained \cite{kruger}
and compared to the correlation obtained from the solution 
of Eq.~(\ref{mctdave}).
In standard equilibrium MCT, the response function is obtained 
{\it a posteriori} using the FDT. 
Complementary approaches, like field theory are plagued 
by difficulties related to maintaining FDT in a self-consistent manner. 
As a consequence, the status of the approximation in \cite{kruger} is not 
entirely clear at this stage.
Moreover, an important difference is found with mean-field results
since a nearly constant value of the FDR, $X=1/2$,  
is predicted \cite{kruger} in contrast to mean-field models where the FDR
is generically dependent of the state point, with $X \sim T$ at low
temperatures~\cite{reviewLeticia}, as also 
found in numerical work~\cite{angelani,angelani2,prlsilica}.
This suggests that still more work is needed to clarify the 
status (derivation and physical behaviour) 
of non-equilibrium effective temperatures in glassy 
systems under shear.

\subsection{Current status of nonequilibrium studies}

Aging studies have a long history in the field of the 
glass transition \cite{struik}, since glasses are all by definition
nonequilibrium aging materials. It is only since the 90's, however,
that a large body of work has been performed to describe 
nonequilibrium glasses at a more fundamental level, using 
tools and concepts from statistical mechanics, as briefly reviewed 
above. Despite much conceptual progress, it is not clear whether 
these theoretical advances have led to a better description 
of the physical properties of, say, polymer glasses 
under complex thermo-mechanical histories, and thus
have had concrete experimental consequences \cite{mckenna2}. It certainly
remains true that a large number of experiments have been devised
to analyze two-time response and correlations functions in a number
of materials, in order to test theoretical predictions, although
it appears that less studies of aging 
materials were performed in more recent years.  

We wish to suggest that recent theoretical and experimental 
work to characterize and understand memory
and rejuvenation effects in spin glasses could be fruitfully 
revisited in the field of structural glasses, see 
the experimental chapters in \cite{youngbook}. 
In particular, a clear link between memory effects and typical 
lengthscales over which the slow dynamics takes place was 
established in the context of spin glasses 
\cite{microT,berthierbouchaud,chaos2}.
We have discussed in previous sections that dynamic  
lengthscales also depend very sensitively on temperature
in structural glasses, as illustrated for instance in Fig.~\ref{cecile}. 
At present, it is not really known how these ideas 
apply to aging of structural glasses and whether they would yield 
memory effects similar to the ones observed in spin glasses. 
In the same vein, very little work was performed to understand
how the mosaic picture of RFOT \cite{rfotaging,bbreview} or 
the frustration-limited domains discussed in Sec.~\ref{sectiongilles} 
would respond to complex thermal histories, nor has 
mode-coupling theory be fully developed to handle non-stationary dynamics
far from equilibrium except in the context of 
mean-field glassy systems \cite{CugKur3}.

More work should also certainly be performed to characterize
dynamic heterogeneities in nonequilibrium systems. Indeed, 
when experiments were devised to detect for instance 
two-time correlation functions in aging materials, it became
immediately clear that slow dynamics in these far from 
thermal equilibrium conditions 
was also characterized by intriguingly large
dynamic fluctuations \cite{lucalaurence} 
showing up in the form of intermittent
time signals \cite{Bellon1,Buisson1,luca,lequeux}, 
very large spatial correlations \cite{lucaaging2}, 
peculiar forms of relaxation functions \cite{lucaaging1,bellour,leheny2},
or superdiffusive processes \cite{leheny1,onion}.
Although well-documented experimentally and displaying 
some form of `universality', these non-trivial dynamics 
have received too little attention from the theoretical
community, see however \cite{pitard}.
It is in particular intriguing that these effects are apparently not
observed in computer simulations where dynamic heterogeneity
seems to proceed roughly as under equilibrium conditions 
\cite{giorgioaging,castillo,horacio,djamelaging}.
In this context, an interesting development is the extension 
of some of the concepts derived from mean-field spin glass models, 
in particular the notion of time reparametrization invariance
of the dynamical equations of motion, to finite dimensional 
aging materials \cite{malcom,chamon}. In particular, this approach
naturally explains the appearance of nontrivial 
mesoscopic dynamic fluctuations in aging materials, and provides
specific new tools to analyze a number of physical 
quantities, such as distributions of time correlation functions
\cite{reviewleticia}.

In striking contrast with aging studies, the rheology of soft glassy materials 
has been increasingly actively studied over the last decade, and 
has developed as a research field on its own that will certainly continue
to expand in the coming years. 
Thus, we close this prospective section 
with just a few selected issues on the rheology of glasses
that are currently the object of 
intense research. First, since all the above mentioned theoretical 
modeling of glassy rheology are somehow `mean-field' in nature, 
it is not clear how spatial heterogeneities or correlations 
can be described theoretically.  
Yet, experiments~\cite{coussot,becu,ianni} 
and simulations~\cite{varnik,johnson,falk1}
clearly reveal for instance that 
soft glassy materials commonly display the phenomenon of 
shear-banding. Namely,
when submitted to a macroscopic shear force, the system 
spontaneously `phase separates' between a flowing state
supporting the shear, and an immobile state with no flow. 
This observation means that at least two dynamical states 
exist for a given level of external forcing, so that
the flow curve $\sigma(\dot\gamma)$ is multi-valued, but there
is presently no agreement on the microscopic origins
of this observation~\cite{varnik,pisa,manning,fielding,bandsHS}.
A second relevant question related to spatial aspects
is the subject of dynamic heterogeneity. We have underlined in 
Sec.~\ref{dh} the importance of spatial fluctuations of the 
dynamic at thermal equilibrium, and have described in 
Sec.~\ref{theory} how different theories describe these
fluctuations. Much less is known under shear, although 
numerical simulations have revealed the existence 
of large scale heterogeneities under flow~\cite{furukawa} that 
are in fact strongly reminiscent of the plastic 
flow of very low temperature glasses \cite{falk,tanguy,plastic3}.  
Further studies of these issues could also shed some light on the flow
of soft glassy materials under confinement, a situation
of great interest due to the rapid development of microfluidic
techniques and applications \cite{goyon,besseling}.

\section{Some general and concluding remarks}

\label{nofuture}

We conclude this long review with few remarks on some key topics that 
are often debated in the literature and on which we would like to present
our point of view.

\subsection{Growing lengthscale(s)} 

For a long time, the research on the 
glass transition has been focused on timescales more than lengthscales. 
The reason is simple: the former are clearly increasing 
rapidly approaching 
$T_g$, whereas the latter remained elusive for  a long time. Recently, 
this state of 
affairs has changed. First, the whole
topic of dynamical heterogeneity made it clear that a complete theory 
of the glass transition has to be able to explain growing dynamical 
lengthscales. 
Furthermore, there are general theoretical, and not just phenomenological
reasons to believe that growing lengthscales should play an important role. 
In fact, a system with a finite number, say $N$, of degrees of freedom is 
expected to 
have a relaxation time no larger than $\exp(KN)$, where $K$ is independent of N and 
possibly dependent on temperature, but in a non singular way except at $T=0$. 
A system, whose largest correlation length is $\xi$, can be viewed as a collection of independent
sub-systems of linear size $\xi$. The relaxation time $\tau$ is therefore equal to the one
of a given sub-system and therefore cannot be larger than $\exp(K\xi^d)$.
This implies that a large relaxation times should be related to large 
numbers of spatially correlated degrees of freedom. 
This intuition was made rigorous in 
\cite{MontanariSemerjian},
where it is shown 
that indeed the point-to-set length defined in Sec.~\ref{scalingrfot} has to 
grow at least as fast 
as $c(\log \tau_\alpha)^{1/d}$, where $d$ is the spatial dimension and 
$c$ a proportionality 
constant. Since $c$ actually depends on temperature (as $K$ in the 
previous expression)
this inequality makes a growing length a necessity only if the 
relaxation timescale diverges
at a finite temperature. If not, one should evaluate the constant 
and check whether the inequality
indeed implies a lengthscale that indeed becomes `large'
at low temperature.
In any case, all these results, and many others, have been so 
influential in stressing the importance
of possibly growing lengthscales that by now many -- possibly too many -- 
lengthscales have been
defined and studied in the literature 
\cite{leidenbook}. The more pressing open questions 
in this field concern the relation between the well-studied 
dynamical lengthscales (e.g. $\xi_4$), and the more 
recently devised static point-to-set lengthscales.  
Are these two types of spatial correlations related? Do
structural correlations drive dynamical ones? 
What is their precise relation with the overall increase of the 
viscosity? Is there a unique way of defining 
static correlation lengthscales, and are point-to-set correlations
equivalent to alternative, more 
geometric ways of defining static lengthscales 
\cite{gillesreview,coslovich,jorgelength,tanaka}?
We believe that the intensive study of dynamical 
lengthscales in the past decade will be continued by a similar
intensive search of static correlations in future work
to disentangle all these issues.

\subsection{Glass and jamming transitions}

We introduced in Sec.~\ref{jammingsubsection}
the idea that many different materials undergo
a fluid to amorphous solid transition reminiscent of the glass transition 
of molecular liquids, and indeed we included experimental or numerical
results on colloids or granular materials without further discussion
in the rest of the article.  
It is time to discuss more critically the assumption that 
all these materials `jam' in a similar way.

From a practical point of view, one should distinguish 
two distinct `solidity' transitions that can both be observed 
in the system of hard spheres, which we mentioned several times 
in this review. At thermal equilibrium at temperature 
$k_B T$~\footnote{For hard spheres, the temperature simply sets the 
microscopic timescale since it only affects the microscopic 
timescale. As long as the 
system is in equilibrium, the thermodynamic and dynamical properties 
at different temperatures are identical up
to a trivial rescaling.}, hard spheres undergo in three dimensions 
a glass transition in the regime $\varphi_g \approx 0.57 - 0.59$. 
Above this transition, the system appears as a solid, at least
on experimental timescales. However, this system 
is compressible (it is a hard sphere glass), its pressure
is finite and its equation of state,  $Z \equiv  P(\varphi) / \rho k_B T =
Z(\varphi)$ is a smooth function of $\varphi$ across $\varphi_g$.
At present, there no indication that the 
physics of this first fluid - solid transition in the hard sphere system
at finite temperature 
is any different from the glass transition observed, say, in a Lennard-Jones
liquid. This means that all the concepts and theories reviewed in this 
paper are actually relevant to this situation. For instance, 
experimental evidence suggests that
hard spheres undergo a change from MCT-like behaviour to an 
activated regime when increasing $\varphi$ \cite{LucaaboveMCT}, they display
a similar spatially heterogeneous dynamics \cite{weeks}, 
and display in analytical calculations 
a mode-coupling \cite{barraths} and Kauzmann
transitions \cite{ParisiZamponireview} completely 
analogous to model liquids. Interestingly, the same 
phenomenology is found in dense granular materials {\it driven} for instance
by cyclic shear \cite{dauchot} or air flow \cite{durian}. In these cases, 
a non-equilibrium steady state is reached thanks to 
the mechanical driving, which plays a role similar to Brownian forces
in colloids. When increasing the density or decreasing the strength 
of the driving, these systems appear to display a `granular glass 
transition' with properties
which are again very similar to the ones observed for supercooled liquids, 
even at the most microscopic level \cite{candelier1}.  

A second, distinct 
`solidity' transition occurs in hard spheres out-of-equilibrium. 
As discussed above, it is not possible to compress hard spheres
above a certain density and keep the system in equilibrium. This does not mean 
the system cannot be compressed anymore, it can, but in a non-equilibrium and
protocol dependent manner. This second `solidity' transition takes place 
when the 
system cannot be compressed anymore. This is mainly a geometrical 
problem and thermal energy plays no role in this transition. 
For a three dimensional system, this occurs near `random close packing' 
$\varphi_{\rm rcp} = 0.63-0.65$ \cite{bernal}. 
At this density, the compressibility vanishes, the reduced pressure 
$Z(\varphi)$ diverges, and the number of contacts per particle 
is exactly equal to the minimal number required for
the system to behave mechanically as a solid 
\cite{alexander}. This second transition 
is thus directly relevant to understand the static properties of 
granular materials. It is interesting also for systems made of 
large (athermal) particles such as foams and emulsions: because 
the particles are soft, these systems can be compressed 
above $\varphi_{\rm rcp}$ \cite{jamming}. 
In these soft systems, the (osmotic)
pressure in the solid phase is now proportional 
to the particle surface tension, and thus solidity is driven by the elasticity
of the particles rather than temperature. From a dynamical point of view,
much less is known about this second transition. Connections with the 
glass problem 
are still rather speculative but they are the focus on an intense 
research activity, 
triggered by the seminal contribution of \cite{liunagel}. For instance,
dynamic heterogeneity at $T=0$ near random close packing has only been 
studied very recently \cite{lechenault2,trappe4,claus}, 
and important differences
with the dynamics of viscous liquids were noted. Therefore, although 
glass and jamming transitions might be observed in 
the same system (say, hard spheres), they likely correspond to two distinct
ways for the system to become a solid.

\subsection{Metastability and the role of the crystal} 

In this review we 
have discussed very little the role of the crystalline state. 
As many (but not all) researchers, 
we have assumed that the crystal does not play an important role for the 
glass transition phenomenon, apart from the fact 
that crystal nucleation 
has to be avoided  by supercooling. 
(For all known glass-formers, the melting temperature is larger than 
the experimental glass transition.)
This may be questionable for several 
reasons. 
A first objection is that equilibrium thermodynamic theories of the glass 
transition are problematic because the true
thermodynamic phase is indeed the crystal. 
This is not a real concern: supercooled 
liquids are in a long living metastable state. 
As long as the nucleation time is much larger than the structural relaxation 
time, $\tau_\alpha$, they can be considered as 
{\it bona fide} equilibrium states, at least using  
Feynman's definition: ``When all fast things have 
happened and all slow things have not, then the system is in equilibrium'' 
\cite{feynmanbook}. After all, the people who value diamonds  
do not worry that they will turn into graphite anytime soon. 
A more serious concern is that the relaxation time 
of the supercooled liquid cannot really diverge at a 
finite temperature if the thermodynamically stable state
is a crystal, because  the nucleation time is 
necessarily finite at finite temperature if 
we assume a finite Gibbs free energy difference between the crystal and the 
supercooled liquid\footnote{This is due to the fact that the maximum 
barrier to form the critical nucleus and 
also for its subsequent growth is necessarily finite, thus nuclei will 
form and expand even though these processes can be extremely slow. In 
practice, crystal nucleation can be much slower than what is naively 
expected, see \cite{cavagna-review}.}. Thus, $\tau_\alpha$
will inevitably hit the (finite) nucleation time before diverging. Below this 
temperature the supercooled liquid
is no more a metastable state because it nucleates the crystal before 
actually being able to relax its structure.     
This, however, is also not necessarily 
a serious problem. First, not all theories 
are based on a divergence at finite temperature. Second, 
several theories explain the slowing down 
of the dynamics by the proximity to a phase transition but none of them 
needs that the transition actually takes place to be proven correct.
Thus, although
important conceptually, the existence of the crystal does not imply 
that theories with no singularity should be preferred
to describe the physically relevant temperature 
regime where nucleation is unimportant.
A final, more physical, reason to take into account the crystal in 
explaining the glass transition would be that the slow
dynamics in the supercooled state is due to the existence
of some local order reminiscent of the crystal structure,
as in frustration-limited domain approaches \cite{gillesreview}, 
and reemphasized
in several recent numerical papers \cite{coslovich,tanaka,dyrepeter}.

\subsection{The `ideal' glass transition} 

This is certainly a recurrent, probably utopian, but nevertheless
      fascinating, topic for discussion. Several theories explain the
      slowing down of the dynamics by the proximity to a phase transition.
      None of them requires that the transition actually takes place.
      This holds 
  either because the transition is avoided by construction as in the
      frustration-limited domain theory, or because is not accessible
      experimentally as in dynamical facilitation theory or random first
      order transition theory. What happens at lower temperatures, where
      no real system can be equilibrated, is after all just a matter of
      curiosity. It is however interesting, just for a short while, to
      dwell on what an ideal glass transition could possibly be. It should
      correspond to a true divergence of the relaxation time and the
      viscosity at a finite temperature $T_{\rm ideal}$. Because of the
      proof by Montanari and Semerjian for lattice models, 
      we now strongly suspect that static
      correlations, actually point-to-set correlations, would generally
diverge
      at the transition\footnote{If the relaxation timescale only diverges at zero temperature but faster than Arrhenius then 
      the ideal glass transition would take place at $T=0$. Again, on the basis of the bound of Montanari and Semerjian, we  would expect a diverging length in this case too.} \cite{MS0}. 
      Thus, the transition would correspond to the
      development of some long-range order (likely amorphous order), since
      suitable boundary conditions fix the density field in an amorphous
      configuration in the entire (infinite) sample. 

It is also important
      to discuss what properties an ideal glass transition would not
      display. Actually, it is sometimes assumed that at an ideal glass
      transition, as discussed in the context of theories in Sec.~\ref{theory},
 the dynamics would be completely arrested. This is not
      necessary and, actually, impossible if $T_{\rm ideal}>0$. For example, a probe particle will still be able to move at
      $T_{\rm ideal}$ and below. The situation would be similar to the
      crystalline state, where particles diffuse, although very slowly,
      leaving the crystalline order intact. What would diverge at $T_{\rm
      ideal}$ is only the time to destroy the amorphous correlation in the
      density field.
For example, at the ideal glass transition advocated in RFOT theory, 
the density field
orders in one of the possible amorphous low temperature configurations, so that
the time to relax the structure becomes infinite whereas, instead, the
self-diffusion coefficient stays finite. Also, it is sometimes believed that 
the ideal
glass transition is related to a fragmentation of the configuration space for
{\it finite} size systems, i.e. that below a certain temperature or above a
certain density it is no more possible to
go from any typical equilibrium configuration to another one. This is clearly
not true, as it can be seen easily for soft particle models, and actually
even for hard spheres. 
This can be harder to prove for some effective models \cite{procaccia}. In
any case,
this is not what the ideal glass transition would be: such kind of dynamical
arrests can only take place in effective theories whose domain of applicability
must break down at a some point (maybe not accessible in experiments).

Thus, the theoretical possibility of a finite temperature 
`ideal' glass transition towards a genuine `glass' state exist,
and will no doubt continue to obsess many physicists in
coming years. 

\subsection{Final words}

The problem of the glass transition, already very exciting in itself,
has ramifications well beyond the physics of supercooled liquids. 
Glassy systems figure among the even larger class of 
`complex systems'. These are formed by a set of interacting 
degrees of freedom that
show an emergent behaviour: as a whole they exhibit properties
not obvious from the properties of the individual parts.
As a consequence the study of glass-formers as statistical mechanics models
characterized by frustrated interactions is a fertile ground to develop new
concepts and techniques that will likely be applied to other physical, 
and more generally, scientific situations.
The glass transition in supercooled liquids can in fact be considered 
as one of the simplest situations where frustration, geometry, ergodicity,
and disorder compete to produce a glassy state. Hence, the concepts 
reviewed in this article should be directly applicable to
scores of more complex systems, such as for instance 
physical gels, liquids in confined geometries, diffusion in crowded 
biological environments, dense granular media, 
self-assembly, or microfluidic flows of dense emulsions 
and colloidal suspensions.
Thus, we certainly expect more progress to emerge
in the future along these interdisciplinary routes. 

\acknowledgments

The views and analyses we have exposed in this paper are 
the results of interactions with a large number of 
researchers in this field. We thank in particular our collaborators  
on these subjects: we believe our joint work 
largely contributed in one way or another to this contribution.
Our work was supported by the ANR Grants CHEF, TSANET and DYNHET.
GB wishes to thank AA for her (almost) infinite patience without 
which this work and many others would not have been completed.

\end{document}